\documentclass[letterpaper,11pt,twoside]{report}
\usepackage{mathptmx}  
\usepackage{graphicx,
}
\usepackage{latexsym, epsfig, theorem, amsmath, amsxtra}
\usepackage{amsbsy, amssymb, 
amsfonts, euscript, longtable}
\usepackage[latin1]{inputenc}  
\usepackage[T1]{fontenc}       
\usepackage{yfonts}
\usepackage[dvips]{color}
\usepackage[rigidchapters]{titlesec}
\usepackage{times}

\RequirePackage{amssymb}                

 \font \nchap = cmti10 at 80pt  

 \titleformat{\chapter}[display]
  {\bfseries\Large}
  {\filleft\MakeUppercase{}
   \nchap\thechapter }
  {4ex}
  {\titlerule[1pt]%
   \vspace{2pt}%
   \titlerule
   \vspace{1pc}%
   \filright}
  [\vspace{2ex}%
   \titlerule
   \titlerule]

 \titleformat{\section}[block]
 {\large\bfseries\filright}{\fbox{\itshape\thesection}}{0.5em}{}

 \titleformat{\subsection}
  {\normalsize\bfseries\itshape\filright}{{\itshape\thesubsection}}{1em}{}

\renewcommand{\chaptermark}[1]%
               {\markboth{#1}{}}

\renewcommand{\sectionmark}[1]%
              {\markright{\thesection\ #1}}

\begin{document}




\setcounter{page}{1}
\renewcommand{\thepage}{\roman{page}}

\bigskip
\bigskip
\bigskip
\bigskip

\centerline{\Large \bf $\quad$ INSTITUTO POTOSINO DE
INVESTIGACI\'ON}
\bigskip
\centerline{\Large \bf CIENT\'IFICA Y TECNOL\'OGICA}

\bigskip
\bigskip
\bigskip
\bigskip
\bigskip

\centerline{\Large \bf $\quad$ DIVISI\'ON DE MATEM\'ATICAS
APLICADAS}
\bigskip
\centerline{\Large \bf Y SISTEMAS COMPUTACIONALES}

\bigskip
\bigskip
\bigskip
\bigskip
\bigskip

\centerline{\huge $\quad \,\,$  MATHEMATICAL METHODS OF
FACTORIZATION}
\bigskip \centerline{\huge $\quad$ AND A FEEDBACK APPROACH} \bigskip
\centerline{\huge $\quad \,$ FOR BIOLOGICAL SYSTEMS}

\bigskip
\bigskip
\bigskip
\bigskip
\bigskip

\centerline{\Large \bf $\quad \,\,$ PH. D. THESIS IN APPLIED
SCIENCES}

\bigskip
\bigskip
\bigskip
\bigskip
\bigskip

\centerline{\Large \bf $\quad \,\,$ OCTAVIO CORNEJO-P\'EREZ}

\bigskip
\bigskip
\bigskip
\bigskip
\bigskip

\centerline{\Large \bf $\quad \,\,$ SUPERVISORS:}
\bigskip \centerline{\Large \bf $\quad \,\,$ DR. HARET
CODRATIAN ROSU BARBUS}
\bigskip \centerline{\Large \bf $\quad \,\,$ DR. ALEJANDRO RICARDO FEMAT-FLORES}

\bigskip
\bigskip
\bigskip

\centerline{\Large \bf $\quad \,\,$ SAN LUIS POTOS\'I, S. L. P.,
MEXICO}
\bigskip \centerline{\Large \bf $\quad \,\,$ SEPTEMBER 20th, 2005}

\chapter*{Acknowledgments}

I am grateful to my parents, brothers and close relatives for
their permanent support not only during my doctoral studies but
also during all my life till now.

\bigskip
I am also very grateful to my thesis advisors, Dr. Haret C. Rosu
Barbus and Dr. Ricardo Femat for everything I have learned from
them and for their support, collaboration and friendship.

\bigskip
I thank Drs. J. Socorro Garc\'{\i}a-D\'{\i}az, Rom\'an
L\'opez-Sandoval, El\'{\i}as P\'erez-L\'opez and Marco A.
Reyes-Santos, for their kindness and availability for reviewing
this document, as well as for their comments and useful remarks on
the present work that helped me to improve it.

\bigskip
I would like to acknowledge the authorities of IPICYT for the
excellent working conditions that allowed me to achieve good
progress in my doctoral investigations.

\bigskip
Last but not the least, I would like to thank all my IPICYT
friends from all the areas of research. Special mentions go to
Eugenia (Maru), Luis Adolfo (my Brother), P\'anfilo (the Sevillian
Panfilote) and Vrani (the Dane).

\bigskip
And of course nothing would have been possible without the
financial support from CONACYT.

\bigskip
To all the people and institutions I mentioned here, once again
Thank You.

\bigskip
\bigskip
\hfill Octavio


\chapter*{Abstract}

This thesis presents the original results I have obtained during
the three-year doctoral period in the Divisi\'on de Matem\'aticas
Aplicadas y Sistemas Computacionales (DMASC) of the Instituto
Potosino de Investigaci\'on Cient\'{\i}fica y Tecnol\'ogica
(IPICYT), in San Luis Potos\'{\i}, M\'exico. These results have
been obtained under supervision and collaboration of Dr. Haret C.
Rosu in what refers to the first part of the thesis, and of Dr.
Ricardo Femat for the second part.

\medskip
The first part deals with some types of factorization methods that
we were able to develop and that lead us to particular solutions
of travelling kink type for reaction-diffusion equations and also
to more general nonlinear differential equations of interest in
biology and nonlinear physics. We also applied supersymmetric
approaches in the context of biological dynamics of microtubules
and the related transport properties associated to their domain
walls. In addition, a complex supersymmetric extension of the
classical harmonic oscillator by which we obtain new oscillatory
modes has been developed; results that could be extended to
physical optics and the physics of cavities. Moreover, an
application to chemical physics of diatomic molecules using
supersymmetric and factorization procedures is developed.

\medskip
The second part contains a detailed study on the synchronization
of the chaotic dynamics of two Hodgkin-Huxley neurons, by means of
the mathematical tools belonging to the geometrical control
theory. Despite using different parameters for each of the two
neurons our analysis shows that synchronization states are
achieved. The synchronization is attained by the feedback
structure of the interconnection (coupling). Numerical results for
the obtained neuronal dynamical states are displayed.




\chapter*{Resumen}

Esta tesis presenta los resultados originales que he obtenido
durante los tres a\~nos de periodo doctoral en la Divisi\'on de
Matem\'aticas Aplicadas y Sistemas Computacionales (DMASC) del
Instituto Potosino de Investigaci\'on Cient\'{\i}fica y
Tecnol\'ogica (IPICYT), en San Luis Potos\'{\i}, M\'exico. Estos
resultados se han obtenido bajo la supervisi\'on y colaboraci\'on
del Dr. Haret C. Rosu en lo referente a la primera parte de tesis,
y del Dr. Ricardo Femat para la segunda parte.

\medskip
La primera parte trata con algunos m\'etodos de factorizaci\'on
que fuimos capaces de desarrollar y que nos condujeron a
soluciones particulares del tipo \textit{kink} viajeras para
ecuaciones de reacci\'on-difusi\'on y tambi\'en para ecuaciones
diferenciales no lineales m\'as generales de inter\'es en
biolog\'{\i}a y f\'{\i}sica no lineal. Se aplicaron tambi\'en
t\'ecnicas de supersimetr\'{\i}a en el contexto de din\'amica
biol\'ogica de microt\'ubulos y las propiedades de transporte
asociadas a sus paredes de dominio. En adici\'on, se desarroll\'o
una extensi\'on supersim\'etrica compleja del oscilador arm\'onico
cl\'asico por el cual obtuvimos nuevos modos de oscilaci\'on;
resultados que pueden extenderse a \'optica f\'{\i}sica y la
f\'{\i}sica de cavidades. Adem\'as, se desarroll\'o una
aplicaci\'on a la fisico-qu\'{\i}mica de mol\'eculas diat\'omicas
usando procedimientos de supersimetr\'{\i}a y de factorizaci\'on.

\medskip
La segunda parte contiene un estudio referente a sincronizaci\'on
de la din\'amica ca\'otica de dos neuronas Hodgkin-Huxley, en
donde se han aplicado los m\'etodos matem\'aticos pertenecientes a
la teor\'{\i}a de control geom\'etrico. Aunque se han utilizado
diferentes par\'ametros para cada una de las dos neuronas, nuestro
estudio muestra que se obtienen estados din\'amicos de
sincronizaci\'on. La sincronizaci\'on se logra por la estructura
de retroalimentaci\'on de la interconexi\'on (acoplamiento). Se
muestran los resultados num\'ericos para los estados de din\'amica
neuronal obtenidos.

%



\chapter*{Preface}

Scientific research and technological progress are important
characteristics of the modern world. They represent fundamental
activities that can help mankind to understand and transform
nature with the purpose of improving standards of life.

Almost three years have past since I started my doctoral degree
activity with the hope to contribute myself to the worldwide
scientific knowledge. The lines of research I chose were on the
border between mathematics and biology because I was convinced
that the interdisciplinary activity is very rewarding and could
give me better perspectives.

The doctoral thesis consists of four parts, of which the first
contains five chapters and is devoted to factorization methods of
differential equations and their applications in biology and
physics,
whereas the second part is divided in two chapters and deals with
the synchronization phenomena as studied in neuronal ensembles.
The thesis ends up with a final conclusion and the bibliography
presented in Parts III and IV, respectively.

\medskip
The first chapter is a general presentation of the factorization
methods for linear second order differential equations. Also, the
organization for Part I of the thesis is presented.

\medskip
The second chapter contains an original result for performing
factorizations of second order differential equations with
polynomial nonlinearities that has been reported in a paper
published in Physical Review E in 2005. At the same time the novel
procedure allows to obtain particular solutions of travelling kink
type in a very efficient way.

\medskip
The third chapter presents more applications of the method to more
complicated nonlinear differential equations. The results of this
chapter are published in Progress of Theoretical Physics in 2005.

\medskip
In the fourth chapter, I included the results of a supersymmetric
factorization model in the context of microtubules that we
published in Physics Letters A in 2003.

\medskip
The fifth chapter refers to the original results that have been
published in Journal of Physics A in December of 2004. A complex
extension to the classical harmonic oscillator based on a
supersymmetric factorization procedure that has been applied
before in particle physics is introduced in this chapter. The
application of the same method to the case of Morse potential, a
well-known exactly solvable problem in quantum mechanics with many
applications in the physics and chemistry of diatomic molecules is
also included here; these results are published in Revista
Mexicana de F\'{\i}sica, 2005.

\medskip
With the sixth chapter starts the second part of the thesis. Some
remarks on the kink type results obtained through factorization
methods in the first part for pulse propagation along neuron
axons, and the connection with the synchronization dynamics of a
minimal ensemble of two neurons, employing nonlinear control
theory are presented.

\medskip
In the seventh chapter, we focus first on synchronization
phenomena from the standpoint of their role and importance in
natural and technical systems. The concept of chaos and the
presence of chaotic behavior in nature are also described. Next,
synchronization methods for the control of chaos and their
applications in biological systems are shortly reviewed. The
problem of the synchronization of two Hodgkin-Huxley (HH) neurons
is emphasized because of its possible implications in the
dynamical
processes of the brain. 
A brief discussion of the widely known HH mathematical model of
the neuron is given. Also, in the Introduction section, the
organization of Chapters 7 and 8 belonging to Part II of the
thesis is presented.


\medskip
In the eighth chapter, numerical results for the synchronized
dynamics of two HH neurons are presented. The mathematical methods
employed belong to the theory of geometrical nonlinear control and
are used with the goal of studying the synchronization of two HH
neurons that are unidirectionally coupled. These results are
published in Chaos, Solitons and Fractals in July 2005.

\bigskip

The order of published papers in this thesis is the following:

\medskip

{\bf Chapter 2}.  H.C. Rosu, O. Cornejo-P\'erez,
\textit{Supersymmetric pairing of kinks for polynomial
nonlinearities}, Phys. Rev. E {\bf 71}, 046607 (2005).\\

{\bf Chapter 3}.  O. Cornejo-P\'erez, H.C. Rosu, \textit{Nonlinear
second order ODE's: factorizations and particular solutions},
Prog. Theor. Phys. {\bf 114}, 533 (2005).\\

{\bf Chapter 4}.  H.C. Rosu, J.M. Mor\'an-Mirabal, O. Cornejo,
\textit{One-parameter nonrelativistic supersymmetry for
microtubules}, Phys. Lett. A {\bf 310}, 353 (2003).\\

{\bf Chapter 5}.  H.C. Rosu, O. Cornejo-P\'erez, R.
L\'opez-Sandoval, \textit{Classical harmonic oscillator with
Dirac-like parameters and possible applications}, J. Phys. A {\bf
37}, 11699 (2004). O. Cornejo-P\'erez, R. L\'opez-Sandoval, H.C.
Rosu, \textit{Riccati nonhermiticity with application to the
Morse potential}, Rev. Mex. F\'{\i}s. {\bf 51}, 316 (2005).\\

{\bf Chapter 8}.  O. Cornejo-P\'erez, R. Femat,
\textit{Unidirectional synchronization of Hodgkin-Huxley neurons},
Chaos, Solitons and Fractals {\bf 25}, 43 (2005).\\




\tableofcontents  \newpage




\setcounter{page}{1}
\renewcommand{\thepage}{\arabic{page}}         




\listoffigures    


{\bf Fig. 2.1}: The front of mutant genes (Fisher's wave of
advance) in a population and the partner susy kink propagating
with the same
velocity. The axis are in arbitrary units.\\

{\bf Fig. 2.2}: The polymerization kink of Portet, Tuszy\'nski and
Dixon \cite{p}
and  the susy kink propagating with the same velocity.\\

{\bf Fig. 3.1}: Real part for the factorization curve of the
parameter $a_{1_{+}}=a_{1_{+}}(\alpha,\beta)$ that allows the
factorization of Eq. (\ref{ec10}). $a_{1}\neq 0$.
$\alpha\in[-10,10]$ and $\beta\in[-10,10]$.\\

{\bf Fig. 3.2}: Imaginary part for the factorization curve of the
parameter $a_{1_{+}}=a_{1_{+}}(\alpha,\beta)$ that allows the
factorization of Eq. (\ref{ec10}). $a_{1}\neq 0$.
$\alpha\in[-10,10]$ and $\beta\in[-10,10]$.\\

{\bf Fig. 3.3}: Real part for the factorization curve of the
parameter $E_{+}=E_{+}(G,A)$ that allows the factorization of Eq.
(\ref{ec28}). Note that $a_{1}=-\frac{E}{3}$; $E\neq 0$.
$G\in[-10,10]$ and $A\in[-10,10]$.\\

{\bf Fig. 3.4}: Imaginary part for the factorization curve of the
parameter $E_{+}=E_{+}(G,A)$ that allows the factorization
of Eq. (\ref{ec28}). $E\neq 0$. $G\in[-10,10]$ and $A\in[-10,10]$.\\

{\bf Fig. 3.5}: Factorization curve of the parameter
$\nu=\nu(\mu)$ that allows the factorization of Eq. (\ref{ec35}).
$a_{1}=-\frac{\mu}{\sqrt{2}}$.\\

{\bf Fig. 3.6}: Real part for the factorization curve of the
parameter $a_{1_{+}}=a_{1_{+}}(\alpha,\beta,\delta=1)$ that allows
factorization of Eq. (\ref{ec40}) with $\delta=1$.
$a_{1}\neq 0$. $\alpha\in[-20,20]$ and $\beta\in[-20,20]$.\\

{\bf Fig. 3.7}: Imaginary part for the factorization curve of the
parameter $a_{1_{+}}=a_{1_{+}}(\alpha,\beta,\delta=1)$ that allows
factorization of Eq. (\ref{ec40}) with $\delta=1$. $a_{1}\neq 0$.
$\alpha\in[-20,20]$ and $\beta\in[-20,20]$.\\

{\bf Fig. 3.8}: Real part for the factorization curve of the
parameter $e_{1_{+}}=e_{1_{+}}(\alpha,\beta,\delta=1)$ that allows
factorization of Eq. (\ref{ec40}) with $\delta=1$. $e_{1}\neq 0$.
$\alpha\in[-20,20]$ and $\beta\in[-20,20]$.\\

{\bf Fig. 3.9}: Imaginary part for the factorization curve of the
parameter $e_{1_{+}}=e_{1_{+}}(\alpha,\beta,\delta=1)$ that allows
factorization of Eq. (\ref{ec40}) with $\delta=1$. $e_{1}\neq 0$.
$\alpha\in[-20,20]$ and $\beta\in[-20,20]$.\\

{\bf Fig. 4.1}: The Montroll asymmetric double-well potential
(MDWP) calculated using Eq.~(\ref{u}) for $\epsilon _0=0$. In all
figures $\alpha _1=1$, $\alpha _2=-1.5$, $\beta
=-2.5/\sqrt{2}$, $\gamma =-0.5$, $\epsilon =0.1$.\\

{\bf Fig. 4.2}: The Montroll ground state wave function cf
Eq.~(\ref{phi0}) for $\phi _0(0)=1$.\\

{\bf Fig. 4.3}: The one-parameter Darboux modified MDWP for
$\lambda =1$.\\

{\bf Fig. 4.4}: The low-scale left hand side of the
singularity.\\

{\bf Fig. 4.5}: The low-scale right hand side of the
singularity.\\

{\bf Fig. 4.6}: The wave functions for $\lambda =1$.\\

{\bf Fig. 4.7}: One parameter Darboux-modified MDWP for
$\lambda =10$.\\

{\bf Fig. 4.8}: The bottom of the potential at the right
hand side.\\

{\bf Fig. 4.9}: The ground state wave function
corresponding to $\lambda =10$.\\

{\bf Fig. 4.10}: Plot of the integral $I_M(\xi)$ that
produces the deformation of the potential and wave functions.\\

{\bf Fig. 5.1}: The real part of the bosonic mode ${\rm
w_2^{+}(y;\frac{1}{2},\frac{1}{2}})$ for ${\rm t}\in [0,10]$ and
${\rm K\in[0,4]}$.\\

{\bf Fig. 5.2}: The imaginary part of the bosonic mode ${\rm
w_2^{+}(y;\frac{1}{2},\frac{1}{2}})$ for ${\rm t}\in [0,10]$
and ${\rm K\in[0,4]}$.\\

{\bf Fig. 5.3}: The real part of the bosonic mode ${\rm
w_2^{+}(y;\frac{1}{2},\frac{1}{2}})$ for ${\rm t}\in [0,20]$ and
${\rm K}=0.01$.\\

{\bf Fig. 5.4}: The imaginary part of the bosonic mode ${\rm
w_2^{+}(y;\frac{1}{2},\frac{1}{2}})$ for ${\rm t}\in [0,20]$
and ${\rm K}=0.01$.\\

{\bf Fig. 5.5}: The real part of the bosonic mode ${\rm
w_2^{+}(y;\frac{1}{2},\frac{1}{2}})$ for ${\rm t}\in [0,20]$ and
${\rm K}=2$.\\

{\bf Fig. 5.6}: The real part of the bosonic mode ${\rm
w_2^{+}(y;\frac{1}{2},\frac{1}{2}})$ for ${\rm t}\in [0,20]$ and
${\rm K}=2$ in the vertical strip [-0.5, 0.5].\\

{\bf Fig. 5.7}: The imaginary part of the bosonic mode ${\rm
w_2^{+}(y;\frac{1}{2},\frac{1}{2}})$ for ${\rm t}\in [0,20]$
and ${\rm K}=2$.\\

{\bf Fig. 5.8}: The fermionic zero mode $-1/\cos{\rm  t}$, (red
curve), and the real part of $-1/{\rm w_2^{+}}$, (blue
curve), for ${\rm K}=0.01$.\\

{\bf Fig. 5.9}: The fermionic zero mode $-1/\cos{\rm  t}$, (red
curve), and the imaginary part of $-1/{\rm w_2^{+}}$, (blue
curve), for ${\rm K}=2$.\\

{\bf Fig. 5.10}: Real part of the bosonic wave function $w_{2}$ in
the range
$x\in[0,3]$ and ${\rm K\in[0,2]}$.\\

{\bf Fig. 5.11}: Imaginary part of the bosonic wave function
$w_{2}$ in the range
$x\in[0,3]$ and ${\rm K\in[0,2]}$.\\

{\bf Fig. 5.12}: Real part of the fermionic wave function $w_{1}$
in the range
$x\in[0,3]$ and ${\rm K\in[0,2]}$.\\

{\bf Fig. 5.13}: Imaginary part of the fermionic wave function
$w_{1}$ in the range
$x\in[0,3]$ and ${\rm K\in[0,2]}$.\\

{\bf Fig. 8.1}: Spiking patterns of the master (solid line) and
slave (dashed line) systems for the action potentials in
desynchronized and synchronized states. The forcing functions
amplitud and frequency parameters as specified in the text:
$I_{ext_{M}}(t)=-2.58\textrm{sin}(.245t)$,
$I_{ext_{S}}(t)=-3.15\textrm{sin}(.715t)$.\\

{\bf Fig. 8.2}: Dynamical response of the implemented
control action of Fig 8.1.\\

{\bf Fig. 8.3}: Phase locking of the synchronized action
potentials of Fig 8.1.\\

{\bf Fig. 8.4}: Spiking patterns of the master (solid line) and
slave (dashed line) systems for the action potentials in
desynchronized state and the transition to a robust
synchronization state when the modified feedback control law is
implemented. The forcing functions are
$I_{ext_{M}}(t)=-2.58\textrm{sin}(.245t)$,
$I_{ext_{S}}(t)=-3.15\textrm{sin}(.715t)$.\\

{\bf Fig. 8.5}: Dynamical response of the implemented
modified control law of Fig 8.4.\\

{\bf Fig. 8.6}: Phase locking of the action potentials in robust
synchronization state of Fig 8.4.





\part{FACTORIZATION METHODS}

\chapter{Factorization techniques for linear second order differential equations}

\bigskip
\bigskip

\section{Introduction}

Factorization methods are powerful yet simple algebraic procedures
to find eigenspectra and eigenfunctions of differential operators
that avoid "cumbersome transformations, recourse to the ready-made
equipment of the mathematical warehouse or expansion into power
series", to cite from the very first paragraph of the 1940's
papers of Schr\"odinger \cite{schr1}. At the present time, one can
find in the literature very good informative review papers on the
factorization topics \cite{review1,review2}. It is now well known
that for second-order linear differential operators, the
factorizations are equivalent to their Darboux isospectrality (or
covariance) and also they represent a simple form of intertwining
\cite{review2}. In this introduction, we will touch upon both
these issues.

In the case of Sturm-Liouville operators, E. Schr\"odinger first
developed a factorization method he called "that of adjoint first
order operators" in 1940-1941 \cite{schr1}, during the period he
lived in Dublin. In his very first paper on the method,
Schr\"odinger deals with four cases: the Planck (harmonic)
oscillator, the nonrelativistic hydrogen atom, the spherical
harmonics in the three-dimensional hypersphere, and the Kepler
motion in the hypersphere.

For the quantum harmonic oscillator, he wrote the amplitude equation
\begin{equation}\label{t_s1}
\frac{d^2\psi}{dx^2}-x^2\psi + \lambda \psi=0~,
\end{equation}
and noticed that it can be written in two different factorized forms
\begin{eqnarray}
\left(\frac{d}{dx}-x\right)\left(\frac{d}{dx}+x\right)\psi +(\lambda -1)\psi &=&0~, \nonumber\\
\left(\frac{d}{dx}+x\right)\left(\frac{d}{dx}-x\right)\psi +(\lambda +1)\psi &=&0~.\nonumber
\end{eqnarray}
Operating on one of these equations with the second of the two
first order differential operators which occur in it, one gets for
the function which results from $\psi$ by applying that operator
an equation of the {\em other} type, but with $\lambda +2$ or
$\lambda -2$, respectively, instead of $\lambda$. The mutual
adjointness of the two first order linear operators maintains the
quadratic integrability of the solutions and furthermore the whole
spectrum can be obtained by repeated application of the adjoint
operator to the $\lambda =1$ solutions of the partner operators,
e.g.,
\begin{equation}\label{t_s2}
\left(\frac{d}{dx}+x\right)\psi _{v}^{+} =0 \quad \,\Rightarrow
\quad \psi _{v}^{+}= \textrm{e}^{-\frac{x^2}{2}}
\end{equation}
leads to the odd eigenfunctions in the form
\begin{equation} \label{t_s3}
\psi _{2n-1} = \left(\frac{d}{dx}-x\right)^{n}\psi _{v}^{+}~,
\end{equation}
whereas the even eigenfunctions are obtained similarly from the function $\psi _{v}^{-}$.

In his last paper on the method \cite{schr2}, Schr\"odinger
factorized the hypergeometric equation, finding that there are
several ways of factorizing it. His factorization procedure
originated "from a, virtually, well-known treatment of the
oscillator", i.e., an approach that can be traced back to Dirac's
creation and annihilation operators for the harmonic oscillator
\cite{diracbook} and to older factorization ideas in a paper of
Pauli \cite{pauli26} and in Weyl's treatment of spherical
harmonics with spin \cite{weyl}. It should be noted that whereas
Dirac's first-order operators were considered only as a trick (or
`stratagem'), too insignificant to replace the Sturm-Liouville
theory, Schr\"odinger speaks neatly about a method and applies it
in a systematic way. However, Schr\"odinger's works were not very
much taken into account perhaps because of the war years.

A decade later, in 1951, Infeld and Hull \cite{ih} wrote an
influential paper in which they introduced a different
factorization method that became widely known. They studied
equations of the form
$$
[\hat{M}(x,m)+\lambda _{n}^{0}]y_{n-m}^{m}(x)=0~,
$$
where $\hat{M}(x,m)$ is an operator of the form
$$
\hat{M}(x,m)=\frac{d^2}{dx^2}+r(x,m)
$$
and $m=1,2,...,n$ plays the role of a parameter in the potential,
whereas the specific feature of their method is that the
eigenvalue $\lambda _{n}^{0}$ is the same for all values of $m$.
Infeld and Hull noticed that such equations can be written in two
factorized forms
$$
[-\hat{O}_{+}(m,m+1)\hat{O}_{-}(m+1,m)-L(m+1)+\lambda _{n}^{0}] y_{n-m}^{m}(x)=0
$$
and
$$
[-\hat{O}_{-}(m,m-1)\hat{O}_{+}(m-1,m)-L(m)+\lambda _{n}^{0}]
y_{n-m}^{m}(x)=0~.
$$
The eigenfunctions of the neighboring operators $\hat{M}(x,m)$ and
$\hat{M}(x,m\pm 1)$ are connected by the following relations
$$
\hat{O}_{+}(m-1,m)y_{n-m}^{m}(x)= [\lambda _{n}^{0}-L(m+1)]y_{n-m}^{m}(x)
$$
and
$$
\hat{O}_{-}(m+1,m)y_{n-m}^{m}(x)= y_{n-m+1}^{m-1}(x)~.
$$
In addition, the condition
$$
\hat{O}_{-}(n+1,n)y_{0}^{n}(x)=0
$$
is satisfied leading to
$$
\lambda _{n}^{0}=L(n+1)~.
$$
The eigenfunctions $y_{n}^{0}(x)$ of the operator $\hat{M}(x,0)$ can be obtained from $y_{0}^{n}(x)$ multiplicatively
$$
y_{n}^{0}(x)=\hat{O}_{+}(0,1)\,\hat{O}_{+}(1,2)...\hat{O}_{+}(n-1,n)y_{0}^{n}(x)~.
$$

Nothing noteworthy happened for thirty years until Witten
\cite{w81} wrote a paper on dynamical breaking of supersymmetry,
in which supersymmetric quantum mechanics (SUSYQM) was introduced
as a toy model for supersymmetry breaking in quantum field
theories.

The SUSY breaking is presented by Witten as a sort of "phase
transition" with the order parameter being the Witten index,
defined as the grading operator $\tau=(-1)^{\hat N _{f}}$, where
$\hat N _{f}$ is the fermion number operator. For the case of
one-dimensional SUSYQM, Witten's index operator is the third Pauli
matrix $\sigma _{3}$, which is +1 for the bosonic sector and -1
for the fermionic sector of the one dimensional quantum problem at
hand. It became also quite common to call a particular Riccati
solution as a (Witten) "superpotential". Papers that now are
standard references are published during 1982-1984.
For example, a breakthrough algebraic result has been obtained in
1983 by Gendenshtein \cite{gend83} who introduced the important
concept of {\em shape invariance} (SI) in SUSYQM. The SI property
is displayed by some classes of potentials with respect to their
parameter(s), say $a_n$, and reads
$$
V_{n+1}(x,a_{n})=V_{n}(x,a_{n+1})+R(a_{n})~,
$$
where $R$ should be a remainder independent of $x$. This property
assures a fully algebraic scheme for the spectrum and wave
functions. Fixing $E_{0}=0$, the excited spectrum is given by the
algebraic formula
$$
E_{n}=\sum _{k=2}^{n+1} R(a_{k})~,
$$
and the wave functions are obtained from
$$
\psi _{n}(x, a_1)=\prod _{k=1}^{n} A^{+}(x, a_{k})  \psi _{0}(x,
a_{n+1})~.
$$
Another remarkable result of that period is due to Mielnik
\cite{m84}, who provided the first application of the {\em
general} Riccati solution to the harmonic oscillator, obtaining a
harmonic potential with an additive tail of the type $D^2 [\ln
{\rm erf} + {\rm const.}]$ similar to the Abraham-Moses class of
isospectral potentials in the area of inverse scattering. D.
Fern\'andez gave a second application to the atomic hydrogen
spectrum, whereas M.M. Nieto clarified further the inverse
scattering aspects of Mielnik's construction. Mielnik's procedure
may be seen as a double Darboux transformation in which the
general Riccati (superpotential) solution is involved.
In addition, Andrianov and his collaborators \cite{andr84}
discovered the relation between SUSYQM and Darboux Transformations
(DT) or Darboux covariance while playing with matrix Hamiltonians
in SUSYQM.

\medskip

\section{Darboux covariance}

\medskip

\noindent The {\em Darboux covariance} of a Sturm-Liouville
equation is clearly stated by Matveev and Salle \cite{MS91}.
Consider the equation
$$
-\psi _{xx}+u\psi=\lambda \psi~,
$$
and perform the following DT (denoted by $\psi [1],\, u[1]$)
$$
\psi \rightarrow \psi [1]=(D-\sigma _{1})\psi=\psi _{x}-\sigma
_{1}\psi= \frac{W(\psi _{1},\psi)}{\psi _{1}}~,
$$
$$
u\rightarrow u[1]=u-2\sigma _{1x}=u-2D^{2}\ln \psi _{1}~,
$$
where
$$\sigma _{1}=\psi _{1x}\psi _{1}^{-1}
$$
is the sigma notation of Matveev and Salle for the logarithmic
derivative, and $W$ is the Wronskian determinant. Then, the
Darboux-transformed equation becomes
$$
-\psi _{xx}[1] +u[1]\psi[1]=\lambda\psi[1]~,
$$
i.e., the spectral parameter $\lambda$ does not change (a result
known as Darboux isospectrality). When DTs are applied iteratively
one gets Crum's result. One can also say that the two SL equations
are related by a DT.

Following Matveev and Salle, in order to demonstrate the
equivalence of SUSYQM with a single DT we consider two
Schr\"odinger equations
$$
-D^2 \psi +u\psi=\lambda \psi~,
$$
$$
-D^2 \phi +v\phi=\lambda \phi~,
$$
related by DT, i.e., $v=u[1]$ and $\phi =\psi [1]$, and notice
that the function $\phi _{1}=\psi ^{-1}_{1}$ satisfies the
Darboux-transformed equation for $\lambda =\lambda _{1}$.

\noindent If now one uses the second (transformed) equation as
initial one and perform the DT with the generating function $\phi
_{1}$, one just goes back to the initial $u$ equation. That is why
one can think of the latter procedure as a sort of {\em inverse
DT} that can be obtained from the direct one as follows: 
$$
u=v-2D^{2}\ln \phi _{1}=v[-1]=v-2D^{2}\ln \psi ^{-1}_{1}~,
$$
$$
\psi =\left(\phi _{x}-\frac{\phi _{1x}}{\phi _{1}}\phi\right)
(\lambda _{1}-\lambda)
=\left(\phi _{x}+\frac{\psi _{1x}}{\psi _{1}}\phi\right)
(\lambda _{1}-\lambda)~.
$$
Using the sigma notation,
$$
\sigma =\frac{\psi _{1x}}{\psi _{1}}=-\frac{\phi _{1x}}{\phi _{1}}
$$
the Riccati (SUSYQM) representation of the Darboux pair of Schr\"odinger
potentials is obtained
$$
u=v[-1]=\sigma _{x}+\sigma ^{2}+\lambda _{1}~,
$$
$$
v=u[1]=-\sigma _{x}+\sigma ^{2}+\lambda _{1}~.
$$
It is now easy to enter the issue of SUSYQM concept of supercharge operators.
For that, one employs the factorization operators
$$
B^{+}=-D+\sigma,\qquad B^{-}=D+\sigma~.
$$
They effect the wave function part of the direct and inverse DT,
respectively. Moreover,
$$
B^{+}B^{-}=-D^{2}+v-\lambda _{1}~,
$$
$$
B^{-}B^{+}=-D^{2}+u-\lambda _{1}~.
$$
Thus, the commutator $[B^{+},B^{-}]=v-u=-2D^{2}\ln \psi _{1}$ gives the
Darboux difference in the shape of the Darboux-related
potentials. Introducing the Hamiltonian operators
$$
H^{+}= B^{-}B^{+}+\lambda _{1}~,
$$
$$
H^{-}=B^{+}B^{-}+\lambda _{1}~,
$$
one can also interpret the $B$ operators as factorization ones and
write the famous matrix representation of SUSYQM, as well as the
simplest possible superalgebra.

\noindent
The factorizing operators in matrix representation are called
{\em supercharges} in SUSYQM, and are nilpotent operators
$$ Q^{-}= A_-\sigma _+ =
\left( \begin{array}{cc}
0 & 0 \\
A^{-} & 0
\end{array} \right)~,\qquad \left(Q^{-}\right)^2=0~,$$
and
$$Q^{+}= A_+\sigma _-  =
\left( \begin{array}{cc}
0 & A^{+} \\
0 & 0
\end{array} \right)~, \qquad \left(Q^{+}\right)^2=0~.$$

\noindent
$\sigma _-= \left( \begin{array}{cc}
0 & 1 \\
0 & 0
\end{array} \right)$ and
$\sigma _+=\left( \begin{array}{cc}
0 & 0 \\
1 & 0
\end{array} \right)$ are Pauli matrices.
In this realization, the matrix form of the
Hamiltonian operator reads
$$ {\cal H} =
\left( \begin{array}{cc}
A^+ A^- & 0 \\
0 & A^- A^+
\end{array} \right) =
\left( \begin{array}{cc}
H_- & 0 \\
0 & H_+
\end{array} \right)~,
$$
defining the partner Hamiltonians as diagonal elements of ${\cal H}$.
They are partners in the sense that they are isospectral, apart from
the ground state $\phi _{gr,-}$ of $H_-$, which is not included in the
spectrum of $H_+$.


\medskip

\section{The Mielnik construction}
\vskip 2mm \noindent An interesting possibility to build families
of potentials {\em strictly} isospectral with respect to the
initial (bosonic) one arises if one asks for the most general
superpotential (i.e., the general Riccati solution) such that $\rm
V_+(x)=  w_{g}^2 + \frac{d w_{g}}{dx}$, where ${\rm V_+}$ is the
fermionic partner potential. It is easy to see that one particular
solution to this equation is ${\rm w_p= w(x)}$, where w(x) is the
common Witten superpotential. One is led to consider the following
Riccati equation ${\rm  w_{g}^2 + \frac{d w_{g}}{dx}=w^2_p
+\frac{d w_p}{dx}}$, whose general solution can be written in the
form ${\rm w_{g}(x)= w_p(x) + \frac{1}{v(x)}}$, where ${\rm v(x)}$
is an unknown function. Using this ansatz, one obtains for the
function ${\rm v(x)}$ the following Bernoulli equation
\begin{equation}\label{miel1}
{\rm \frac{dv(x)}{dx} - 2 \, v(x)\, w_p(x) = 1},
\end{equation}
that has the solution
\begin{equation}\label{miel2}
{\rm v(x)= \frac{{\cal I}_0(x)+ \mu}{u_{0}^{2}(x)}},
\end{equation}
where ${\rm {\cal I}_0(x)= \int _{c}^{x} \, u_0^2(y)\, dy}$,
($c=-\infty$ for full line problems and $c=0$ for half line
problems, respectively), and $\mu$ is an integration constant
thereby considered as a free parameter. Thus, ${\rm w_{g}(x)}$ can
be written as follows
\begin{eqnarray}\label{miel3}
{\rm w_{g}(x;\mu)}&=&{\rm w_p(x) + \frac{d}{dx}} \Big[ {\rm ln}({\cal I}_0(x) + \mu) \Big] \nonumber\\
&=&{\rm w_p(x)+\sigma_{0}(\lambda)} \nonumber\\
&=&{\rm - \frac{d}{dx} \Big[ ln \left(\frac{u_0(x)}{{\cal I}_0(x)+
\mu}\right)\Big]}~.
\end{eqnarray}

Finally, one easily gets the $V_-(x;\mu)$ family of potentials
\begin{eqnarray}
{\rm  V_-(x;\mu)} &=& {\rm w_{g}^2(x;\mu) -
\frac{d w_{g}(x;\mu)}{dx}}\nonumber \\
&=& {\rm V_-(x) - 2 \frac{d^2}{dx^2} \Big[ ln({\cal I}_0(x) + \mu)}
\Big] \nonumber \\
&=& {\rm V_-(x) -2\sigma _{0,x}(\mu)} \nonumber \\
&=& {\rm V_-(x) - \frac{4 u_0(x) u_0^\prime (x)}{{\cal I}_0(x)
+ \mu}
+ \frac{2 u_0^4(x)}{({\cal I}_0(x) + \mu)^2}~.}
\end{eqnarray}
All ${\rm  V_-(x;\mu)}$ have the same supersymmetric partner potential
${\rm V_+(x)}$ obtained by deleting the ground state.
They are asymmetric double-well potentials that may be considered as a sort
of intermediates between the bosonic potential ${\rm V_-(x)}$ and
the fermionic partner ${\rm V_+(x)=V_-(x)-2\sigma _{0,x}(x)}$.
From the last rhs of Eq.~(\ref{miel3}) one can infer the ground state wave functions
for the potentials ${\rm V_-(x;\mu)}$ as follows
\begin{equation}\label{mielX}
{\rm u_0(x;\mu)= f(\mu)
\frac{u _0(x)}{{\cal I}_0(x) + \mu}},
\end{equation}
where ${\rm f(\mu)}$ is a normalization factor that can be shown
to be of the form ${\rm f(\mu)= \sqrt{\mu(\mu +1)}}$. One can now
understand the double Darboux feature of this construction by
writing the parametric family in terms of their unique "fermionic"
partner potential
\begin{equation}\label{mielY}
{\rm V_{-} (x;\mu)=V_{+}(x)
-2\frac{d^2}{dx^2}\ln\left(\frac{1}{u_{0}(x;\mu)}\right)},
\end{equation}
which shows that the Mielnik transformation is of the inverse
Darboux type, allowing at the same time a two-step (double
Darboux) interpretation, namely, in the first step one goes to the
fermionic system and in the second step one returns to a deformed
bosonic system.

An application of this construction to microtubules is presented
in Chapter 4.


\bigskip

\noindent
\section{The connection with intertwining}

\noindent
Intertwining has been introduced by the French mathematician J. Delsarte  in 1938 \cite{d38}
as an operatorial relationship involving so-called transformation (or transmutation) operators
but the second World War delayed the detailed mathematical studies
that came only in the 1950's.
By definition,
two operators $L_{0}$ and $L_{1}$ are said to be intertwined by an operator
$T$ if
\begin{equation} \label{intw1}
L_{1}T=TL_{0}~.
\end{equation}
If the eigenfunctions $\varphi _{0}$ of $L_0$ are known, then from
the intertwining relation one can show that the (unnormalized)
eigenfunctions of $L_1$ are given by $\varphi _{1}=T\varphi _{0}$.
The main problem in the intertwining transformations is to
construct the transformation operator $T$. One-dimensional quantum
mechanics is one of the simplest examples of intertwining
relations since Witten's transformation operator $T_{qm}=T_1$ is
just a first spatial derivative plus a differentiable coordinate
function (the superpotential) that should be a logarithmic
derivative of the true bosonic zero mode (if it exists), but of
course higher-order transformation operators can be constructed
without much difficulty.

\noindent
Thus, within the realm of the one-dimensional quantum mechanics, writing $T_1=
D-\frac{u^{'}}{u}$, where $u$ is a true bosonic zero mode, one can
infer that the
adjoint operator $T^{\dagger}_{1}=-D-\frac{u^{'}}{u}$ intertwines in
the opposite direction, taking solutions of $L_{1}$ to those of $L_{0}$
\begin{equation}  \label{15}
\varphi _{0}=T_{1}^{\dagger}\varphi _{1}~.
\end{equation}
In particular, for standard one-dimensional quantum mechanics,
$L_0=H_{-}$ and $L_1=H_{+}$ and although the true zero mode of
$H_{-}$ is annihilated by $T_1$, the corresponding (unnormalized)
eigenfunction of $H_{+}$ can nevertheless be obtained by applying
$T_1$ to the other independent zero energy solution of $H_{-}$. It
is only in the last decade or so, that the intertwining approach
becomes well-known to the SUSYQM factorization community and some
authors start to play
with higher-order generalizations. But, as always, the most
important (at least for standard quantum mechanics) are the
simplest cases, namely the Darboux first-order intertwining
operators.\\

The first part of this thesis deals with factorization methods,
among which an original factorization of nonlinear second order
ordinary differential equations (ODE) and supersymmetric
techniques, as applied to some biological and physical systems.
Chapters 2 and 3 contain explicitly the new factorization
procedure developed by us to obtain kink type solutions for
nonolinear second order ODE that describe several important
processes, for instance, the tubulin polymerization in
microgravity conditions and the pulse propagation along nerve
axons. In Chapter 4, supersymmetric approaches are applied in the
framework of biological dynamics of microtubules (MTs); the latter
results are related to transport properties associated to the MT
domain walls. In Chapter 5, applications of supersymmetric
factorization procedures in some physical systems are presented. A
complex extension for the classical harmonic oscillator by means
of a direct relationship between the Dirac and Schr\"oedinger
equations is obtained. In addition, the same procedure is applied
to a molecular physics problem in connection with the dissociation
of diatomic molecules.

\chapter{A new factorization technique for differential equations with polynomial nonlinearity}

\bigskip

{\small \noindent {\bf Abstract}. In this chapter, it is shown how
one can obtain kink solutions of ordinary differential equations
with polynomial nonlinearities by an efficient factorization
procedure \textit{directly} related to the factorization of their
nonlinear polynomial part. This is different of previous
factorization procedures of differential equations of this type
that have been performed by only a few authors, most notably by
Berkovich \cite{ber}. Of main interest here because of their
numerous applications are the reaction-diffusion equations in the
travelling frame and the damped-anharmonic-oscillator equations.
In addition, interesting pairing of the kink solutions, a result
obtained by reversing the factorization brackets in the
supersymmetric quantum mechanical style, are reported. In this
way, one gets ordinary differential equations with a different
polynomial nonlinearity possessing kink solutions of different
width but propagating at the same velocity as the kinks of the
original equation. This pairing of kinks could have many
applications. The mathematical procedure is illustrated with
several important cases, among which the generalized Fisher
equation, the FitzHugh-Nagumo equation, and the polymerization
fronts of microtubules (MTs). In the latter case, a new
polymerization front is predicted that can show up in solutions
containing MTs borne on satellites. Because of the microgravity
conditions the polymerization rates could deviate from the normal
ones and this could lead to a change of the width of the
polymerization front.}






\bigskip

\section{Introduction}

Factorization of second-order linear differential equations, such
as the Schr\"odinger equation, is a well established method to get
solutions in an algebraic manner \cite{schr2,ih,yu1}. We are
interested in factorizations of ordinary differential equations
(ODE) of the type
\begin{equation}\label{e0}
u^{\prime\prime} +\gamma u^{\prime} + F(u)
= 0~,
\end{equation}
where $F(u)$ is a given polynomial in $u$. If the independent
variable is the time then $\gamma$ is a damping constant and we
are in the case of nonlinear damped oscillator equations. Many
examples of this type are collected in the Appendix of a paper of
Tuszy\'nski et al. \cite{tod}. However, the coefficient $\gamma$
can also play the role of the constant velocity of a travelling
front if the independent variable is a travelling coordinate used
to reduce a reaction-diffusion (RD) equation to the ordinary
differential form as briefly sketched in the following. These RD
travelling fronts or kinks are important objects in low
dimensional nonlinear phenomenology describing
topologically-switched configurations in many areas of biology,
ecology, chemistry and physics.


Consider a scalar RD equation for $u(x,t)$
\begin{equation}\label{e1}
\frac{\partial u}{\partial t}={\cal D}\frac{\partial ^2u}{\partial
x^2}+sF(u)~,
\end{equation}
where ${\cal D}$ is the diffusion constant
and $s$ is the strength of the reaction process. Eq~(\ref{e1}) can be
rewritten as
\begin{equation}
\label{e2}
\frac{\partial u}{\partial t}=\frac{\partial ^2u}{\partial x^2}+F(u)~,
\end{equation}
where the coefficients have been eliminated by the rescalings
$\tilde{t}=st$ and $\tilde{x}=(s/{\cal D})^{1/2}x$, and dropping
the tilde.  Usually, the scalar RD equation possesses travelling
wave solutions $u(\xi)$ with $\xi =x-{\rm v} t$, propagating at
speed ${\rm v}$. For this type of solutions the RD equation turns
into the ODE
\begin{equation}\label{4}
u^{\prime\prime} +{\rm v} u^{\prime} + F(u) = 0~,
\end{equation}
where $'=D=\frac{d}{d\xi}$. 
Eq. (\ref{4}) has the same form as nonlinear damped oscillator
equations with the velocity playing the role of the friction
constant.

For applications in physical optics and acoustics it is convenient
to write the travelling coordinate in the form $\xi = kx-\omega
t=k(x-{\rm v}t)$ with $k{\rm v} = \omega$. This is a simple
scaling by $k$ of the previous coordinate turning Eq.~(\ref{4})
into the form
\begin{equation}\label{4bis}
u^{\prime\prime} +\frac{{\rm v}}{k} u^{\prime} + \frac{1}{k^2}F(u) = 0
\end{equation}
that can be changed back to the form of Eq.~(\ref{e0}) by redefining
$\tilde{\gamma}=\frac{{\rm v}}{k}$ and $\tilde{F}(u)=\frac{1}{k^2}F(u)$.

In general, performing the factorization of Eq.~(\ref{e0}) means the following
\begin{equation}\label{6}
\Big[D-f_{2}(u)\Big]\Big[D-f_{1}(u)\Big]u=0~.
\end{equation}
This leads to the equation
\begin{equation}\label{7}
u^{\prime\prime}-\frac{df_{1}}{du}uu^{\prime}-f_{1}u^{\prime}-f_{2}u^{\prime}+f_{1}f_{2}u=0~.
\end{equation}

The following groupings of terms are possible related to different factorizations:

{\em  a) Berkovich grouping}: In 1992, Berkovich \cite{ber} proposed to group the terms as follows
\begin{equation}\label{b1}
u^{\prime\prime}-\left(f_{1}+f_{2}\right)u^{\prime}
+\left(f_{1}f_{2}-\frac{df_{1}}{du}u^{\prime}\right)u=0~,
\end{equation}
and furthermore discussed a theorem according to which any
factorization of an ODE of the form given in Eq.~(\ref{6}) allows
to find a class of solutions that can be obtained from solving the
first-order differential equation
\begin{equation}
u^{\prime}-f_{1}(u)u=0.\label{b1a}
\end{equation}
Substituting the first-order ODE (\ref{b1a}) in the Berkovich
grouping one gets
\begin{equation}\label{b2}
u^{\prime\prime}-\left(f_{1b}+f_{2b}\right)u^{\prime}
+\left(f_{1b}f_{2b}-\frac{df_{1b}}{du}f_{1b}u\right)u=0,
\end{equation}
where we redefined $f_1(u)=f_{1b}(u)$ and $f_2(u)=f_{2b}(u)$ to
distinguish this case from our proposal following next. For the
specific form of the ODEs we consider here, Berkovich's conditions
read
\begin{equation}\label{bco2}
f_{1b}\left(-\gamma -f_{1b}-\frac{df_{1b}}{du}u\right)=\frac{F(u)}{u}~,
\end{equation}
\begin{equation}\label{bco1}
f_{1b}+f_{2b}=-\gamma ~.
\end{equation}

{\em b) Grouping of this work}: We propose here the different grouping of terms
\begin{equation}\label{og1}
u^{\prime\prime}-\left(\frac{d\phi _{1}}{du}u
+\phi _{1}+\phi _{2}\right)u^{\prime}+\phi _{1}\phi _{2}u=0
\end{equation}
that can be considered the result of changing the Berkovich
factorization by setting $f_{1b}(u)=\phi _1(u)$ and
$f_{2b}(u)\rightarrow \phi _{2}(u)$ under the conditions
\begin{equation}\label{co1}
\phi _{1}\phi _{2}=\frac{F(u)}{u}~,
\end{equation}
\begin{equation}\label{co2}
\phi _{1}+\phi _2+\frac{d\phi _{1}}{du}u=-\gamma~.
\end{equation}

The following simple relationship exists between the
factoring functions:
$$
\phi_{2}(u)=f_{2b}(u)-\frac{df_{1b}(u)}{du}u$$
and further (third, and so forth) factorizations can be obtained through linear combinations
of the functions $f_{1b}$, $f_{2b}$ and $\phi _{2}$.

Based on our experience, we think that the grouping we propose is more
convenient than that of Berkovich and also of other people employing
more difficult procedures. The main advantage resides in the fact
that whereas in Berkovich's scheme Eq.~(\ref{bco2}) is still a
differential equation to be solved, in our scheme we make a choice
of the factorization functions by merely factoring polynomial
expressions according to Eq.~(\ref{co1}) and then imposing
Eq.~(\ref{co2}) leads easily to an $n$-depending $\gamma$
coefficient for which the factorization works. This fact makes our
approach extremely efficient in finding particular solutions of
the kink type as one can see in the following.

In the next section, it is shown on the explicit case of the
generalized Fisher equation all the mathematical constructions
related to the factorization brackets and their supersymmetric
quantum mechanical like reverse factorization. In less detail, but
within the same approach, damped nonlinear oscillators of
Dixon-Tuszy\'nski-Otwinowski type and the FitzHugh-Nagumo
equation, are studied in Sections 2.3 and 2.4, respectively.

\medskip

\section{Generalized Fisher equation} 

Let us consider the generalized Fisher equation given by
\begin{equation}
\label{c1-1}
u^{\prime\prime}
+\gamma u^{\prime} + u(1-u^n) = 0,
\end{equation}
The case $n=1$ refers to the common Fisher
equation and it will be shortly discussed as a subcase.
Eq. (\ref{co1}) allows to factorize the polynomial function
\begin{equation} \label{c1-2}
\phi _{1}\phi
_{2}=\frac{F(u)}{u}=(1-u^n)=(1-u^{n/2})(1+u^{n/2}),~\nonumber
\end{equation}
Now, by choosing
\begin{equation}\label{c1-3}
\phi _{1}=a_{1}(1-u^{n/2}),\;\; \phi
_{2}=\frac{1}{a_{1}}(1+u^{n/2})~,\quad a_1\neq 0~,\nonumber
\end{equation}
the explicit forms of $a_{1}$ and
$\gamma$ can be obtained from Eq. (\ref{co2})
\begin{equation} \label{c1-4}
\frac{d\phi _{1}}{du}u+\phi _{1}+\phi
_{2}=-\frac{n}{2}a_{1}u^{n/2}+a_{1}(1-u^{n/2})+(1/a_{1})(1+u^{n/2})=-\gamma\,
.\nonumber
\end{equation}
Introducing the notation $h_n=(\frac{n}{2}+1)^{1/2}$ one gets:
$a_{1}=\pm h_{n}^{-1}~$, $\gamma=\mp\left( h_{n} +
h_{n}^{-1}\right)~$.

Then Eq. (\ref{c1-1}) becomes
\begin{equation} \label{c1-6}
u^{\prime\prime}
\pm \left(h_{n} + h_{n}^{-1}\right)   u^{\prime} + u(1-u^n) = 0
\end{equation}
and the corresponding factorization is
\begin{equation} \label{c1-6b}
\Big[ D \pm
h_n(u^{n/2}+1) \Big] \Big[ D \mp
h_{n}^{-1}(u^{n/2}-1) \Big]u=0~.
\end{equation}
It follows that Eq.~(\ref{c1-6}) is
compatible with the first-order differential equation
\begin{equation}\label{c1-7}
u^{\prime}\mp
h_{n}^{-1}\left(u^{n/2}-1\right)u=0~.
\end{equation}
Integration of Eq. (\ref{c1-7}) gives for $\gamma >0$
\begin{equation} \label{c1-8}
u_{>}^{\pm}= \left(1\pm
\exp\Big[\left( h_{n} - h_{n}^{-1}\right) (\xi-\xi _{0})\Big]\right)^{-2/n}~.
\end{equation}
Rewritten in the hyperbolic form, we get
\begin{eqnarray} \label{c1-9}
u_{>}^{+}=\left(\frac{1}{2}-\frac{1}{2}\tanh\Big[\frac{1}{2} \left( h_{n} - h_{n}^{-1}\right) (\xi-\xi _{0})\Big]\right)^{2/n}~, \nonumber\\
u_{>}^{-} = \left(\frac{1}{2}-\frac{1}{2}\coth\Big[\frac{1}{2} \left( h_{n} - h_{n}^{-1}\right) (\xi-\xi _{0})\Big]\right)^{2/n}~.
\end{eqnarray}
The $\textrm{tanh}(\cdot)$ form is precisely the solution obtained
long ago by Wang \cite{wang88} and Hereman and Takaoka \cite{ht90}
by more complicated means.

Moreover, a different solution is possible for $\gamma <0$
\begin{equation}\label{gfi10}
u_{<}^{\pm}= \left(1\pm
\exp\Big[-  \left( h_{n} - h_{n}^{-1}\right) (\xi-\xi _{0})\Big]\right)^{-2/n},
\end{equation}
or
\begin{eqnarray}
u_{<}^{+}=\left(\frac{1}{2}+\frac{1}{2}\tanh\Big[-\frac{1}{2} \left( h_{n} - h_{n}^{-1}\right) (\xi-\xi _{0})\Big]\right)^{2/n},\nonumber\\
u_{<}^{-} = \left(\frac{1}{2}+\frac{1}{2}\coth\Big[-\frac{1}{2} \left( h_{n} - h_{n}^{-1}\right) (\xi-\xi _{0})\Big]\right)^{2/n}~,
\label{gfi11}
\end{eqnarray}
respectively.

\medskip

{\bf  2.2.1 Reversion of factorization brackets without the change
of the scaling factors}

Choosing now $\phi _{1}=a_{1}(1+u^{n/2})$ and $\phi_{2}=\frac{1}{a_{1}}\left(u^{n/2}-1\right)$
leads to the same equation (\ref{c1-6})
but now with the factorization
\begin{equation}
\Big[ D \mp h_{n}(u^{n/2}-1) \Big] \Big[ D \pm
h_{n}^{-1}(u^{n/2}+1) \Big]u=0 \label{gfi14}~,
\end{equation}
and therefore the compatibility is with the different first-order equation
\begin{equation}\label{gfi15}
u^{\prime}\pm h_{n}^{-1}\left(u^{n/2}+1\right)u=0~.
\end{equation}
However, the direct integration gives the solution (for $\gamma >0$)
\begin{eqnarray}\label{gfi16}
u &=&\left(-\frac{1}{1\pm \exp[\left( h_{n} - h_{n}^{-1}\right)
(\xi-\xi _{0})]}\right)^{2/n}\nonumber\\
&=&(-1)^{2/n} \left(1\pm\exp\Big[\left( h_{n} - h_{n}^{-1}\right) (\xi-\xi _{0})\Big]\right)^{-2/n}~,
\end{eqnarray}
which are similar to the known solution Eq.~(\ref{c1-8}). For
$\gamma <0$, solutions of the type given by Eq.~(\ref{gfi10}) are
obtained.

\medskip

{\bf 2.2.2 Direct reversion of factorization brackets}

Let us perform now a direct inversion of the factorization brackets in (\ref{c1-6b}) similar to what is done in supersymmetric quantum mechanics
in order to enlarge the class of exactly solvable quantum hamiltonians
\begin{equation}\label{gfi22}
\Big[ D \mp h_{n}^{-1}(u^{n/2}-1) \Big]\Big[ D \pm
h_n(u^{n/2}+1) \Big] u=0~.
\end{equation}
Doing the product of differential operators the following
RD equation is obtained
\begin{equation}\label{gfi23}
u^{\prime\prime} \pm \left( h_{n} + h_{n}^{-1}\right)  u^{\prime} +
u\left[1+u^{n/2}\right]\left[1-h_{n}^{4}u^{n/2}\right] = 0~.
\end{equation}
Eq. (\ref{gfi23}) is compatible with the equation
\begin{equation}\label{gfi24}
u^{\prime}\pm h_n\left(u^{n/2}+1\right)u=0,
\end{equation}
and integration of the latter gives the kink solution of Eq.~(\ref{gfi23})
\begin{equation}\label{gfi26}
u_{>}^{\pm}=\left(-\frac{1}{1\pm
\exp[(h_n^3-h_n)(\xi-\xi _{0})]}\right)^{\frac{2}{n}}=
\left(1\pm
\exp\Big[(h_n^3-h_n)(\xi-\xi _{0})\Big]\right)^{-\frac{2}{n}}
\end{equation}
for $\gamma >0$. On the other hand,  for $\gamma <0$ the exponent is the same but of opposite sign.
Hyperbolic forms of the latter solutions are easy to write down and are similar up to widths to Eqs.~(\ref{c1-9}) and (\ref{gfi11}), respectively.

Thus, a different RD equation given by (\ref{gfi23}) with modified
polynomial terms and its solution have been found by reverting the
factorization terms of Eq. (\ref{c1-6}). Although the reaction
polynomial is different the velocity parameter remains the same.
The main result, which is a general one, that we find here is the
following: {\em At the velocity corresponding to the travelling
kink of a given RD equation there is another propagating kink
corresponding to a different RD equation that is related to the
original one by reverse factorization}. We can call this kink as
the supersymmetric (susy) kink because of the mathematical
construction.

Finally, one can ask if the process of reverse factorization can be continued with Eq.~(\ref{gfi23}). It can be shown that this is not the case because Eq.~(\ref{gfi23})
has already a discretized (polynomial-order-dependent) $\gamma$ and this fact prevents further solutions of this type. Suppose we consider the following factorization functions
\begin{equation}\label{f1}
\tilde{\phi}_1=\tilde{a}_{1}^{-1}\Big[1-h_n^4u^{n/2}\Big]~, \qquad \tilde{\phi} _2=\tilde{a}_1\left(1+u^{n/2}\right)~.
\end{equation}
Then, one gets $\tilde{a}_1=\pm h_{n}^{3}$ and solve
$\tilde{a}_1^{-1}+\tilde{a}_1=h_{n}^{-1}+h_n$. The solutions are:
$n=0$, which implies linearity, and $n=-4$, which leads to a
Milne-Pinney equation. On the other hand, Eq.~(\ref{gfi23}) with
an arbitrary $\gamma$ can be treated by the inverse factorization
procedure to get the susy partner RD equation and its susy kink.\\

\newpage

{\bf 2.2.3 Subcase $n=1$} 

This subcase is the original Fisher equation describing the propagation of mutant genes
\begin{equation}\label{Fish-1}
\frac{\partial u}{\partial t}=\frac{\partial^2 u}{\partial
x^2}+u(1-u)~.
\end{equation}

In the travelling frame, the Fisher equation has the form
\begin{equation}
\label{Fish0} u^{\prime\prime} +\gamma u^{\prime} + u(1-u) = 0~.
\end{equation}
When the $\gamma$ parameter takes the value $\gamma _1=\frac{5}{6}
\sqrt{6}$ (i.e., $h_1=\frac{\sqrt{6}}{2}$) one can factor Fisher's
equation and employing our method leads easily to the known kink
solution
\begin{equation}\label{Fish1}
u_{\rm F}=\frac{1}{4}\left(1-\tanh\Big[\frac{\sqrt{6}}{12}(\xi-\xi _{0})\Big]\right)^{2}~
\end{equation}
that was first obtained by Ablowitz and Zeppetella \cite{az79} with a series solution method.
On the other hand, the susy kink for this case reads
\begin{equation}\label{Fish2}
u_{\rm F, susy}=\frac{1}{4}\left(1-\tanh\Big[\frac{\sqrt{6}}{8}(\xi-\xi _{0})\Big]\right)^{2}~,
\end{equation}
i.e., it has a width one and a half times greater than the common
Fisher kink and is a solution of the partner equation
\begin{equation} \label{Fish3}
u^{\prime\prime}+\frac{5\sqrt{6}}{6} u^{\prime} + u\left(1-\frac{5}{4}u^{1/2}-\frac{9}{4}u\right) = 0~.
\end{equation}

A plot of the kinks $u_{\rm F}$ and $u_{\rm F,susy}$ is displayed in
Fig.~2.1.
%

\vskip 4ex \centerline{ \epsfxsize=200pt \epsfbox{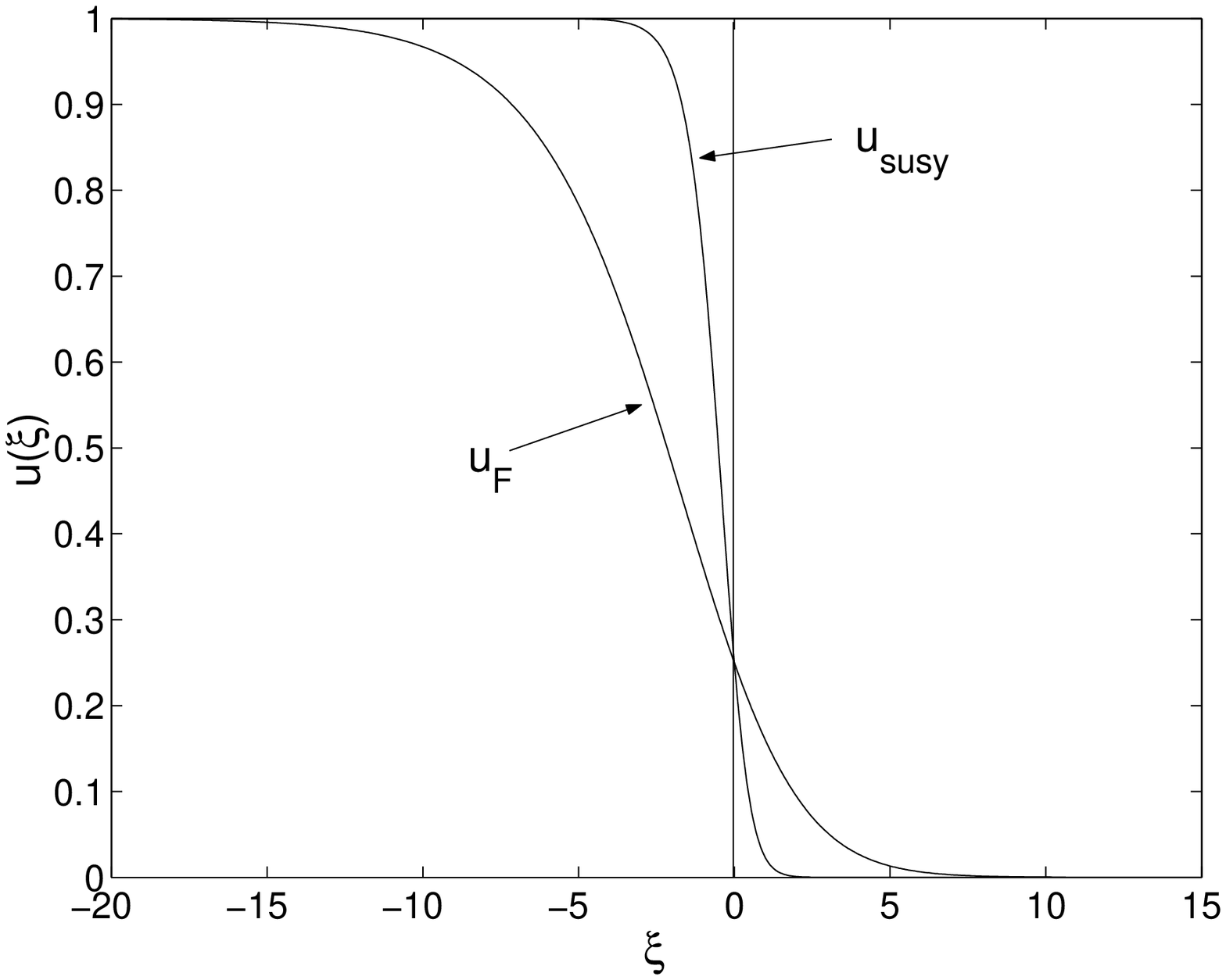}}
\vskip 3ex
\begin{center}
{\small{Fig. 2.1}:$\quad$
The front of mutant genes (Fisher's wave of advance) in a
population and the partner susy kink propagating with the same
velocity. The axis are in arbitrary units.}
\end{center}

\medskip

{\bf  2.2.4 Subcase $n=6$}

This subcase is of interest in the light of experiments on
polymerization patterns of MTs in centrifuges.
It has been discovered that the polymerization of the tubulin dimers
proceeds in a kink-switching fashion propagating with a constant velocity within the sample.
Portet, Tuszynski and Dixon \cite{p} used RD equations to discuss the modification of self-organization patterns of MTs as well as the tubulin polymerization under the influence of reduced gravitational fields. They used the value $n=6$ for the mean critical number of tubulin dimers at which the polymerization process starts and showed that the same nucleation number enters the polynomial term of the RD process
for the number concentration ${\rm c}$ of tubulin dimers
\begin{equation}\label{nconc}
{\rm c}^{\prime\prime} +\frac{5}{2}{\rm c}^{\prime} +
{\rm c}\left(1-{\rm c}^{6}\right)= 0~.
\end{equation}
The polymerization kink in their work reads
\begin{equation}\label{polykink}
{\rm c}_{{\rm PTD}}     
=2^{-\frac{1}{3}}\left(1-\tanh\Big[\frac{3}{4}(\xi-\xi _{0})\Big]\right)^{1/3}~.
\end{equation}
On the other hand, the susy polymerization kink (see Fig.~(2.2))
of the form
\begin{equation}\label{polyk1}
{\rm c}_{\rm susy}   
=
2^{-\frac{1}{3}}\Big(1-\tanh[3(\xi-\xi _{0})]\Big)^{1/3}
\end{equation}
can be taken into account according to the hyperbolic form of Eq.~(\ref{gfi26}). It propagates with the same speed and corresponds to the equation
\begin{equation}\label{nconc1}
{\rm c}^{\prime\prime} \pm\frac{5}{2}{\rm c}^{\prime} +
{\rm c}\left(1-15 {\rm c}^{3}-16{\rm c}^{6}\right)= 0~.
\end{equation}
In principle, this equation could be obtained as a consequence of modifying the kinetics steps in 
the microtubule polymerization process.  


\vskip 4ex \centerline{ \epsfxsize=200pt \epsfbox{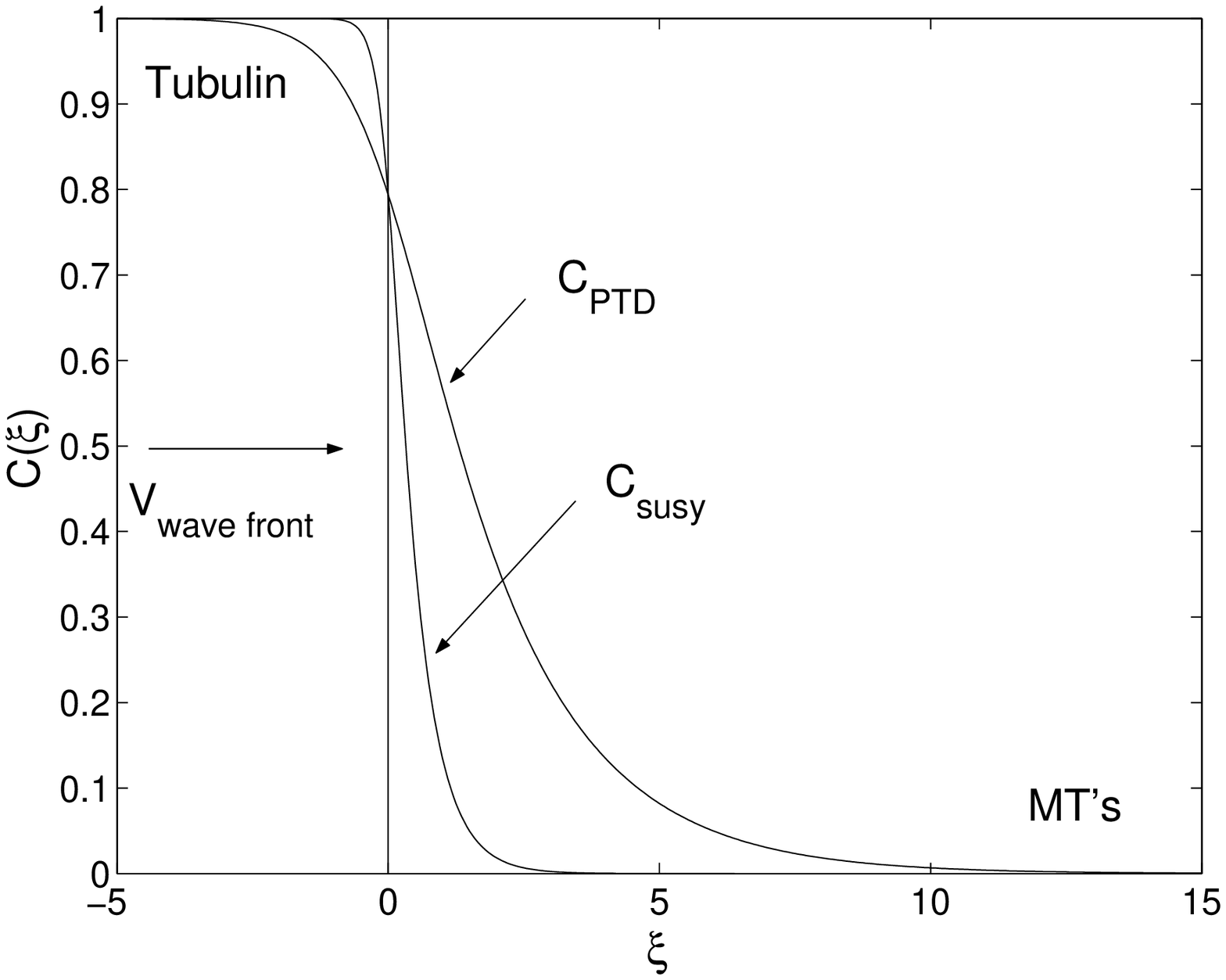}}
\vskip 3ex
\begin{center}
{\small{Fig. 2.2}:$\quad$
The polymerization kink of Portet, Tuszy\'nski and Dixon \cite{p}
and  the susy kink propagating with the same velocity.}
\end{center}

\medskip

\section{Equations of the Dixon-Tuszy\'nski-Otwinowski type}

In the context of damped anharmonic oscillators, Dixon et al.
\cite{dto91} studied equations of the type (in this section, we
use $'=D_{\tau}=\frac{d}{d\tau}$)
\begin{equation}\label{dto1}
u^{\prime\prime}+u^{\prime}+Au-u^{n-1}\equiv u^{\prime\prime}+u^{\prime}+u(\sqrt{A}-u^{\frac{n}{2}-1})(\sqrt{A}+u^{\frac{n}{2}-1}) =0
\end{equation}
and gave solutions for the cases $A=\frac{2}{9}$ and $A=\frac{3}{16}$, with $n=4$ and $n=6$, respectively.
For this case, time is the independent variable. The factorization method works nicely if one uses $g_n=\sqrt{n/2}$ and dealing
with the more general equation
\begin{equation}
u^{\prime\prime}\pm\sqrt{A}(g_n+g_{n}^{-1}) 
u^{\prime}+u(A-u^{n-2})=0, \label{otwi1}
\end{equation}
for which we can employ either the factorization functions
\begin{eqnarray}
\phi _{1}=\mp
g_{n}^{-1}\left(\sqrt{A}-u^{\frac{n}{2}-1}\right),\quad \phi
_{2}=\mp
g_n\left(\sqrt{A}+u^{\frac{n}{2}-1}\right)\label{otwi2}\nonumber
\end{eqnarray}
or
\begin{eqnarray}
\phi _{1}=\mp
g_{n}^{-1}\left(\sqrt{A}+u^{\frac{n}{2}-1}\right),\quad \phi
_{2}=\mp
g_n\left(\sqrt{A}-u^{\frac{n}{2}-1}\right)~.\label{otwi3}\nonumber
\end{eqnarray}
Then, Eq. (\ref{otwi1}) can be factored in the forms
\begin{equation}
\Big[ D_{\tau} \pm g_n(u^{\frac{n}{2}-1}+\sqrt{A})
\Big]\Big[ D_{\tau} \mp g_{n}^{-1}(u^{\frac{n}{2}-1}-\sqrt{A})
\Big] u=0 \label{otwi4}
\end{equation}
and
\begin{equation}
\Big[D_{\tau} \mp g_n(u^{\frac{n}{2}-1}-\sqrt{A})
\Big]\Big[D_{\tau} \pm g_{n}^{-1}(u^{\frac{n}{2}-1}+\sqrt{A})
\Big] u=0~. \label{otwi5}
\end{equation}
Thus, Eq. (\ref{otwi1}) is compatible with the equations
\begin{eqnarray}
 u^{\prime} \mp g_{n}^{-1}\left(u^{\frac{n}{2}-1}-\sqrt{A}\right)
 u=0,\label{otwi6}\\
u^{\prime} \pm g_{n}^{-1}\left(u^{\frac{n}{2}-1}+\sqrt{A}\right) u=0
\label{otwi7}
\end{eqnarray}
that follows from Eq.~(\ref{otwi4}) and Eq.~(\ref{otwi5}).
Integration of Eqs. (\ref{otwi6}), (\ref{otwi7}) gives the
solution of Eq. (\ref{otwi1})
\begin{equation}
u_{>}=\left(\frac{\sqrt{A}}{1\pm
\exp\Big[\sqrt{A}(g_n -g_{n}^{-1})(\tau-\tau_{0})\Big]}\right)^{\frac{2}{n-2}},\quad\quad
\gamma >0 \label{otwi8}
\end{equation}
and
\begin{equation}
u_{<}=\left(\frac{\sqrt{A}}{1\pm
\exp\Big[-\sqrt{A}(g_n - g_{n}^{-1})(\tau-\tau_{0})\Big]}\right)^{\frac{2}{n-2}},
\quad\quad\gamma <0~.\label{otwi9}
\end{equation}
The solutions obtained by Dixon et al. are particular cases of the
latter formulas.

Reversing now the factorization brackets in (\ref{otwi4})
\begin{equation}
\Big[ D_{\tau} \mp g_{n}^{-1}\left(u^{\frac{n}{2}-1}-\sqrt{A}\right)
\Big]\Big[ D_{\tau} \pm g_n(u^{\frac{n}{2}-1}+\sqrt{A})
\Big] u=0 \label{otwi10}
\end{equation}
leads to the following equation
\begin{equation}
u^{\prime\prime}\pm\sqrt{A}(g_n +g_{n}^{-1})
u^{\prime}+u\left(\sqrt{A}+u^{\frac{n}{2}-1}\right)\left(\sqrt{A}-\frac{n^2}{4}u^{\frac{n}{2}-1}\right)=0,
\label{otwi11}
\end{equation}
which is compatible with the equation
\begin{equation}
u^{\prime} \pm g_n\left(u^{\frac{n}{2}-1}+\sqrt{A}\right)u=0
\label{otwi12}
\end{equation}
whose integration gives the solution of Eq.~(\ref{otwi11})
\begin{equation}
u_{>}= \left(\frac{\sqrt{A}}{1\pm
\exp[\sqrt{A}g_n(\tau-\tau_{0})]}\right)^{\frac{2}{n-2}},\quad\quad\gamma >0\label{otwi13}
\end{equation}
and
\begin{equation}
u_{<}= \left(\frac{\sqrt{A}}{1\pm
\exp[-\sqrt{A}g_n(\tau-\tau_{0})]}\right)^{\frac{2}{n-2}},\quad\quad\gamma <0~.\label{otwi14}
\end{equation}

\section{FitzHugh-Nagumo equation}

Let us consider the FitzHugh-Nagumo equation, which is a common
approximation to describe nerve fiber propagation,
\begin{equation}\label{fhn0}
\frac{\partial u}{\partial t}-\frac{\partial^2 u}{\partial
x^2}+u(1-u)(a-u)=0~,
\end{equation}
where $a$ is a real constant. Moreover, if $a=-1$, one gets the
real Newell-Whitehead equation describing the dynamical behavior
near the bifurcation point for the Rayleigh-B\'enard convection of
binary fluid mixtures. The travelling frame form of (\ref{fhn0})
has been discussed in detail by Hereman and Takaoka \cite{ht90}
\begin{equation}
u^{\prime\prime}+\gamma u^{\prime}+u(u-1)(a-u)=0. \label{fhn1}
\end{equation}
The FitzHugh-Nagumo polynomial function allows the following factorizations:
\begin{eqnarray}
\phi _{1}=\pm(\sqrt{2})^{-1}(u-1),\quad \phi
_2=\pm\sqrt{2}(a-u)\label{fhn2}\nonumber
\end{eqnarray}
when the $\gamma$ parameter is equal to $\gamma _{a1}=\pm\frac{-2a+1}{\sqrt{2}}$ that we also write as $\gamma _{a,1}=\pm \sqrt{a}(g_{a1}-g_{a1}^{-1})$, where $g_{a1}=-\sqrt{2a}$.

In addition, we can employ the factorization functions
\begin{eqnarray}
\phi _{1}=\pm(\sqrt{2})^{-1}(a-u),\quad \phi
_2=\pm\sqrt{2}(u-1)\label{fhn3}\nonumber
\end{eqnarray}
when $\gamma _{a,2}=\pm\frac{-a+2}{\sqrt{2}}$, or written again in the more symmetric form $\gamma _{a,2}=\pm \sqrt{a}(g_{a2}-g_{a2}^{-1})$, where $g_{a2}=-\sqrt{a/2}$.  Thus, Eq. (\ref{fhn1}) can
be factored in the two cases
\begin{equation}
u^{\prime\prime}\pm \gamma _{a,1} 
u^{\prime}+u(u-1)(a-u)=0, \label{fhn3a}
\end{equation}
and
\begin{equation}
u^{\prime\prime}\pm \gamma _{a,2} 
u^{\prime}+u(u-1)(a-u)=0~.
\label{fhn3b}
\end{equation}
In passing, we notice that for the Newell-Whitehead case $a=-1$ the two
equations coincide and are the same as the generalized Fisher equation for $n=2$.

In factorization bracket forms, Eqs. (\ref{fhn3a}) and (\ref{fhn3b}) are written as follows
\begin{equation}
\Big[ D \mp\sqrt{2}(a-u)\Big]\Big[ D \pm
(\sqrt{2})^{-1}(1-u)\Big] u=0 \label{fhn4}
\end{equation}
and
\begin{equation}
\Big[ D \mp \sqrt{2}(u-1)\Big]\Big[ D \mp
(\sqrt{2})^{-1}(a-u)\Big] u=0, \label{fhn5}
\end{equation}
and are compatible with the first order differential equations
\begin{eqnarray}
u^{\prime} \pm (\sqrt{2})^{-1}(1-u)u=0,\quad\quad {\rm for} \;\;
\gamma  _{a,1}~,
 \label{fhn6} \\
u^{\prime} \mp (\sqrt{2})^{-1}(a-u)u=0,\quad\quad {\rm for}
\;\;\gamma _{a,2}~.
 \label{fhn7}
\end{eqnarray}
Integration of Eqs. (\ref{fhn6}) and (\ref{fhn7}) gives the
solution of Eq. (\ref{fhn1}) for the two different values of the
wave front velocity $\gamma _{a,1}$ and $\gamma _{a,2}$.

For Eq.~(\ref{fhn3a}) we get
\begin{eqnarray}
u_{>}= \frac{1}{1\pm
\exp[(\sqrt{2})^{-1}(\xi-\xi _{0})]},
\quad u_{<}=
\frac{1}{1\pm
\exp[-(\sqrt{2})^{-1}(\xi-\xi _{0})]},
\end{eqnarray}
for $\gamma _{a,1}$ positive and negative, respectively.

As for Eq.~(\ref{fhn3b}), the solutions are
\begin{eqnarray}
u_{>}= \frac{a}{1\pm
\exp[-(\sqrt{2})^{-1}a(\xi -\xi _{0})]},
\quad u_{<}= \frac{a}{1\pm \exp[(\sqrt{2})^{-1}a(\xi -\xi _{0})]},
\end{eqnarray}
for $\gamma _{a,2}$ positive and negative, respectively.

Considering now the factorizations (\ref{fhn4}) and (\ref{fhn5}), the
change of order of the factorization brackets gives
\begin{equation}
\Big[ D \pm (\sqrt{2})^{-1}(1-u)\Big] \Big[ D
\mp\sqrt{2}(a-u)\Big]u=0 \label{fhn8}
\end{equation}
and
\begin{equation}
\Big[ D \mp (\sqrt{2})^{-1}(a-u)\Big] \Big[ D \mp
\sqrt{2}(u-1)\Big]u=0~. \label{fhn9}
\end{equation}
Doing the product of differential operators (and considering the
factorization term $u^{\prime}- \phi _{2}u=0$) gives the following RD
equations
\begin{equation}
u^{\prime\prime} \pm \gamma _{a1}
u^{\prime} + u(4u-1)(a-u)= 0, \label{fhn10}
\end{equation}
and
\begin{equation}
u^{\prime\prime} \pm \gamma _{a2}    
u^{\prime} +u(u-1)(a-u-3u^2)= 0, \label{fhn11}
\end{equation}
Eqs. (\ref{fhn10}) and (\ref{fhn11}) are compatible with the
equations
\begin{equation}
u^{\prime} \mp\sqrt{2}(a-u)u=0\label{fhn12}
\end{equation}
and
\begin{equation}
u^{\prime} \mp \sqrt{2}(u-1)u=0~, \label{fhn13}
\end{equation}
respectively. Integrations of Eqs. (\ref{fhn12}) and (\ref{fhn13})
give the solutions of Eqs. (\ref{fhn10}) and (\ref{fhn11}), respectively.
The explicit forms are the following:

(i) for (\ref{fhn10})
\begin{eqnarray}
u_{>}= \frac{a}{1\pm
\exp[-\sqrt{2}a(\xi-\xi _{0})]}~, 
\quad u_{<}=
\frac{a}{1\pm
\exp[\sqrt{2}a(\xi -\xi _{0})]}~. 
\end{eqnarray}

(ii) for (\ref{fhn11})  
\begin{eqnarray}
u_{>}= \frac{1}{1\pm \exp[\sqrt{2}(\xi -\xi _{0})]}~, 
\quad u_{<}= \frac{1}{1\pm
\exp[-\sqrt{2}(\xi -\xi _{0})]}~. 
\end{eqnarray}

\section{Conclusion of the chapter}

In this chapter, we have been concerned with stating an efficient
factorization scheme of ODE with polynomial nonlinearities that
leads to an easy finding of analytical solutions of the kink type
that previously have been obtained by far more cumbersome
procedures. The main result is an interesting pairing between
equations with different polynomial nonlinearities, which is
obtained by applying the susy quantum mechanical reverse
factorization. The kinks of the two nonlinear equations are of
different widths but they propagate at the same velocity, or if we
deal with damped polynomial nonlinear oscillators the two kink
solutions correspond to the same friction coefficient. Several
important cases, such as the generalized Fisher and the
FitzHugh-Nagumo equations, have been shown to be simple
mathematical exercises for this factorization technique. The
physical prediction is that for commonly occurring propagating
fronts, there are two kink fronts of different widths at a given
propagating velocity. Moreover, the reverse factorization
procedure can be also applied to the Berkovich scheme with similar
results. It will be interesting to apply the approach of this work
to the discrete case in which various exact results have been
obtained in recent years \cite{discrete}. More general cases in
which the coefficient $\gamma$ is an arbitrary function are also
of much interest because of possible applications. The same
factorization scheme as it works for more complicated ordinary
differential equations is described in the next chapter.

\chapter{Application to more general nonlinear differential equations}





\bigskip

{\small {\bf Abstract}. In the previous chapter we considered the
coefficient in front of the first derivative as a constant
quantity. However, the employed factorization technique can be
used almost unchanged for the more general case when the condition
of constancy of this coefficient is relaxed. In this chapter, we
obtain more kink type solutions through the same factorization
procedure for a number of more general nonlinear ordinary second
order differential equations with important applications in
biology and physics.}


\bigskip

\section{Introduction}

\noindent Considering the following type of differential equation
\begin{equation}
u^{\prime\prime} + g(u)u^{\prime}+ F(u)=0\label{ec1}
\end{equation}
where again as in the previous chapter $\, ^{\prime}$ means the
derivative $D=\frac{d}{d\xi}$ and $\xi=x-vt$; one can factorize
Eq. (\ref{ec1}) in the following form
\begin{equation}
\left[ D-\phi_{2}(u)\right]\left[ D-\phi
_{1}(u)\right]u=0.\label{ec2}
\end{equation}
Performing now the product of differential operators leads to the equation
\begin{equation}
u^{\prime\prime}-\frac{d\phi _{1}}{du}uu^{\prime}-\phi
_{1}u^{\prime} -\phi_{2}u^{\prime}+\phi _{1}\phi_{2}u=0\, ,
\label{ec3}
\end{equation}
for which one way of grouping the terms is as follows
\begin{equation}
u^{\prime\prime}-\left(\phi _{1}+\phi_{2}+\frac{d\phi
_{1}}{du}u\right)u^{\prime} +\phi _{1}\phi_{2}u=0\, .\label{ec4}
\end{equation}
Eqs. (\ref{ec1}) and (\ref{ec4}) are lead to the conditions
\begin{eqnarray}
g(u)=-\left(\phi _{1}+\phi_{2}+\frac{d\phi _{1}}{du}u\right)\label{ec5}
\end{eqnarray}
and
\begin{eqnarray}
F(u)=\phi _{1}\phi_{2}u~.\label{ec6}
\end{eqnarray}
If $F(u)$ is a polynomial function, then $g(u)$ will have the same
order as the bigger of the factorizing functions $\phi_{1}(u)$ and
$\phi_{2}(u)$, and will also be a function of the constant
parameters provided by the function $F(u)$.

In the context of classical mechanics, Eq.~(\ref{ec1}) could be
seen as an anharmonic oscillator with nonlinear damping. The case
$g(u)=\nu$ where $\nu$ is a constant value has been presented in
the previous chapter. There, by means of a simple factorization
method exact particular solutions of the kink type for
reaction-diffusion equations and damped-anharmonic oscillators
with polynomial nonlinearities have been obtained. In addition,
SUSYQM-like reversing of factorization brackets has been performed
providing new kink solutions for equations with different
polynomial nonlinearities.

Based on the given grouping in Eq. (\ref{ec4}) for Eq.
(\ref{ec1}), a simple mathematical procedure is proposed by which
one gets particular solutions through factorization methods that
allows finding solutions satisfying a compatible (nonlinear) first
order differential equation.

The purpose of this chapter is to further apply this mathematical
scheme to a wealth of important cases for which explicit
particular solutions are not easy to find in the literature or are
obtained by more involved techniques. The examples we present
herein are the modified Emden equation, the Generalized Lienard
equation, the convective Fisher equation, the generalized
Burgers-Huxley equation, all of whom have significant applications
in nonlinear physics. Explicit particular solutions are presented.

\bigskip
\section{Modified Emden equation}

\noindent
Let us consider the following modified Emden equation
\begin{equation}
u^{\prime\prime} + \alpha uu^{\prime} + \beta u^{3}=0~.
\label{ec7}
\end{equation}
The polynomial $F(u)=\beta u^{3}$ allows the following
factorizing functions
\begin{eqnarray}
\phi _{1}(u)=a_{1}\sqrt{\beta}u~,\quad \textrm{and}\quad
\phi_{2}(u)=\frac{1}{a_{1}}\sqrt{\beta}u~,\quad a_{1}\neq 0~,
\label{ec8}\nonumber
\end{eqnarray}
where $a_{1}$ is an arbitrary constant. Eq. (\ref{ec5}) is used to
obtain the function $g(u)$,
\begin{equation}
g(u)=-\left(2a_{1}\sqrt{\beta}u + \frac{1}{a_{1}}\sqrt{\beta}u
\right)=-\sqrt{\beta}\left(\frac{2a_{1}^{2}+1}{a_{1}} \right)u~,
\label{ec9}\nonumber
\end{equation}
then identifying
$\alpha=-\sqrt{\beta}\left(\frac{2a_{1}^{2}+1}{a_{1}} \right)$ (or
$a_{1_{+,-}}=\frac{-\alpha\pm\sqrt{\alpha^{2}-8\beta}}{4\sqrt{\beta}}$),
where we use $a_{1}$ as a fitting parameter providing that
$a_{1}<0$ for $\alpha>0$. We note that $g(u)=g(\beta,a_{1};u)$.
Eq. (\ref{ec7}) is now rewritten in the following form
\begin{equation}
u^{\prime\prime} - \sqrt{\beta}\left(\frac{2a_{1}^{2}+1}{a_{1}}
\right) uu^{\prime} + \beta u^{3}=0~,\label{ec10}
\end{equation}
the equation can be factorized as follows
\begin{equation}
\left( D- \frac{1}{a_{1}}\sqrt{\beta}u\right)\left( D-
a_{1}\sqrt{\beta}u\right)u=0\label{ec11}
\end{equation}
and therefore the compatible first order differential equation is
\begin{equation}
u^{\prime}-a_{1}\sqrt{\beta}u^{2}=0~.\label{ec12}
\end{equation}
Integration of Eq. (\ref{ec12}) gives the particular solution of
Eq. (\ref{ec10})
\begin{equation}
u=-\frac{1}{a_{1}\sqrt{\beta}(\xi-\xi_{0})}~,\label{ec13}
\end{equation}
where $\xi_{0}$ is an integration constant. If we consider the
quadratic equation for $a_{1}$, then Eq. (\ref{ec13}) is expressed
as a function of $\alpha$,
\begin{equation}
u=\frac{4}{(\alpha \pm \sqrt{\alpha^2
-8\beta})(\xi-\xi_{0})}.\label{ec13a}
\end{equation}

\bigskip
\vskip 1ex \centerline{ \epsfxsize=200pt \epsfbox{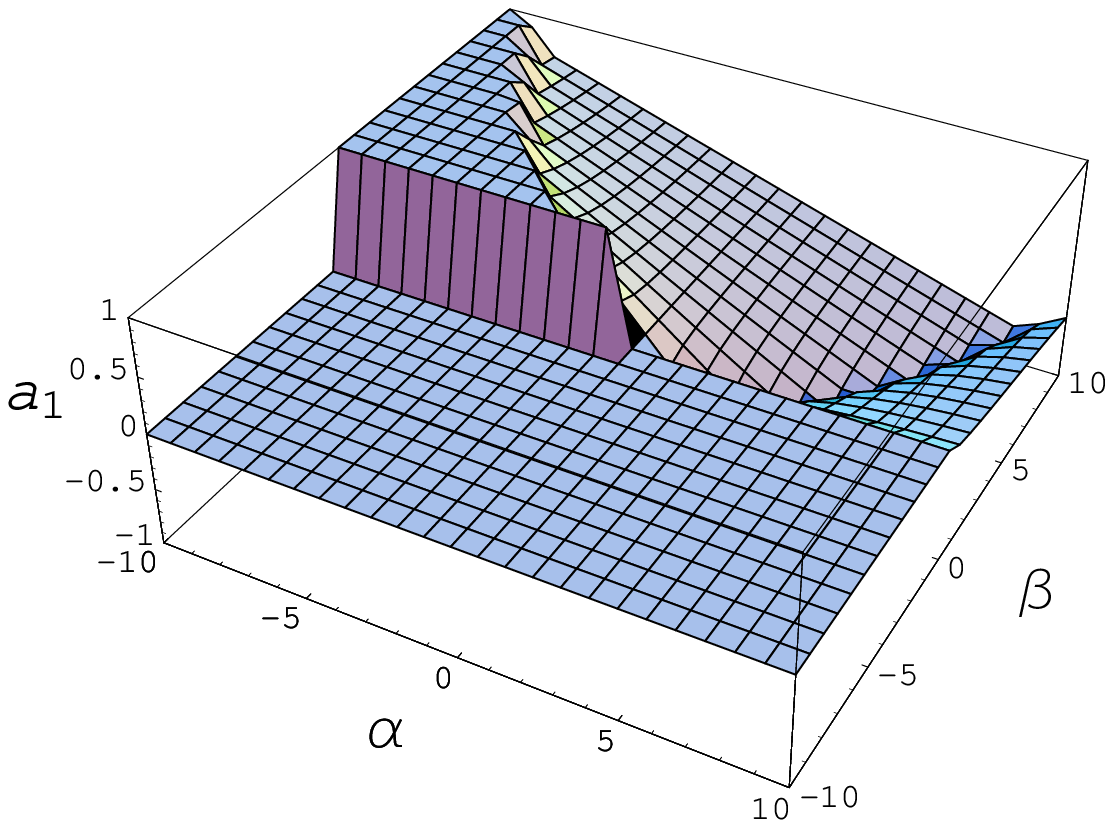}}
\vskip 2ex
\begin{center} {\small{Fig. 3.1:}  Real part for the factorization curve of the parameter
$a_{1_{+}}=a_{1_{+}}(\alpha,\beta)$ that allows the factorization
of Eq. (\ref{ec10}). $a_{1}\neq 0$. $\alpha\in[-10,10]$ and
$\beta\in[-10,10]$.}
\end{center}

\bigskip
\vskip 1ex \centerline{ \epsfxsize=200pt \epsfbox{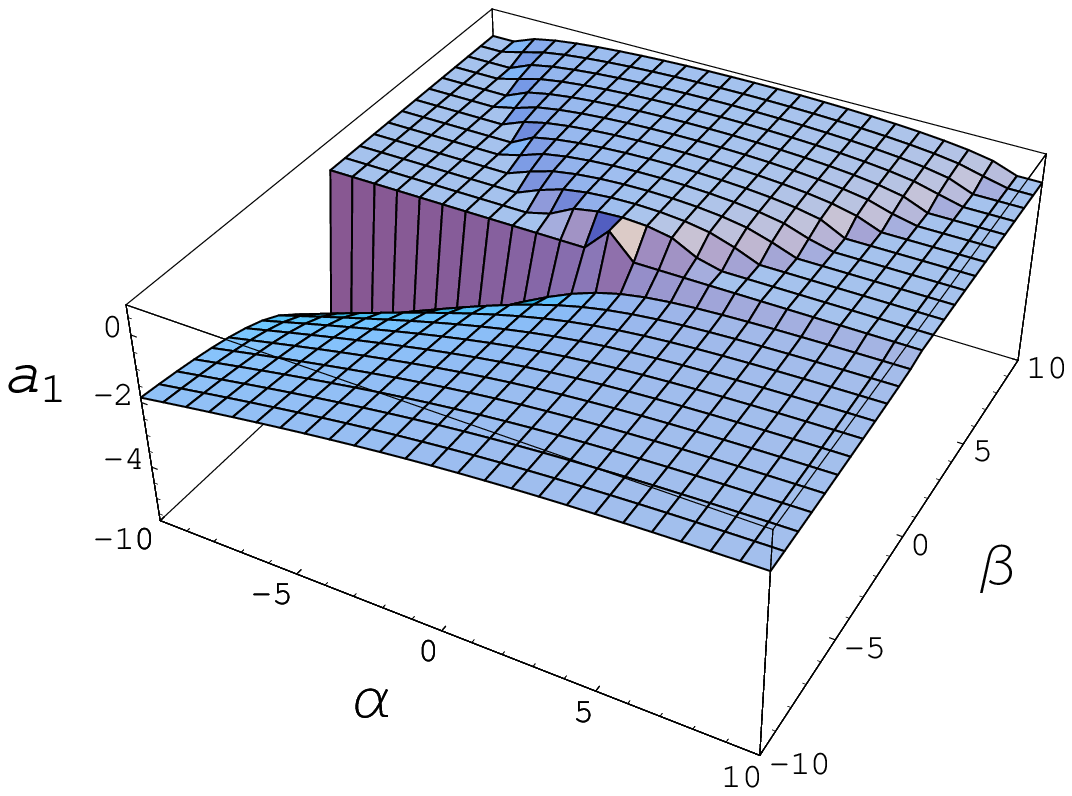}}
\vskip 2ex
\begin{center} {\small{Fig. 3.2:}  Imaginary part for the factorization curve of the parameter
$a_{1_{+}}=a_{1_{+}}(\alpha,\beta)$ that allows the factorization
of Eq. (\ref{ec10}). $a_{1}\neq 0$. $\alpha\in[-10,10]$ and
$\beta\in[-10,10]$.}
\end{center}

\bigskip

Let us consider now another pair of factorizing functions
\begin{eqnarray}
\phi _{1}(u)=a_{1}\sqrt{\beta}u^2,\quad
\phi_{2}(u)=\frac{1}{a_{1}}\sqrt{\beta}~,\label{ec14}\nonumber
\end{eqnarray}
then, using Eq. (\ref{ec5}), the function
$g(u)=-\sqrt{\beta}\left( \frac{1}{a_{1}}+3a_{1}u^{2} \right)$ is
easily obtained. Therefore, the original modified Emden equation
(\ref{ec7}) becomes
\begin{equation}
u^{\prime\prime} - \sqrt{\beta}\left(\frac{1}{a_{1}}+3a_{1}u^{2}
\right)u^{\prime} + \beta u^{3}=0~.\label{ec15}
\end{equation}
This equation allows the factorization
\begin{equation}
\left( D-\frac{1}{a_{1}}\sqrt{\beta} \right)\left( D-
a_{1}\sqrt{\beta}u^{2}\right)u =0~,\label{ec16}
\end{equation}
where from we obtain the compatible first order differential
equation
\begin{equation}
u^{\prime}-a_{1}\sqrt{\beta}u^{3}=0\label{ec17}
\end{equation}
with the solution
\begin{equation}
u=\frac{1}{[-2a_{1}\sqrt{\beta}(\xi-\xi_{0})]^{1/2}}~.\label{ec18}
\end{equation}

The above example shows that different factorizations of $F(u)$
would yield different forms for the function $g(u)$. This is an
important consequence of applying this mathematical technique to
the case $g(u) \neq {\rm const}$.\\

\bigskip
\section{Generalized Lienard equation}
Let us consider now the following generalized Lienard equation
with a cubic polynomial function $F(u)$
\begin{equation}
u^{\prime\prime} + g(u)u^{\prime} +
Au+Bu^{2}+Cu^{3}=0~.\label{ec19}
\end{equation}
The polynomial function $F(u)$ can be factorized in several ways,
we consider first the factorization $F(u)=u(a+b+Cu)(d-e+u)$ where
\begin{equation}
a=B/2,\quad b=\sqrt{B^{2}-4AC}/2,\quad d=B/2C,\quad
e=\sqrt{B^{2}-4AC}/2C,\label{ec20}\nonumber
\end{equation}
and the condition $B^{2}-4AC>0$ holds. If we consider the
factorizing functions as
\begin{equation}
\phi _{1}(u)=a_{1}(a+b+Cu)\quad\textrm{and}\quad
\phi_{2}(u)=\frac{1}{a_{1}}(d-e+u)~,\label{ec21}\nonumber
\end{equation}
where again $a_{1}$ is an arbitrary constant that can be used as a
fitting parameter, the function
$g(u)=\left[\frac{a_{1}^{2}(a+b)+(d-e)}{a_{1}}+\left(
\frac{2a_{1}^{2}C+1}{a_{1}} \right)u\right]$ will be obtained.
Then, Eq. (\ref{ec19}) is rewritten as
\begin{equation}
u^{\prime\prime} + \left[\frac{a_{1}^{2}(a+b)+(d-e)}{a_{1}}+\left(
\frac{2a_{1}^{2}C+1}{a_{1}} \right)u\right]u^{\prime} +
Au+Bu^{2}+Cu^{3}=0~,\label{ec22}
\end{equation}
and the corresponding factorization will be
\begin{equation}
\left[ D- \frac{1}{a_{1}}(d-e+u)\right]\left[ D-
a_{1}(a+b+Cu)\right]u =0\label{ec23}
\end{equation}
where from the compatible first order differential equation is
obtained
\begin{equation}
u^{\prime}-a_{1}(a+b+Cu)u=0~,\label{ec24}
\end{equation}
and whose solution is
\begin{equation}
u=\frac{(a+b)\textrm{exp}[a_{1}(a+b)(\xi-\xi_{0})]}{1-C\textrm{exp}[a_{1}(a+b)(\xi-\xi_{0})]}\label{ec25}
\end{equation}
where $(a+b)=\frac{B+\sqrt{B^{2}-4AC}}{2}$.

Let us consider now the following reduction of terms in Eq.
(\ref{ec19}), $B=0$ and $C=1$ in order to calculate a particular
solution for the so-called autonomous Duffing-van der Pol
oscillator equation \cite{chandrasekar},
\begin{equation}
u^{\prime\prime} + (G+Eu^2)u^{\prime} + Au + u^{3}=0~,\label{ec26}
\end{equation}
where $G$ and $E$ are arbitrary constant parameters. The
polynomial function allows the following factorizing functions
\begin{equation}
f_{1}(u)=a_{1}(A+u^{2})\quad \textrm{and} \quad
\phi_{2}(u)=\frac{1}{a_{1}},\label{ec27}\nonumber
\end{equation}
then $g(u)=-\left( \frac{a_{1}^{2}A+1}{a_{1}} + 3a_{1}u^{2}
\right)$. Eq. (\ref{ec26}) is now rewritten
\begin{equation}
u^{\prime\prime} - \left( \frac{a_{1}^{2}A+1}{a_{1}} + 3a_{1}u^{2}
\right)u^{\prime} + Au + u^{3}=0~.\label{ec28}
\end{equation}
The corresponding factorization of Eq. (\ref{ec28}) is given as follows
\begin{equation}
\left[D -\frac{1}{a_{1}} \right] \left[
D-a_{1}(A+u^{2})\right]u=0~,\label{ec29}
\end{equation}
and the obtained compatible first order equation
\begin{equation}
u^{\prime}-a_{1}(A+u^{2})u=0~.\label{ec30}
\end{equation}
Integration of Eq. (\ref{ec30}) gives the particular solution of
Eq. (\ref{ec28})
\begin{equation}
u=\sqrt{A}\left(
\frac{\textrm{exp}[2a_{1}A(\xi-\xi_{0})]}{1-\textrm{exp}[2a_{1}A(\xi-\xi_{0})]}
\right)^{1/2}.\label{ec31}
\end{equation}
Comparing Eqs. (\ref{ec26}) and (\ref{ec28}), $a_{1}=-\frac{E}{3}$
and $G=\frac{AE^2+9}{3E}$ are obtained. Solution (\ref{ec31}) is
now written as a function of $A$ and $E$,
\begin{equation}
u=\pm\sqrt{A}\left(
\frac{\textrm{exp}[-\frac{2}{3}AE(\xi-\xi_{0})]}{1-\textrm{exp}[-\frac{2}{3}AE(\xi-\xi_{0})]}
\right)^{1/2}.\label{ec32}
\end{equation}

This is a more general result for the particular solution than
that obtained by Chandrasekar et al. in \cite{chandrasekar} by
other means, in fact, it is recuperated
when $E=\beta$ and $A=\frac{3}{\beta^2}$.\\

\bigskip
\vskip 1ex \centerline{ \epsfxsize=200pt \epsfbox{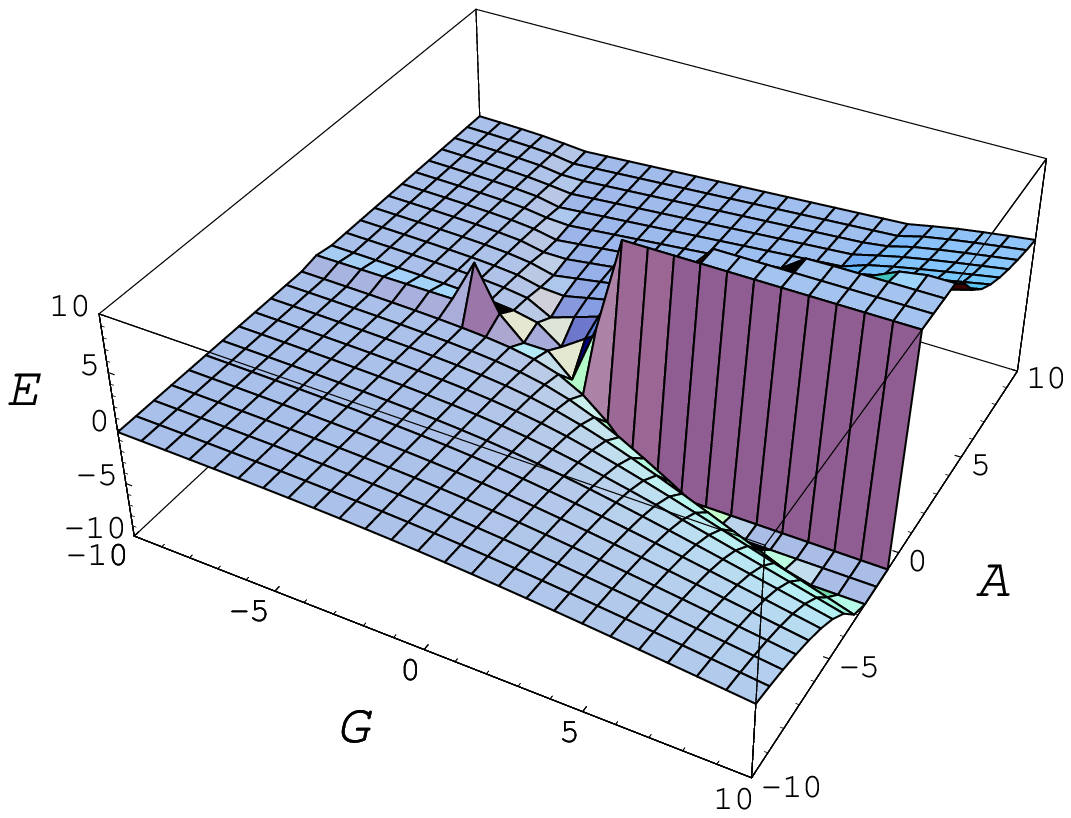}}
\vskip 2ex
\begin{center} {\small{Fig. 3.3:}  Real part for the factorization curve of the parameter
$E_{+}=E_{+}(G,A)$ that allows the factorization of Eq.
(\ref{ec28}). Note that $a_{1}=-\frac{E}{3}$; $E\neq 0$.
$G\in[-10,10]$ and $A\in[-10,10]$.}
\end{center}
\bigskip
\vskip 1ex \centerline{ \epsfxsize=200pt \epsfbox{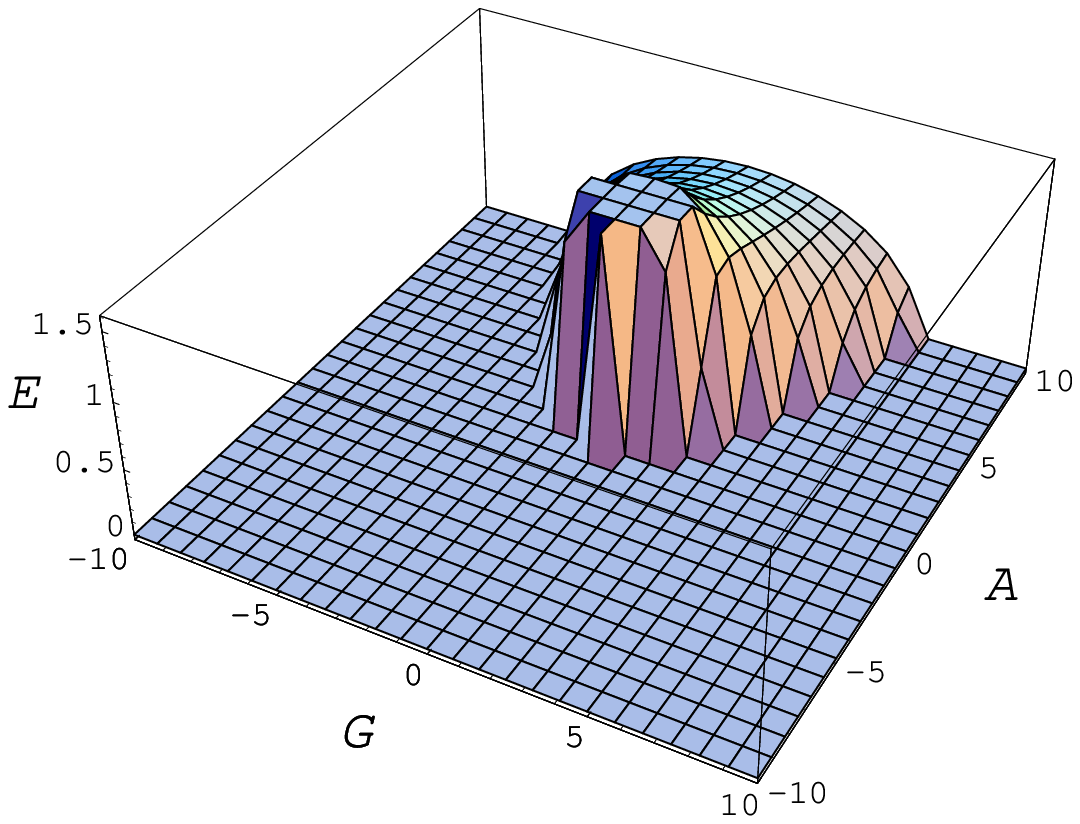}}
\vskip 2ex
\begin{center} {\small{Fig. 3.4:}  Imaginary part for the factorization curve of the parameter
$E_{+}=E_{+}(G,A)$ that allows the factorization of Eq.
(\ref{ec28}). $E\neq 0$. $G\in[-10,10]$ and $A\in[-10,10]$.}
\end{center}

\bigskip
\section{Convective Fisher equation}

Let us consider the convective Fisher equation given in the
following form \cite{schonborn},
\begin{equation}
\frac{\partial u}{\partial
t}=\frac{1}{2}\frac{\partial^{2}u}{\partial x^{2}}+u(1-u)-\mu
u\frac{\partial u}{\partial x}\label{ec32}
\end{equation}
where $\mu$ is a positive parameter that serves to tune the
relative strength of convection. If the variable transformation
$\xi=x-\nu t$ is performed, then we obtain the following ordinary
differential equation
\begin{equation}
u^{\prime\prime} + 2(\nu -\mu u)u^{\prime} +
2u(1-u)=0~.\label{ec33}
\end{equation}
The polynomial function allows the factorizing functions
\begin{equation}
\phi_{1}(u)=\sqrt{2}a_{1}(1-u)\quad\textrm{and}\quad\phi_{2}(u)
=\frac{\sqrt{2}}{a_{1}}~,\label{ec34}\nonumber
\end{equation}
and Eq. (\ref{ec5}) gives the function
$g(u)=-\sqrt{2}\left(\frac{a_{1}^{2}+1}{a_{1}} -2a_{1}u\right)$.
Eq. (\ref{ec33}) is rewritten as follows
\begin{equation}
u^{\prime\prime} + 2\left(-\frac{a_{1}^{2}+1}{\sqrt{2}a_{1}}
+\sqrt{2}a_{1}u \right)u^{\prime} + 2u(1-u)=0~.\label{ec35}
\end{equation}
If we set the fitting parameter $a_{1}=-\frac{\mu}{\sqrt{2}}$,
then we obtain $\nu=\frac{\mu^{2}+2}{2\mu}$. Eq. (\ref{ec35}) is
factorized in the following form
\begin{equation}
\left[D -\frac{\sqrt{2}}{a_{1}} \right] \left[
D-\sqrt{2}a_{1}(1-u)\right]u=0~,\label{ec36}
\end{equation}
that provides the compatible first order equation
\begin{equation}
u^{\prime}-\sqrt{2}a_{1}u(1-u)=u^{\prime}+\mu u(1-u)=0\label{ec37}
\end{equation}
whose integration gives
\begin{equation}
u=\frac{1}{1\pm \textrm{exp}[\mu (\xi-\xi_{0})]}~.\label{ec38}
\end{equation}\\

\bigskip
\vskip 1ex \centerline{ \epsfxsize=200pt
\epsfbox{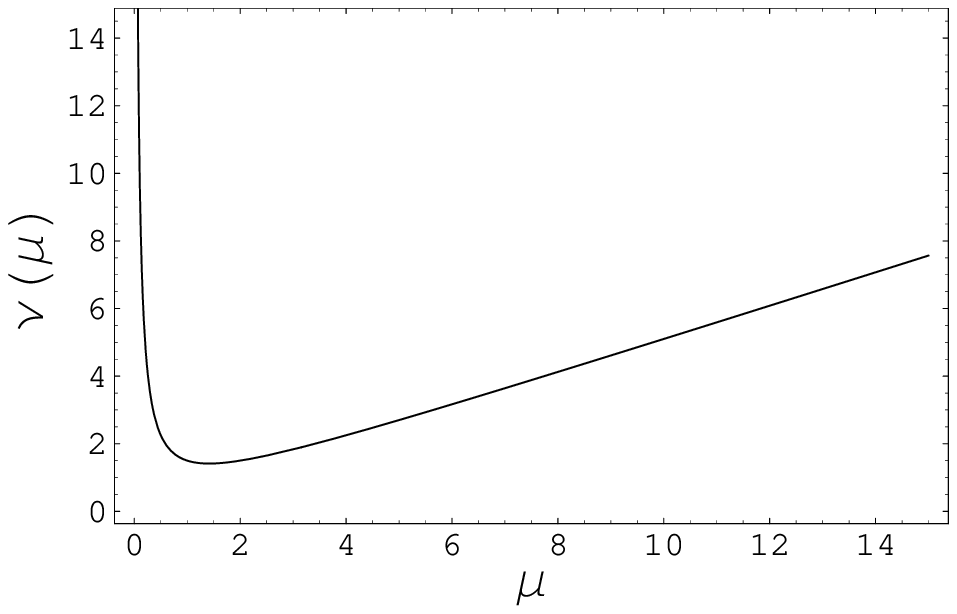}} \vskip 2ex
\begin{center} {\small{Fig. 3.5:}  Factorization curve of the parameter
$\nu=\nu(\mu)$ that allows the factorization of Eq. (\ref{ec35}).
$a_{1}=-\frac{\mu}{\sqrt{2}}$.}
\end{center}

\section{Generalized Burgers-Huxley equation}

In this section we obtain particular solutions for the generalized
Burgers-Huxley equation discussed by Wang et al. in \cite{wang2}
\begin{equation}
\frac{\partial u}{\partial t}-\alpha u^{\delta}\frac{\partial
u}{\partial x} - \frac{\partial^{2}u}{\partial x^{2}}=\beta
u(1-u^{\delta})(u^{\delta}-\gamma)~.\label{ec39}
\end{equation}

If the coordinates transformation $\xi=x-\nu t$ is performed then
Eq. (\ref{ec39}) is rewritten in the following form
\begin{equation}
u^{\prime\prime} + (\nu + \alpha u^{\delta})u^{\prime}+\beta
u(1-u^{\delta})(u^{\delta}-\gamma)=0~,\label{ec40}
\end{equation}
and the polynomial function allows the choice for the factorizing
terms
\begin{equation}
\phi_{1}(u)=\sqrt{\beta}a_{1}(1-u^{\delta})\quad\textrm{and}\quad\phi_{2}(u)
=\frac{\sqrt{\beta}}{a_{1}}(u^{\delta}-\gamma)~.\label{ec41}\nonumber
\end{equation}
Eq. (\ref{ec5}) provides $g(u)=\sqrt{\beta}\left(
\frac{\gamma-a_{1}^{2}}{a_{1}}
+\frac{a_{1}^{2}(1+\delta)-1}{a_{1}}u^{\delta}\right)$, and we can
do the following identification of constant parameters
\begin{equation}
\nu=\sqrt{\beta}\left(\frac{\gamma-a_{1}^{2}}{a_{1}}\right),\quad
\alpha=\sqrt{\beta}\left(\frac{-a_{1}^{2}(1+\delta)+1}{a_{1}}\right)~.\label{ec42}\nonumber
\end{equation}
Writing Eq. (\ref{ec40}) in factorized form
\begin{equation}
\left[D -  \frac{\sqrt{\beta}}{a_{1}}(u^{\delta}-\gamma)\right]
\left[ D - \sqrt{\beta}a_{1}(1-u^{\delta})
\right]u=0~,\label{ec43}
\end{equation}
the solution
\begin{equation}
u=\left( \frac{1}{1\pm
\textrm{exp}[-a_{1}\sqrt{\beta}\delta(\xi-\xi_{0})]} \right)^{1/
\delta} \label{ec45}
\end{equation}
of the compatible first order equation
\begin{equation}
u^{\prime}-\sqrt{\beta}a_{1}u(1-u^{\delta})=0~,\label{ec44}
\end{equation}
is also a particular kink solution of Eq. (\ref{ec40}). Solving
the quadratic equation (\ref{ec42}) for
$a_{1}=a_{1}(\alpha,\beta,\delta)$ we obtain
\begin{equation}
a_{1_{+,-}}=\frac{-\alpha\pm\sqrt{\alpha^2+4\beta(1+\delta)}}{2\sqrt{\beta}(1+\delta)}~,\label{ec45a}\nonumber
\end{equation}
then Eq. (\ref{ec45}) becomes a function
$u=u(\alpha,\beta,\delta;\tau)$, and
$\nu=\nu(\alpha,\beta,\gamma,\delta)$. If we set $\delta=1$ in Eq.
(\ref{ec39}), then we obtain the following particular
Burgers-Huxley solution
\begin{equation}
u= \frac{1}{1\pm
\textrm{exp}[-a_{1}\sqrt{\beta}(\xi-\xi_{0})]}~,\label{ec45b}
\end{equation}
and $a_{1_{+,-}}=\frac{-\alpha\pm\sqrt{\alpha^2+8\beta}}
{4\sqrt{\beta}}$, $\nu=\nu(\alpha,\beta,\gamma)$.

\bigskip
\vskip 1ex \centerline{ \epsfxsize=200pt \epsfbox{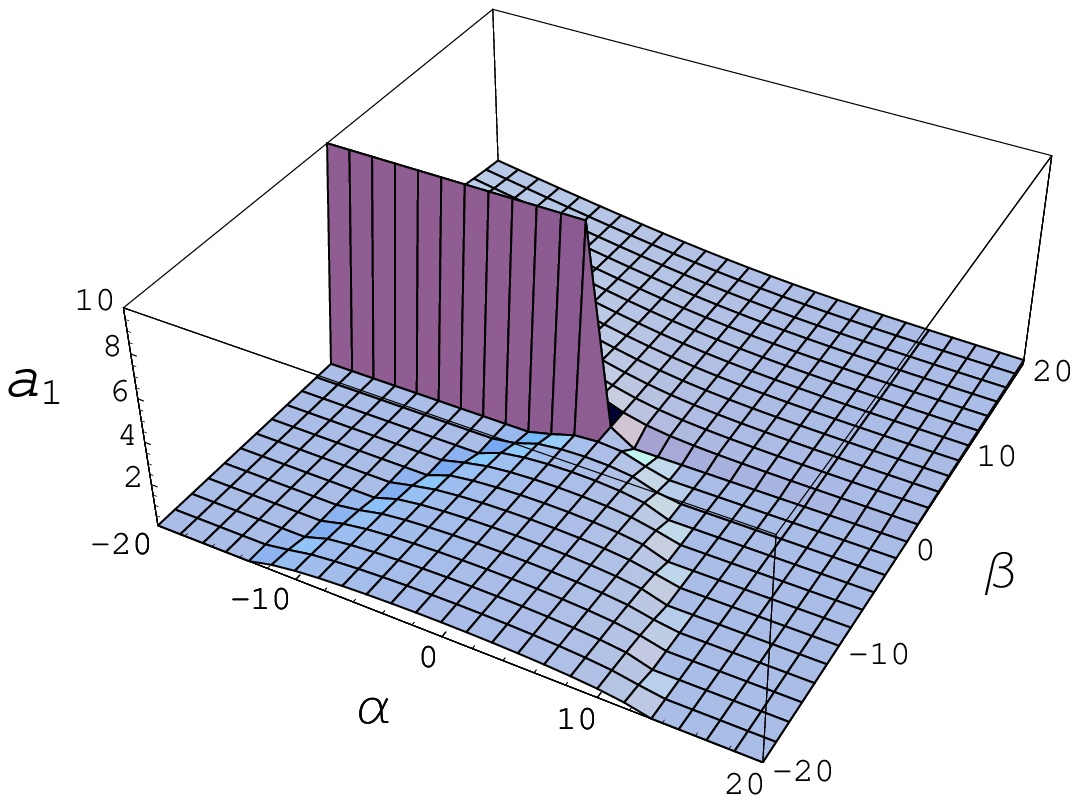}}
\vskip 2ex
\begin{center}
{\small{Fig. 3.6:} Real part for the factorization curve of the
parameter $a_{1_{+}}=a_{1_{+}}(\alpha,\beta,\delta=1)$ that allows
factorization of Eq. (\ref{ec40}) with $\delta=1$. $a_{1}\neq 0$.
$\alpha\in[-20,20]$ and $\beta\in[-20,20]$.}
\end{center}
\bigskip
\vskip 1ex \centerline{ \epsfxsize=200pt \epsfbox{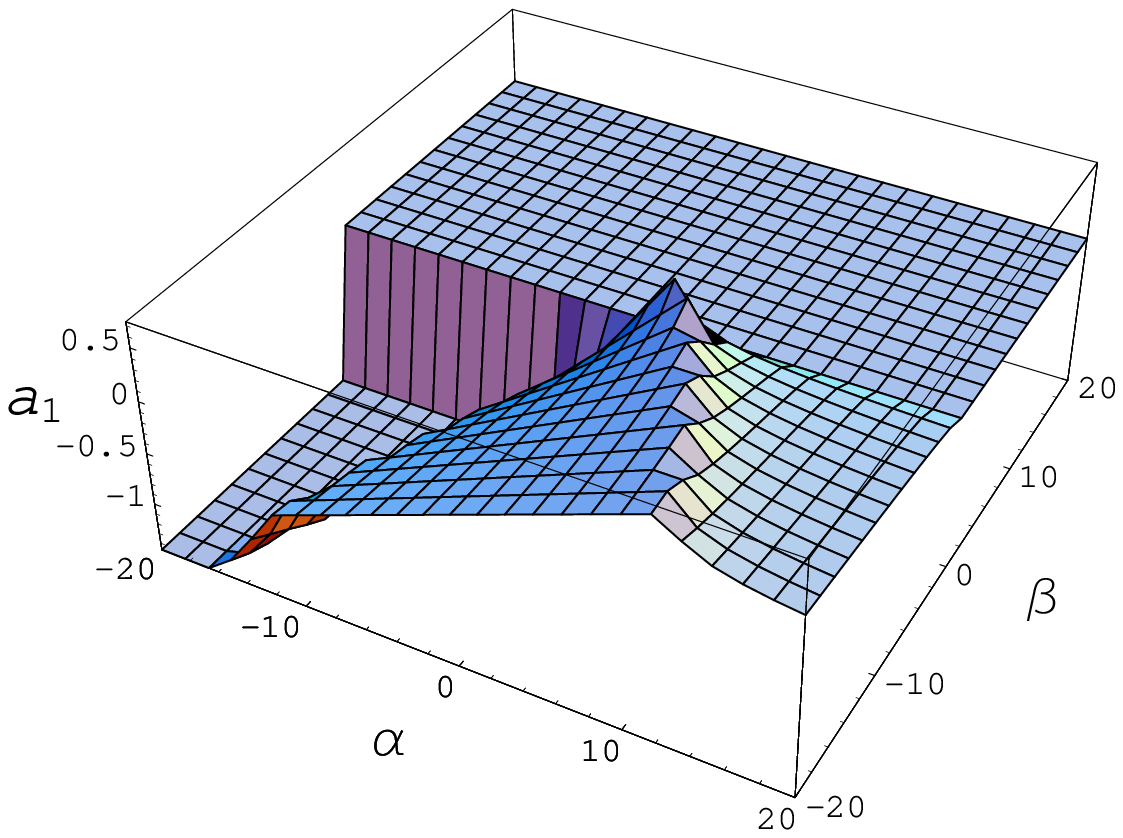}}
\vskip 2ex
\begin{center}
{\small{Fig. 3.7:} Imaginary part for the factorization curve of
the parameter $a_{1_{+}}=a_{1_{+}}(\alpha,\beta,\delta=1)$ that
allows factorization of Eq. (\ref{ec40}) with $\delta=1$.
$a_{1}\neq 0$. $\alpha\in[-20,20]$ and $\beta\in[-20,20]$.}
\end{center}


\bigskip

If we chose now the factorizing terms as
\begin{equation}
\phi_{1}(u)=\sqrt{\beta}e_{1}(u^{\delta}-\gamma)\quad\textrm{and}\quad\phi_{2}(u)
=\frac{\sqrt{\beta}}{e_{1}}(1-u^{\delta})~,\label{ec46}\nonumber
\end{equation}
we obtain $g(u)=\sqrt{\beta}\left( \frac{e_{1}^{2}\gamma
-1}{e_{1}} +\frac{1-e_{1}^{2}(1+\delta)}{e_{1}}u^{\delta}\right)$,
and the following identification of parameters
$\nu=\sqrt{\beta}\left(\frac{e_{1}^{2}\gamma -1}{e_{1}}\right)$
and
$\alpha=\sqrt{\beta}\left(\frac{1-e_{1}^{2}(1+\delta)}{e_{1}}\right)$.
Eq. (\ref{ec40}) is then factorized in the different form
\begin{equation}
\left[D - \frac{\sqrt{\beta}}{e_{1}}(1-u^{\delta}) \right] \left[
D -\sqrt{\beta}e_{1}(u^{\delta}-\gamma)
 \right]u=0~.\label{ec47}
\end{equation}
The corresponding compatible first order equation is now
\begin{equation}
u^{\prime}-\sqrt{\beta}e_{1}u(u^{\delta}-\gamma)=0~,\label{ec48}
\end{equation}
and its integration gives a different particular solution for Eq.
(\ref{ec40}) from that obtained for the first choice of
factorizing terms (\ref{ec41}), however, we point out that the
parameter $\alpha$ has changed for the second choice of
factorizing terms. The solution of Eq. (\ref{ec48}) is given as
follows
\begin{equation}
u= \left( \frac{\gamma}{1\pm \textrm{exp}[\pm
e_{1}\sqrt{\beta}\gamma\delta(\xi-\xi_{0})]} \right)^{1/ \delta}~.
\label{ec49}
\end{equation}
Solving the quadratic equation for
$e_{1}=e_{1}(\alpha,\beta,\delta)$ we obtain
\begin{equation}
e_{1_{+,-}}=\frac{\alpha\pm\sqrt{\alpha^2+4\beta(1+\delta)}}{2\sqrt{\beta}(1+\delta)}~,\label{ec49a}\nonumber
\end{equation}
then Eq. (\ref{ec49}) becomes
$u=u(\alpha,\beta,\gamma,\delta;\tau)$, and
$\nu=\nu(\alpha,\beta,\gamma,\delta)$. If we set $\delta =1$ in
Eq. (\ref{ec39}), then the following Burgers-Huxley solution is
obtained
\begin{equation}
u= \frac{\gamma}{1\pm
\textrm{exp}[e_{1}\sqrt{\beta}\gamma(\xi-\xi_{0})]}~,\label{ec49b}
\end{equation}
and
$e_{1_{+,-}}=\frac{\alpha\pm\sqrt{\alpha^2+8\beta}}{4\sqrt{\beta}}$,
$\nu=\nu(\alpha,\beta,\gamma)$.

\bigskip
\vskip 1ex \centerline{ \epsfxsize=200pt \epsfbox{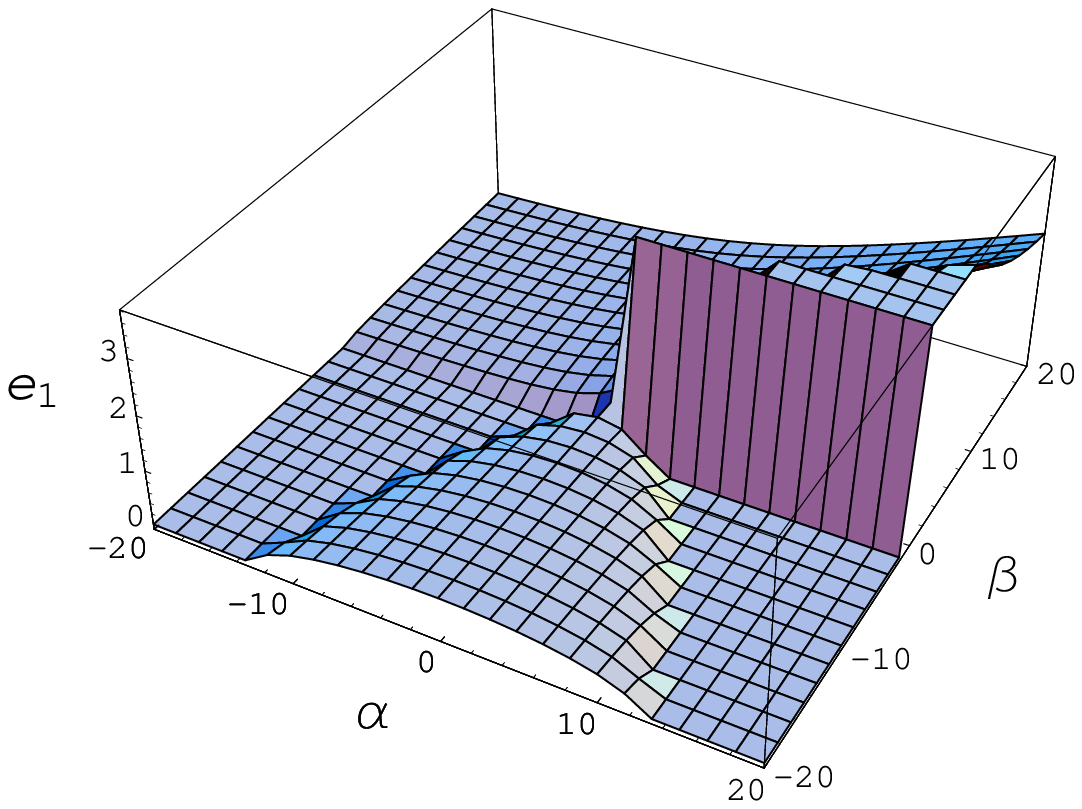}}
\vskip 2ex
\begin{center}
{\small{Fig. 3.8:} Real part for the factorization curve of the
parameter $e_{1_{+}}=e_{1_{+}}(\alpha,\beta,\delta=1)$ that allows
factorization of Eq. (\ref{ec40}) with $\delta=1$. $e_{1}\neq 0$.
$\alpha\in[-20,20]$ and $\beta\in[-20,20]$.}
\end{center}
\bigskip
\vskip 1ex \centerline{ \epsfxsize=200pt \epsfbox{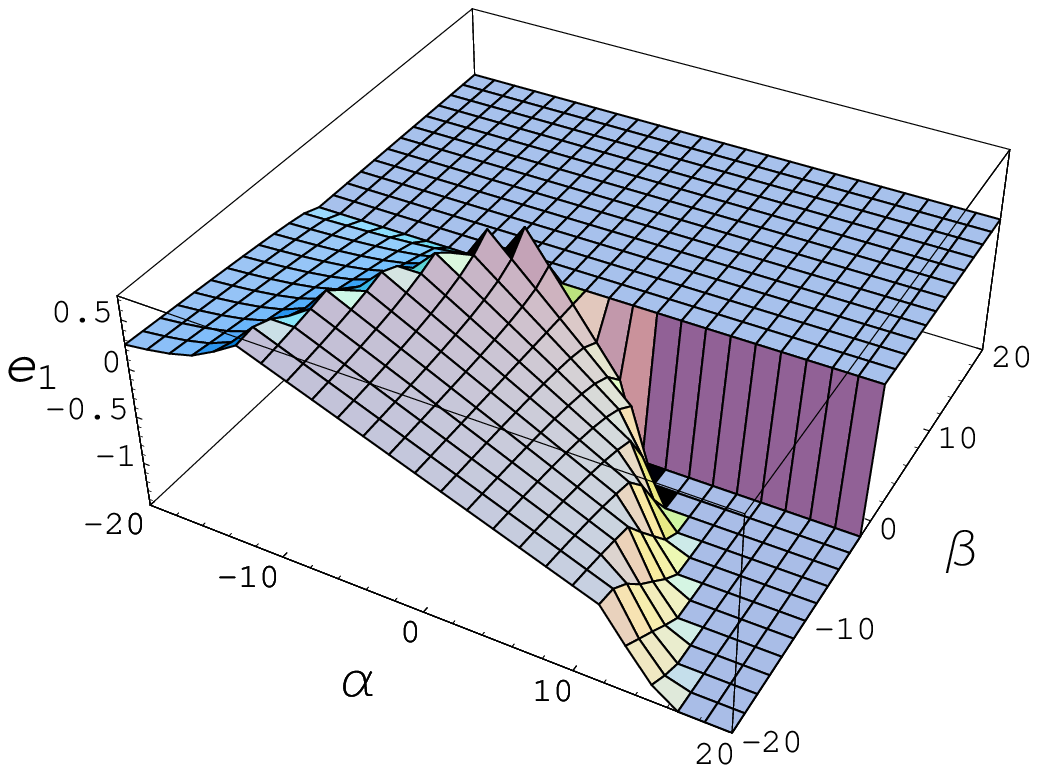}}
\vskip 2ex
\begin{center}
{\small{Fig. 3.9:} Imaginary part for the factorization curve of
the parameter $e_{1_{+}}=e_{1_{+}}(\alpha,\beta,\delta=1)$ that
allows factorization of Eq. (\ref{ec40}) with $\delta=1$.
$e_{1}\neq 0$. $\alpha\in[-20,20]$ and $\beta\in[-20,20]$.}
\end{center}

\bigskip
Eqs. (\ref{ec45}) and (\ref{ec49}) representing particular
solutions for the GBHE, are the same as those obtained
by Wang et al. \cite{wang2}.\\

\bigskip

\section{Conclusion of the chapter}

In this chapter, we apply the same factorization scheme for more
complicated second order nonlinear differential equations as in
Chapter 2. Exact particular solutions have been found for a series
of nonlinear differential equations with applications in physics
and biology: the modified Emden equation, the generalized Lienard
equation, the Duffing-van der Pol equation, the convective Fisher
equation, and the generalized Burgers-Huxley equation. Also, we
display parametric curves along which the differential equations
under consideration could be factorized. We find that the proposed
factorization procedure is easier and more efficient than other
methods used to find particular solutions of second order
differential equations.

\chapter{One-parameter supersymmetry for microtubules}







{\small 

{\bf Abstract}. The simple supersymmetric model of Caticha
\cite{cat} as used by Rosu \cite{rosu} to describe the motion of
ferrodistortive domain walls in microtubules (MTs), is generalized
to the case of Mielnik's one-parameter nonrelativistic
supersymmetry \cite{m84}. By this means, one can introduce
Montroll double-well potentials with singularities that move along
the positive or negative travelling direction depending on the
sign of the free parameter of Mielnik's method. Possible
interpretations of the singularity are microtubule associated
proteins (motors) or structural discontinuities in the arrangement
of the tubulin molecules.

{


\bigskip





\section{Introduction}

Based on well-established results of Collins, Blumen, Currie and
Ross \cite{1} regarding the dynamics of domain walls in
ferrodistortive materials, Tuszy\'nski and collaborators
\cite{mtub1,mtub2} considered MTs to be ferrodistortive and
studied kinks of the Montroll type \cite{mont} as excitations
responsible for the energy transfer within this highly interesting
biological context.

The Euler-Lagrange dimensionless equation of motion of
ferrodistortive domain walls as derived in \cite{1} from a
Ginzburg-Landau free energy with driven field and dissipation
included is of the travelling reaction-diffusion type
\begin{equation} \label{1}
\psi ^{''}+\rho\psi ^{'}-\psi ^3 +\psi+\sigma=0~,
\end{equation}
where the primes are derivatives with respect to a travelling
coordinate $\xi =x-vt$, $\rho$ is a friction coefficient and
$\sigma$ is related to the driven field \cite{1}.

There may be ferrodistortive domain walls that can be identified
with the Montroll kink solution of Eq.~(\ref{1})
\begin{equation}  \label{2}
M(\xi)=\alpha _1+\frac{\sqrt{2}\beta}{1+\exp(\beta\xi)}~,
\end{equation}
where $\beta=(\alpha _2-\alpha _1)/\sqrt{2}$
and the parameters $\alpha _1$ and $\alpha _2$ are two nonequal solutions of the cubic equation
\begin{equation} \label{3}
(\psi -\alpha _1)(\psi -\alpha _2)(\psi -\alpha _3)=\psi ^3 -\psi -\sigma~.
\end{equation}

\section{Caticha's supersymmetric model as applied to MTs}

Rosu has noted that Montroll's kink can be written as a typical
$\tanh$ kink \cite{rosu}
\begin{equation} \label{M}
M(\xi)=
\gamma -\tanh\left(\frac{\beta \xi}{2}\right)~,
\end{equation}
where $\gamma \equiv \alpha _1 +\alpha _2=1+\frac{\alpha _1\sqrt{2}}{\beta}$.
The latter relationship allows one to use a simple construction method of exactly
soluble
double-well potentials in the Schr\"odinger equation proposed by
Caticha \cite{cat}. The scheme is a non-standard application of
Witten's supersymmetric quantum mechanics \cite{w81} having as the
essential assumption the idea of considering the $M$ kink as the
switching function between the two lowest eigenstates of the
Schr\"odinger equation with a double-well potential. Thus
\begin{equation} \label{phiM}
\phi _1=M\phi _0~,
\end{equation}
where $\phi _{0,1}$ are solutions of $\phi ^{''}_{0,1}+[\epsilon _{0,1}-u(\xi)]
\phi _{0,1}(\xi)=0$, and $u(\xi)$ is the double-well potential to be found.
Substituting Eq.~(\ref{phiM}) into the Schr\"odinger equation for
the subscript 1 and substracting the same equation multiplied by the
switching function for the subscript 0, one obtains
\begin{equation} \label{phiR}
\phi ^{'}_{0}+R_M\phi _0=0~,
\end{equation}
where $R_M$ is given by
\begin{equation} \label{R}
R_M=\frac{M^{''}+\epsilon M}{2M^{'}}~,
\end{equation}
and $\epsilon=\epsilon _1-\epsilon _0$ is the lowest energy splitting in
the double-well Schr\"odinger equation.
In addition, notice that Eq.~(\ref{phiR}) is the basic equation introducing the superpotential $R$ in
Witten's supersymmetric quantum mechanics, i.e., the Riccati solution.
For Montroll's kink the corresponding Riccati solution reads
\begin{equation} \label{RM}
R_M(\xi)=-\frac{\beta}{2}{\rm tanh}\left(\frac{\beta}{2}\xi\right)+\frac{\epsilon}{2\beta}\Bigg[\sinh(\beta\xi)+
2\gamma\cosh ^2\left(\frac{\beta}{2}\xi\right)\Bigg]
\end{equation}
and the ground-state Schr\"odinger function is found by means of Eq.~(\ref{phiR})
\begin{eqnarray} \label{phi0}
\phi _{0,M}(\xi) &=& \phi
_0(0)\cosh\left(\frac{\beta}{2}\xi\right)
\exp\left(\frac{\epsilon}{2\beta ^2}\right)
\exp\left(-\frac{\epsilon}{2\beta ^2}\Big[ \cosh (\beta \right.
\xi)\nonumber\\
&& \left. -\gamma \beta\xi-\gamma\sinh(\beta\xi)\Big]\right)~,
\end{eqnarray}
while $\phi _1$ is obtained by switching the ground-state wave function by means of $M$.
This ground-state wave function is of supersymmetric type
\begin{equation} \label{phi}
\phi _{0,M}(\xi)=\phi _{0,M}(0)\exp\Bigg[-\int_0^{\xi} R_M(y)dy\Bigg]~,
\end{equation}
where $\phi _{0,M}(0)$ is a normalization constant.

The Montroll double well potential is determined up to the additive constant $\epsilon _0$ by the `bosonic' Riccati equation
\begin{eqnarray} \label{u}
u_M(\xi)&=&R_M^2-R_M^{'}+\epsilon _0 = \frac{\beta
^2}{4}+\frac{(\gamma ^2 -1)\epsilon ^2}{4\beta
^2}+\frac{\epsilon}{2}+\epsilon _0 \nonumber\\
&&+ \frac{\epsilon}{8\beta^2}\Big[\left(4\gamma ^2\epsilon
+2(\gamma^2 +1)\epsilon{\rm cosh} (\beta \xi )-8\beta^2\right){\rm
cosh} (\beta \xi ) \nonumber\\
&&-4\gamma \left(\epsilon +\epsilon {\rm cosh}(\beta \xi)-2\beta
^2 \right) {\rm sinh } (\beta \xi)\Big].
\end{eqnarray}
Plots of the asymmetric Montroll potential and ground state wave
function are given in Figs.~4.1 and 4.2 for a particular set of
the parameters. If, as suggested by Caticha, one chooses the
ground state energy to be
\begin{equation} \label{eps}
\epsilon _0=-\frac{\beta ^2}{4}-\frac{\epsilon}{2}+\frac{\epsilon ^2}{4\beta ^2}
\left(1-\gamma ^2\right)~,
\end{equation}
then $u_M(\xi)$ turns into a travelling, asymmetric Morse
double-well potential of depths depending on the Montroll
parameters $\beta$ and $\gamma$ and the splitting $\epsilon$
\begin{equation} \label{U}
U_{0, m}^{L,R}=\beta ^2\Bigg[1\pm \frac{2\epsilon \gamma}{(2\beta)^2}\Bigg]~,
\end{equation}
where the subscript $m$ stands for Morse and the superscripts $L$
and $R$ for left and right well, respectively. The difference in
depth, the bias, is $\Delta _m\equiv U_0^L-U_0^R=2\epsilon\gamma$,
while the location of the potential minima on the travelling axis
is at
\begin{equation}  \label{xiLR}
\xi _{m}^{L,R}=\mp\frac{1}{\beta}\ln \Bigg[\frac{(2\beta)^2\pm
2\epsilon\gamma}{\epsilon(\gamma\mp 1)}\Bigg]~,
\end{equation}
that shows that $\gamma \neq \pm 1$.
\bigskip

\section{The Mielnik extension}

We now discuss shortly in this context the Mielnik extension of
these results \cite{m84,review2}. The point is that $R_M$ in
Eq.~(\ref{RM}) is only the particular solution of Eq.~(\ref{u}).
The general solution is a one-parameter function of the form
\begin{equation} \label{genR}
R_M(\xi ;\lambda)=R_M(\xi)+\frac{d}{d\xi}\Big[\ln(I_M(\xi)+\lambda)\Big]
\end{equation}
and the corresponding one-parameter Montroll potential is given by
\begin{equation} \label{genu}
u_M(\xi ;\lambda)=u_M(\xi)-2\frac{d^2}{d\xi ^2}\Big[\ln(I_M(\xi)+\lambda)\Big]~.
\end{equation}
In these formulas, $I_M(\xi)=\int ^{\xi}\phi _{0,M}^2(\xi)d\xi$
and $\lambda$ is an integration constant that is used as a
deforming parameter of the potential and is related to the
irregular zero mode. The one-parameter Darboux-deformed ground
state wave function can be shown to be
\begin{equation} \label{wfl}
\phi_{0,M}(\xi ;\lambda)=\sqrt{\lambda(\lambda+1)}\frac{\phi _{0,M}}{I_M(\xi)+\lambda}~,
\end{equation}
where $\sqrt{\lambda(\lambda+1)}$ is the normalization factor
implying that $\lambda \notin [0,-1]$. Moreover, the Mielnik
parametric potentials and wave functions display singularities at
$\lambda_s=-I_M(\xi_s)$. Plots of $u_M(\xi ;\lambda)$ and $\phi
_{0,M}(\xi; \lambda)$ for $\lambda =1, 10, 100$ are presented in
Figs.~4.3-4.12 and are useful to see the behavior of the
singularities and the deformation effect of the $\lambda$
parameter. For large values of $\pm\lambda$ the singularity moves
towards $\mp \infty$ and the potential and ground state wave
function recover the shapes of the non-parametric potential and
wave function as can be seen in Figs.~4.11 and 4.12, respectively.
The one-parameter Morse case corresponds formally to the change of
subscript $M\rightarrow m$ in Eqs.~(\ref{genR}) and (\ref{genu}).
For the single well Morse potential  the one-parameter procedure
has been studied by Filho \cite{filho} and Bentaiba et al
\cite{bentaiba}.

Mielnik's approach leads to singularities in the double-well
potential and the corresponding wave functions. If the parameter
$\lambda$ is positive the singularity is to be found on the
negative $\xi$ axis, while for negative $\lambda$ it is on the
positive side. Potentials and wave functions with singularities
are not so strange as it seems \cite{cs} and could be quite
relevant even in nanotechnology \cite{fulop} where quantum
singular interactions of the contact type are appropriate for
describing nanoscale quantum devices. We interpret the singularity
as representing the effect of an impurity moving along the MT in
one direction or the other depending on the sign of the parameter
$\lambda$. The impurity may represent a protein attached to the MT
or a structural discontinuity in the arrangement of the tubulin
molecules. This interpretation of impurities has been given by
Trpi\v{s}ov\'a and Tuszy\'nski in non-supersymmetric models of
nonlinear MT excitations \cite{tt}.

\bigskip
\section{Conclusion of the chapter}

In conclusion, the supersymmetric approaches allow for a number of
interesting {\em exact} results in this biological framework and
point to a direct connection between Schr\"odinger double-well
potentials and nonlinear kinks encountered in nonequilibrium
chemical processes. MTs are an important application but the
procedures described here can be used in many other applications.
Moreover, the supersymmetric constructions can be used as a
background for clarifying further details of the exact models.
Although it is not so clear why one should take a certain type of
kink as switching function between the Schr\"odinger split modes,
it is interesting that proceeding in this way one will be led to
some familiar double-well potentials in chemical physics.
%

\newpage

\vskip 1ex \centerline{ \epsfxsize=250pt
\epsfbox{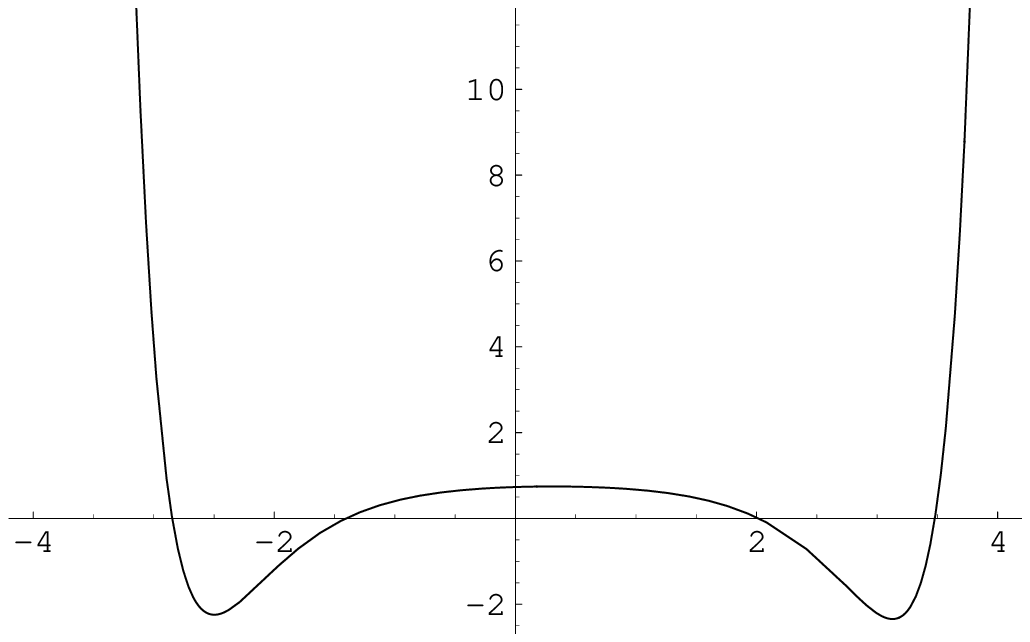}} \vskip 3ex
\begin{center}
{\small{Fig. 4.1:}$\quad$
 The Montroll asymmetric double-well potential (MDWP) calculated using Eq.~(\ref{u}) for $\epsilon _0=0$.
In all figures $\alpha _1=1$, $\alpha _2=-1.5$, $\beta
=-2.5/\sqrt{2}$, $\gamma =-0.5$, $\epsilon =0.1$. }
\end{center}

\vskip 1ex \centerline{ \epsfxsize=250pt
\epsfbox{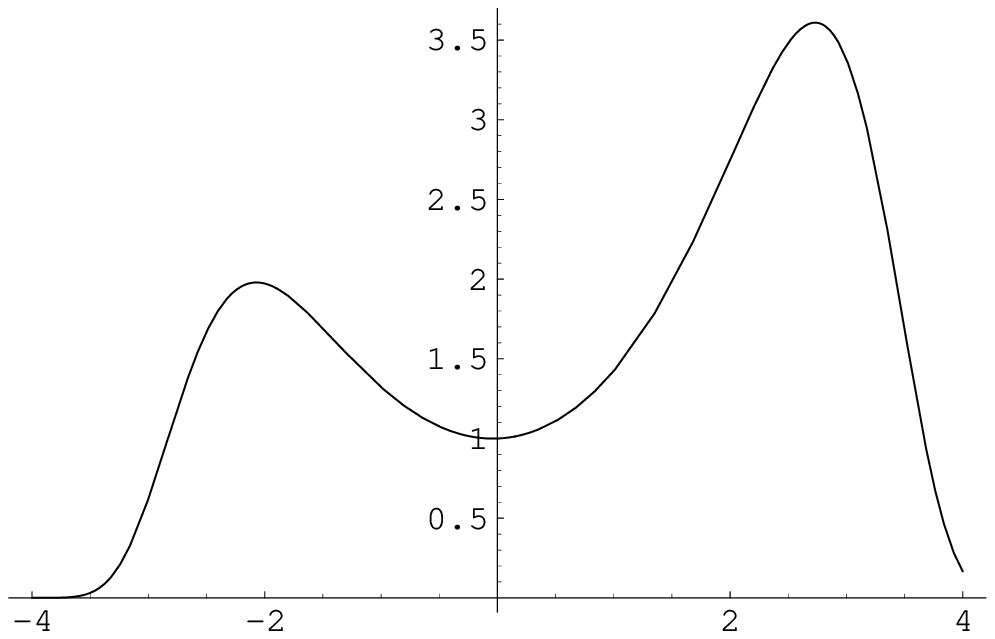}} \vskip 3ex
\begin{center}
{\small{Fig. 4.2:}$\quad$
 The Montroll ground state wave function cf Eq.~(\ref{phi0}) for $\phi _0(0)=1$.
}
\end{center}


\vskip 1ex \centerline{ \epsfxsize=250pt
\epsfbox{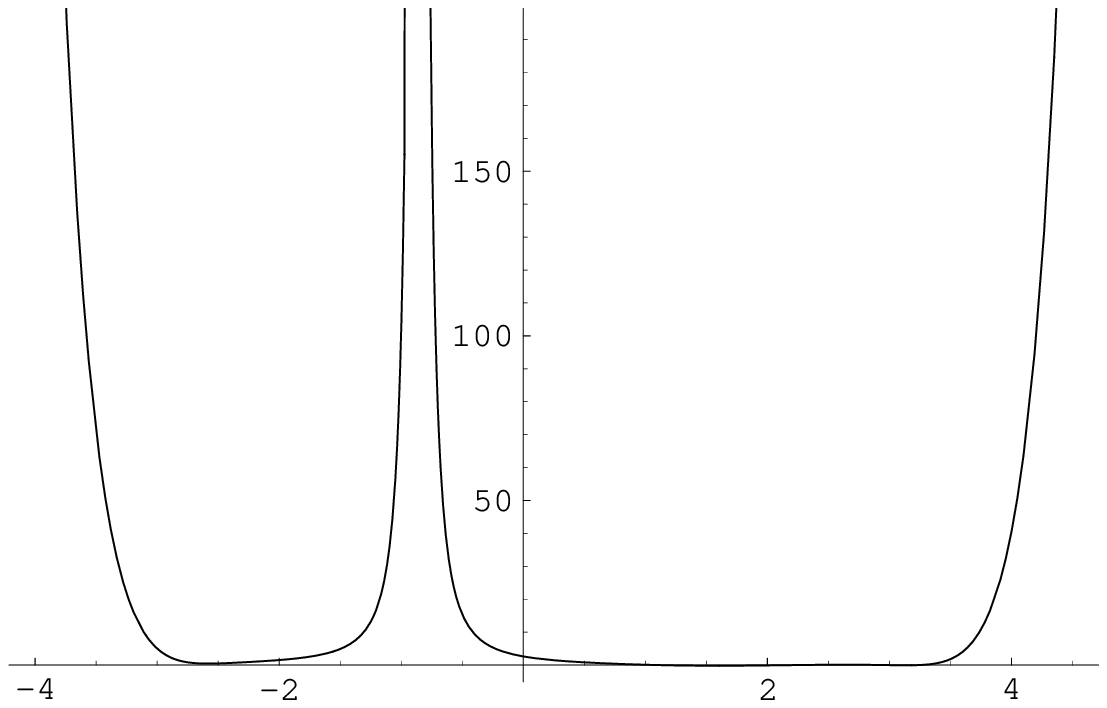}}   
\vskip 3ex
\begin{center}
{\small{Fig. 4.3:}$\quad$
 The one-parameter Darboux modified MDWP for $\lambda =1$.
}
\end{center}

\vskip 1ex \centerline{ \epsfxsize=250pt \epsfbox{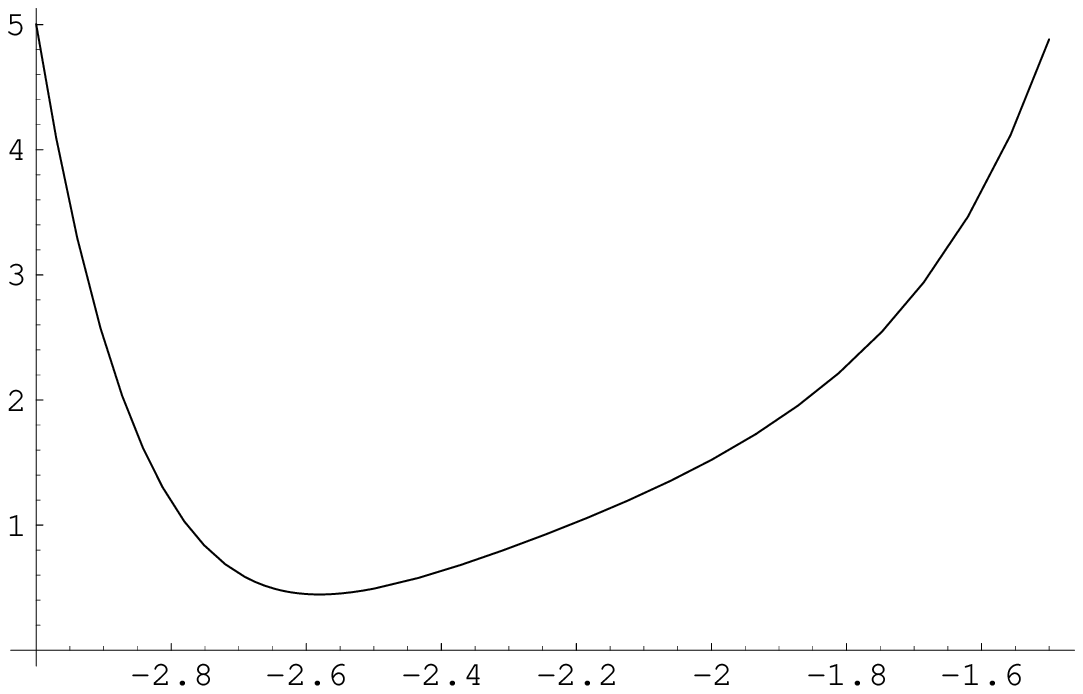}}
\vskip 3ex
\begin{center}
{\small{Fig. 4.4:}$\quad$
 The low-scale left hand side of the singularity.
}
\end{center}

\vskip 1ex \centerline{ \epsfxsize=250pt \epsfbox{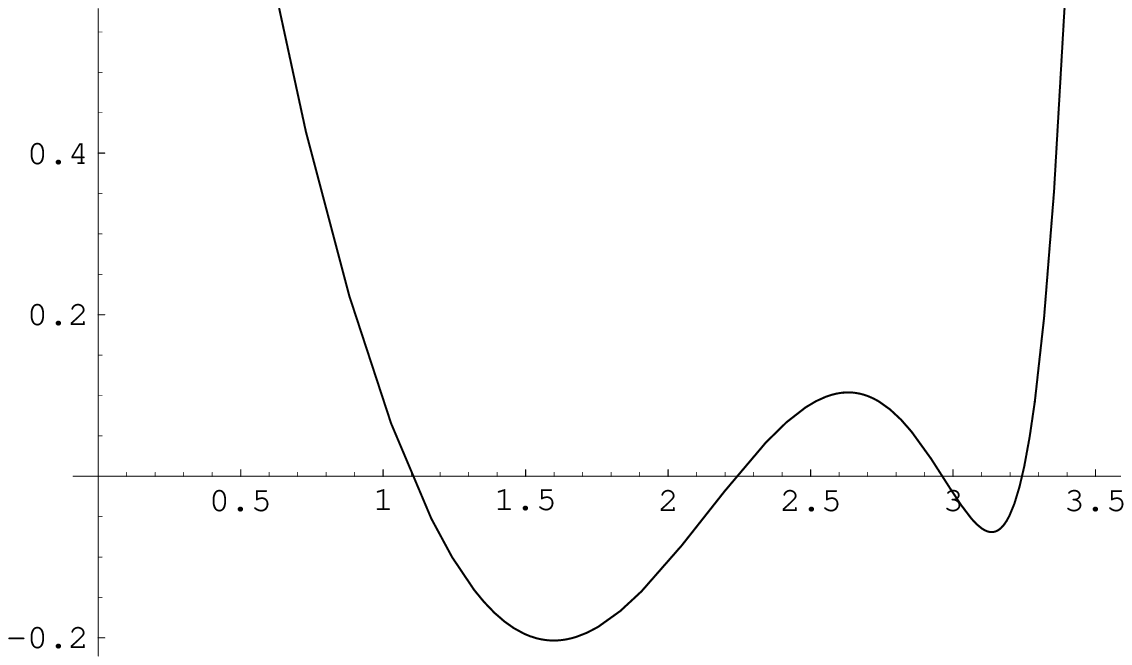}}
\vskip 3ex
\begin{center}
{\small{Fig. 4.5:}$\quad$ The low-scale right hand side of the
singularity. }
\end{center}

\vskip 1ex \centerline{ \epsfxsize=250pt
\epsfbox{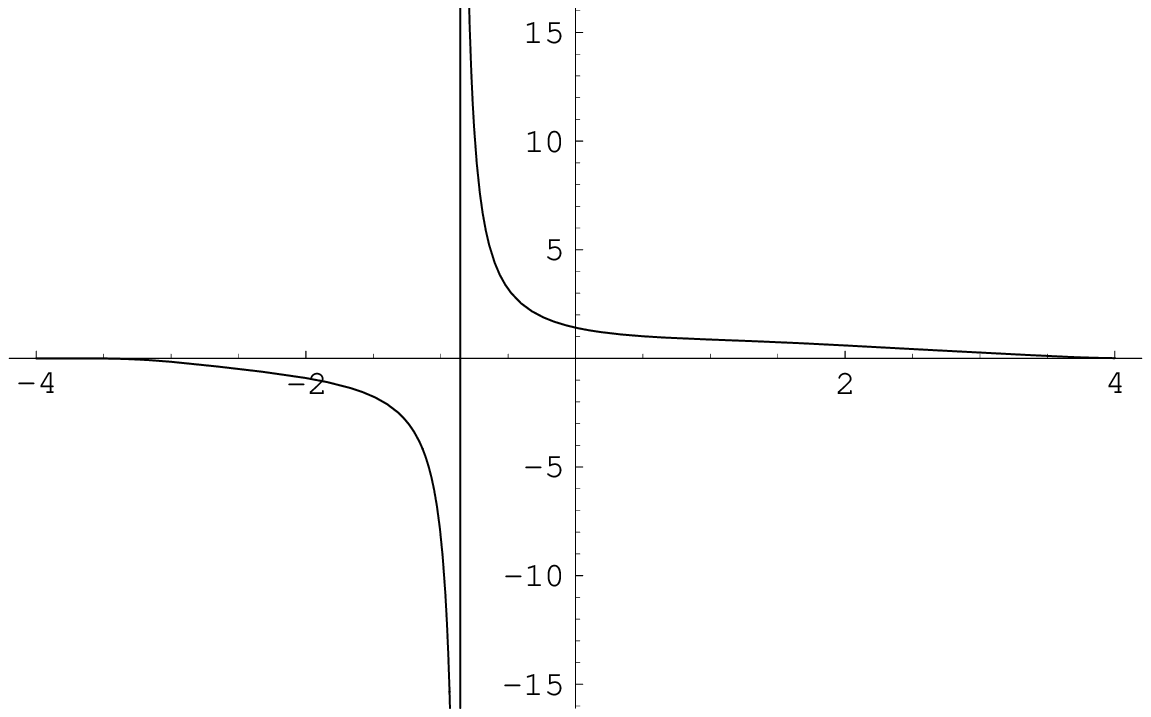}}    
\vskip 3ex
\begin{center}
{\small{Fig. 4.6:}$\quad$
 The wave functions for $\lambda =1$.
}
\end{center}





\vskip 1ex \centerline{ \epsfxsize=250pt \epsfbox{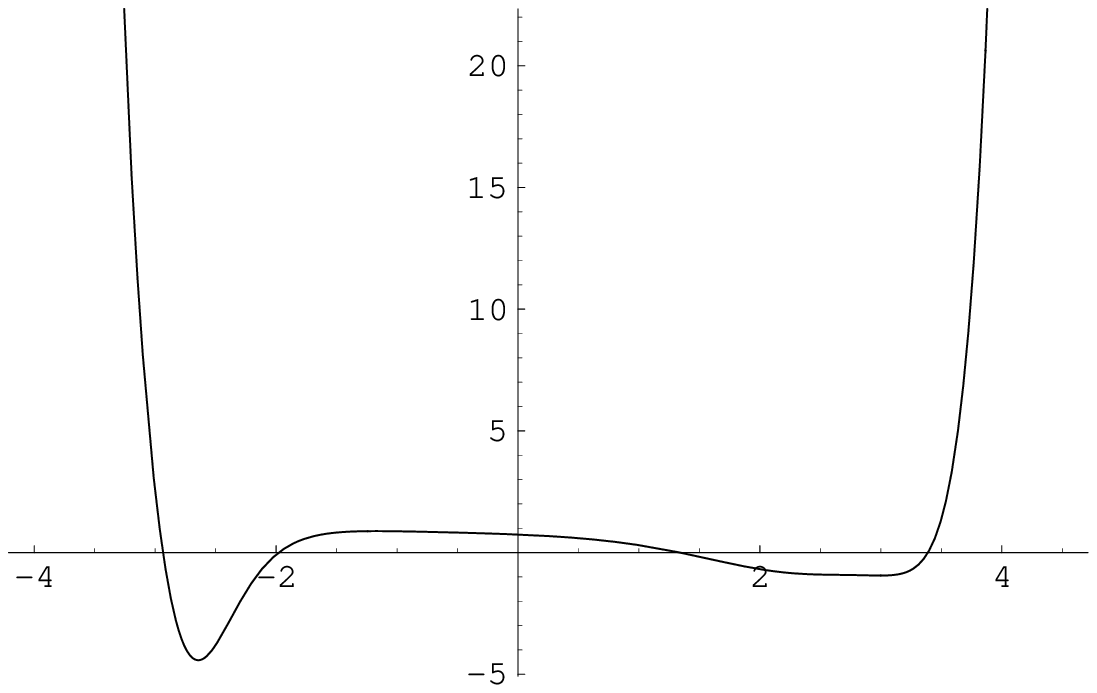}}
\vskip 3ex
\begin{center}
{\small{Fig. 4.7:}$\quad$
 One parameter Darboux-modified MDWP for $\lambda =10$ .}
\end{center}

\vskip 1ex \centerline{ \epsfxsize=250pt \epsfbox{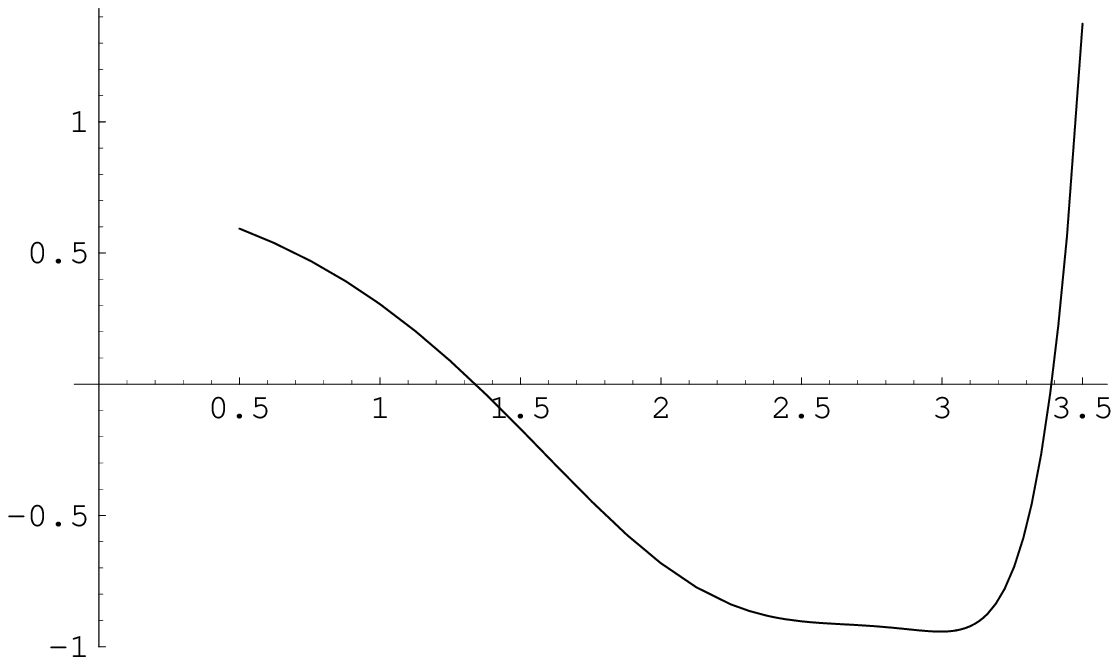}}
\vskip 3ex
\begin{center}
{\small{Fig. 4.8:}$\quad$ The bottom of the potential at the right
hand side. }
\end{center}


\vskip 1ex \centerline{ \epsfxsize=250pt \epsfbox{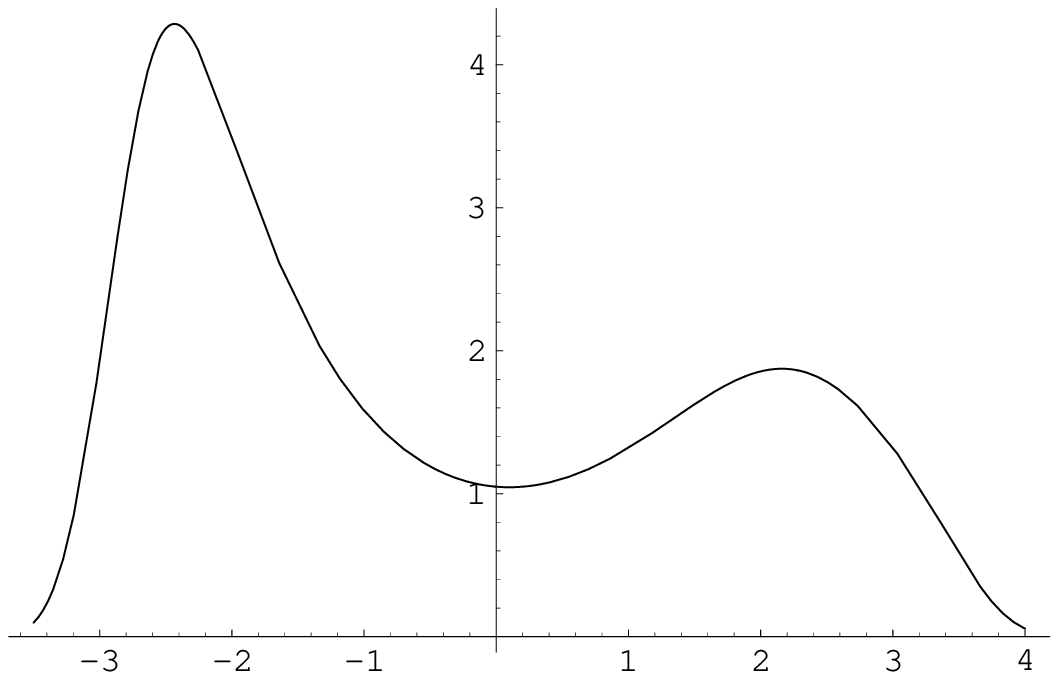}} \vskip
3ex
\begin{center}
{\small{Fig. 4.9:}$\quad$ The ground state wave function
corresponding to $\lambda =10$. }
\end{center}





\vskip 1ex \centerline{ \epsfxsize=250pt
\epsfbox{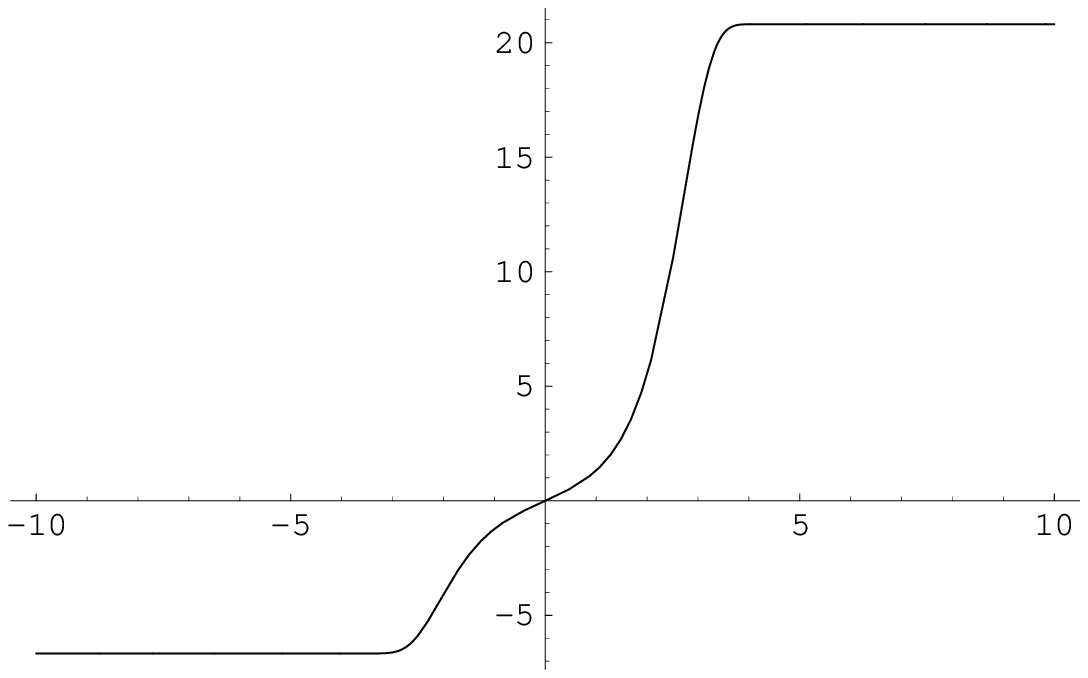}} \vskip 3ex
\begin{center}
{\small{Fig. 4.10:}$\quad$ Plot of the integral $I_M(\xi)$ that
produces the deformation of the potential and wave functions. }
\end{center}

\chapter{Supersymmetric method with Dirac parameters}



{\small {\bf Abstract}. In this chapter we first describe a
"supersymmetric" one-dimensional matrix procedure similar to
relationships of the same type between Dirac and Schr\"odinger
equations in particle physics that we apply to two problems in
classical mechanics and quantum mechanics, respectively. In the
first case, we obtain a class of parametric oscillation modes that
we call K-modes with damping and absorption that are connected to
the classical harmonic oscillator modes through this
supersymmetric procedure that is characterized by coupling
parameters. When a single coupling parameter, denoted by K, is
used, it characterizes both the damping and the dissipative
features of these modes. Generalizations to several K parameters
are also possible and lead to analytical results. If the problem
is passed to the physical optics (and/or acoustics) context by
switching from the oscillator equation to the corresponding
Helmholtz equation, one may hope to detect the K-modes as
waveguide modes of specially designed waveguides and/or cavities.
In the second case, the same method is presented in a style
appropriate for truly quantum mechanical problems and an
application to the Morse potential is performed. We obtain the
corresponding nonhermitic Morse problem with possible applications
to the diffraction on optical lattices.}

{


\bigskip

\section{Introduction} $\quad$ \\
Factorizations of differential operators describing simple
mechanical motion have been only occasionally used in the past,
although in quantum mechanics the procedure led to a vast
literature under the name of supersymmetric quantum mechanics
initiated by a paper of Witten \cite{w81}. However, as shown by
Rosu and Reyes \cite{rr98}, for the damped Newtonian free
oscillator the factorization method could generate interesting
results even in an area settled more than three centuries ago. In
this chapter, we apply some of the supersymmetric schemes to the
basic classical harmonic oscillator. In particular, we show how a
known connection in particle physics between Dirac and
Schr\"odinger equations could lead in the case of harmonic motion
to chirped (i.e., time-dependent) frequency oscillator equations
whose solutions are a class of oscillatory modes depending on one
more parameter, denoted in the following by K, besides the natural
circular frequency $\omega _0$. The parameter K characterizes both
the damping and the losses of these "supersymmetric" partner
modes. Moreover, we do not limit this study to one K parameter
extending it to several such parameters still getting analytic
results. Guided by mathematical equivalence, possible applications
in several areas of physics are identified. Moreover, in the final
part of the chapter the same supersymmetric scheme is used in the
context of exactly solvable quantum problem of the Morse
potential. A nonhermitic version of the Morse problem is
introduced in this way.

\bigskip


\section{Classical harmonic oscillator: The Riccati approach}

The harmonic oscillator can be described by one of the simplest Riccati equation
\begin{equation} \label{ricc}
{\rm u^{'}+u^2+\kappa\omega _0^2=0, \qquad  \kappa =\pm 1~,}
\end{equation}
where the plus sign is for the normal case whereas the minus sign
is for the up side down case. Indeed, employing ${\rm
u=\frac{w^{'}}{w}}$ one gets the harmonic oscillator differential
equation
\begin{equation} \label{schr1}
{\rm w^{''}+\kappa\omega_{0}^2w=0~,}
\end{equation}
with the solutions
$$
{{\rm w_{b}}=
\left\{ \begin{array}{ll}
{\rm W_+\cos(\omega _0 t +\varphi _{+})} & \mbox{if $\kappa =1$}\\
{\rm W_{-}{\rm sinh}(\omega _0 t +\varphi _{-})} & \mbox{if $\kappa =-1$~,}
\end{array} \right.}
$$
where ${\rm W_{\pm }}$ and $\varphi _{\pm}$
are amplitude and phase parameters, respectively, which can be ignored in the following.

The particular Riccati solution
of Eq.~(\ref{ricc}) are
$$
{\rm u_{p}=
\left\{ \begin{array}{ll}
-\omega _0{\rm tan (\omega _0 t)} & \mbox{if $\kappa =1$}\\
\omega _0{\rm coth (\omega _0 t)} & \mbox{if $\kappa =-1$~.}
\end{array} \right.}
$$
It is well known that the particular Riccati solutions enter as
nonoperatorial part in the common factorizations of the
second-order linear differential equations that are directly
related to the Darboux isospectral transformations \cite{MS91}.

Thus, for Eq.~ (\ref{schr1}) one gets (${\rm D_{t}=\frac{d}{dt}}$)
\begin{equation} \label{w3}
{\rm \left(D_{t}+u_{p}\right)
\left(D_{t}-u_{p}\right)w=
w^{''}+(-u_{p}^{'}-u_{p}^{2})w=0}~.
\end{equation}
To fix the ideas, we shall use the terminology of Witten's
supersymmetric quantum mechanics and call Eq.~(\ref{w3}) the
bosonic equation. We stress here that the supersymmetric
terminology is used only for convenience and should not be taken
literally. Thus, the supersymmetric partner (or fermionic)
equation of Eq.~(\ref{w3}) is obtained by reversing the
factorization brackets
\begin{equation} \label{f}
{\rm
\left(D_{t}-u_{p}\right)
\left(D_{t}+u_{p}\right)w_f=
w^{''}+(u_{p}^{'}-u_{p}^2)w={\rm w^{''}
+\omega ^2_{f}(t)w=0}}~,
\end{equation}
which is related to the fermionic Riccati equation
\begin{equation} \label{fR}
{\rm u^{'}-u^2-\omega ^2_{{\rm f}}(t)=0~,}
\end{equation}
where the free term $\omega _{\rm  f}^2$ is the following function of time
$$
{\rm \omega ^2_{f}(t)=u_{p}^{'}-u_{p}^2=
\left\{ \begin{array}{ll}
{\rm \omega _0^2(-1-2{\rm tan}^2 \omega _0 t)} & \mbox{if $\kappa =1$}\\
{\rm \omega _0^2(1-2{\rm coth}^2 \omega _0 t)} & \mbox{if $\kappa =-1$~.}
\end{array} \right.}
$$
The solutions (fermionic zero modes) of Eq.~(\ref{f}) are given by
$$
{{\rm w_{f}}=
\left\{ \begin{array}{ll}
{\rm \frac{-\omega _0}{\cos (\omega _0 t)}} & \mbox{if $\kappa =1$}\\
{\rm \frac{\omega _0}{sinh (\omega _0 t)}} & \mbox{if $\kappa =-1$~,}
\end{array} \right.}
$$
and thus present strong periodic singularities in the first case
and just one singularity at the origin in the second case. These
`partner' oscillators, as well as those to be discussed in the
following, are parametric oscillators, i.e., of time-dependent
frequency. Moreover, their frequencies can become infinite
(periodically). In general, signals of this type are known as
chirps. "Infinite" chirps could be produced, in principle, in very
special astrophysical circumstances, e.g., close to black hole
horizons \cite{J96}.

\bigskip

\noindent
\section{Matrix formulation}

Using the Pauli matrices
$
\sigma _y=
\left( \begin{array}{cc}
0 & -{\rm i }\\
{\rm i} & 0\end{array} \right )$ and  
$\sigma _x=\left( \begin{array}{cc}
0 & 1\\
1 & 0 \end{array} \right )~,
$
we write the matrix equation
\begin{equation} \label{HD}
\hat{\cal D}_0W\equiv {\rm  [\sigma _y D_{t}+\sigma _x (iu_p)]}W=0~,
\end{equation}
where $W=\left( \begin{array}{cc}
{\rm w_1}\\
{\rm w_2}\end{array} \right )$ is a two component spinor.
Eq.~(\ref{HD}) is equivalent to the following decoupled equations
\begin{eqnarray}
({\rm i D_{t}+i u_p)w_1=0}\\
({\rm -i D_{t}+i u_p)w_2=0}~.
\end{eqnarray}
Solving these equations one gets ${\rm w}_1\propto \omega _0/\cos
({\rm \omega _0}t)$ and ${\rm w}_2\propto \omega _0\cos({\rm
\omega _0}t)$ for the $\kappa =1$ case and ${\rm w}_1\propto
\omega _0/{\rm sinh} ({\rm \omega _0}t)$ and ${\rm w}_2\propto
\omega _0{\rm sinh}({\rm \omega _0}t)$ for the $\kappa =-1$ case.
Thus, we obtain
\begin{equation}\label{W1}
W=\left( {\rm \begin{array}{cc}
{\rm w_1}\\
{\rm w_2}\end{array}} \right )=\left( {\rm \begin{array}{cc}
{\rm w_f}\\
{\rm w_b}\end{array}} \right )~.
\end{equation}
This shows that the matrix 
equation is equivalent to the two second-order linear differential
equations of bosonic and fermionic type, Eq.~(\ref{schr1}) and
Eq.~(\ref{f}), respectively, a result quite well known in particle
physics. Indeed, a comparison with the true Dirac equation with a
Lorentz scalar potential $\rm S(x)$
\begin{equation} \label{tD}
{\rm  [-i\sigma _y D_{x}+\sigma _x (m +S(x))]}W={\rm E}W~,
\end{equation}
shows that Eq.~(\ref{HD}) corresponds to a Dirac spinor of `zero
mass' and `zero energy' in an {\em imaginary} scalar `potential'
$\rm i u_p(t)$. We remind that a detailed discussion of the Dirac
equation in the supersymmetric approach has been provided by
Cooper et al \cite{cooper1} in 1988. They showed that the Dirac
equation with a Lorentz scalar potential is associated with a susy
pair of Schr\"odinger Hamiltonians. This result has been used
later by many authors in the particle physics context
\cite{literature}.

\bigskip

\noindent
\section{Extension through parameter {\rm K}}

We now come to the main issue of this work. Consider the slightly more general Dirac-like equation
\begin{equation} \label{HDM}
\hat{\cal D}_{{\rm K}}W\equiv {\rm [\sigma _y D_{t}+\sigma _x (iu_p +K)]W=KW}~,
\end{equation}
where K is a (not necessarily positive) real constant. On the left hand side of the equation, $\rm K$ stands as
an (imaginary) mass parameter of the Dirac spinor, whereas
on the right hand side it corresponds to the energy parameter.
Thus, we have an equation equivalent to a Dirac equation for a spinor
of mass $\rm i K$ at the fixed energy $E={\rm i K}$. This equation
can be written as the following system of coupled equations
\begin{eqnarray}
{\rm iD_{t}w _1+(iu_p+K)w _1=Kw _2}\label{coupl1}\\
{\rm -iD_{t}w _2+(iu_p+K)w _2=Kw _1}~.\label{coupl2}
\end{eqnarray}

The decoupling can be achieved by applying the operator in
Eq.~(\ref{coupl1}) to Eq.~(\ref{coupl2}). For the fermionic spinor
component one gets
\begin{equation} \label{comp1}
{\rm D^{2}_{t}w_1^{+}-\omega _0^2\Big[(1+2\tan ^2 \omega _0 t)+i\frac{2K}{\omega _0}\tan \omega _0 t\Big] w_1^{+}=0  \quad {\rm for} \,\, \kappa =1}
\end{equation}
\begin{equation} \label{comp1b}
{\rm D^{2}_{t}w_1^{-}+\omega _0^2\Big[(1-2{\rm coth}^2 \omega _0 t)+i\frac{2K}{\omega _0}{\rm coth} \,\omega _0 t\Big] w_1^{-}=0
\quad  {\rm for} \,\, \kappa =-1}~,
\end{equation}
whereas the bosonic component fulfills
\begin{equation} \label{comp2}
{\rm D^{2}_{t}w_2^{+}+\omega _0^2\Big[1-i\frac{2K}{\omega _0}\tan \omega _0t\Big] w_2^{+}=0  \quad {\rm for} \,\,  \kappa =1}
\end{equation}
\begin{equation} \label{comp2b}
{\rm D^{2}_{t}w_2^{-}-\omega _0^2\Big[1-i\frac{2K}{\omega _0}{\rm coth} \,\omega _0t\Big] w_2^{-}=0  \quad {\rm for} \,\, \kappa =-1}~.
\end{equation}

\bigskip
The solutions of the bosonic equations are expressed in terms of
the Gauss hypergeometric functions $_{2}{\rm F}_{1}$
\begin{eqnarray}
{\rm w} _{2}^{+}{\rm (t;\alpha_{+},\beta_{+})=\alpha_{+}z_{1}^{(p-\frac{1}{2})}
z_{2}^{(q-\frac{1}{2})}\,
_{2} F_{1}\left[p+q,p+q-1,2p\,;-\frac{1}{2}z_{1}\right]}\nonumber\\
 - {\rm \beta_{+}e^{-2ip\pi}4^{(p-\frac{1}{2})}z_{1}^{-(p-\frac{1}{2})} z_{2}^{(q-\frac{1}{2})}\,
_{2}F_{1}\left[q-p,q-p+1,2-2p\,;-\frac{1}{2}z_{1}\right]}
\end{eqnarray}
and
\begin{eqnarray}
{\rm w_{2}^{-}(t;\alpha_{-},\beta_{-}) = \alpha_{-}z_{3}^{r} z_{4}^{s}\,
_{2}F_{1}\left[r+s,r+s+1,1+2r;\frac{1}{2}z_{3}\right]}\nonumber\\
 + {\rm \beta_{-}4^r z_{3}^{-r} z_{4}^{s}\,
_{2}F_{1}\left[s-r+1,s-r,1-2r;\frac{1}{2}z_{3}\right]}~,
\end{eqnarray}
where the variables ${\rm z}_{i}$ ($i=1,...,4$) are given in the
following form:
\begin{eqnarray}
{\rm z_{1,2}=i\tan(\omega _0 t)\mp 1, \quad 
z_{3,4}={\rm coth}(\omega _0 t)\pm 1},\nonumber 
\end{eqnarray}
respectively. The parameters are the following:
\begin{eqnarray}
&&{\rm p = \frac{1}{2}\left(1+\sqrt{1-\frac{2K}{\omega
_0}}\right), \quad q = \frac{1}{2}\left(1+\sqrt{1+\frac{2K}{\omega
_0}}\right)}, \nonumber\\
&&{\rm r = \frac{1}{2}\sqrt{1+i\frac{2K}{\omega _0}}, \quad s
=\frac{1}{2}\sqrt{1-i\frac{2K}{\omega _0}}}~.\nonumber
\end{eqnarray}

\bigskip

The fermionic zero modes can be obtained as the inverse of the bosonic ones. Thus
\begin{equation}\label{wf1}
{\rm w_1^+=\frac{1}{w_2^{+}(t;\alpha _+,\beta _+)}}\,,\quad {\rm w_1^-=\frac{1}{w_2^{-}(t;\alpha _-,\beta _-)}}~.
\end{equation}
A comparison of ${\rm w_1^+}$ with the common $1/\cos {\rm t}$
fermionic mode is displayed in Figs.~5.3 and 5.4.

In the small ${\rm K}$ regime, ${\rm K} \ll \omega _0$, one gets
$$
{\rm w_2^{+}(t;\alpha _+,\beta _+) \approx\alpha _{+} z_{1}^{(p-\frac{1}{2})}
z_{2}^{(q-\frac{1}{2})}\, _{2}F_1\Big[2\, ,1\, , 2-\frac{K}{\omega _0} ; -\frac{1}{2}z_{1}(t)\Big]}-
$$
\begin{equation} \label{s1}
{\rm \beta  _+ 
e^{-2ip\pi}4^{(p-\frac{1}{2})}z_{1}^{-(p-\frac{1}{2})} z_{2}^{(q-\frac{1}{2})}\,   _2F_1\Big[\frac{K}{\omega _0}\, , 1+\frac{K}{\omega _0}\, ,\frac{K}{\omega _0}\, ; -\frac{1}{2}z_{1}(t)\Big]}
\end{equation}
and
$$
{\rm w_2^{-}(t;\alpha _-,\beta _-)\approx\alpha _{-}
z_{3}^{r} z_{4}^{s}\, _{2}F_1\Big[1, 2, 2+i\frac{K}{\omega _0}\,; \frac{1}{2}z_{3}(t)\Big]}+
$$
\begin{equation} \label{ss2}
{\rm \beta _{-}
4^r z_{3}^{-r} z_{4}^{s}\, _{2}F_1\Big[1-i\frac{K}{\omega _0},-i\frac{K}{\omega _0}, -i\frac{K}{\omega _0}\,; \frac{1}{2}z_{3}(t)\Big]}~.
\end{equation}

Examining the bosonic equations, one can immediately see that the
resonant frequencies acquired resistive time-dependent losses
whose relative strength is given by the parameter K. The fermionic
equations having time-dependent real parts of the frequency can be
interpreted as parametric oscillators which are also affected by
losses through the imaginary part.

\medskip

\noindent

\section{More {\rm K} parameters}

\noindent
A more general case in this scheme is to consider the following matrix Dirac-like equation
$$
\Bigg[\left( \begin{array}{cc}
0 & -{\rm i }\\
{\rm i} & 0\end{array} \right ){\rm D_{\rm t}}+\left( \begin{array}{cc}
0 & 1\\
1 & 0 \end{array} \right )\left( \begin{array}{cc}
 {\rm iu_p +K_1}& 0\\
0 &{\rm  iu_p+K_2}\end{array} \right )\Bigg]\left( \begin{array}{cc}
{\rm w}_1\\
{\rm w}_2 \end{array} \right )=
$$
\begin{equation} \label{Dg}
\left( \begin{array}{cc}
{\rm K_{1}^{'}}& 0\\
0 &{\rm  K_{2}^{'}}\end{array} \right )\left( \begin{array}{cc}
{\rm w_1}\\
{\rm w_2} \end{array} \right )~.
\end{equation}
The system of coupled first-order differential equations will be now
\begin{eqnarray}
\Big[-{\rm i}{\rm D_{\rm t}}+{\rm iu_p}+{\rm K_2}\Big]{\rm w_2}={\rm K_{1}^{'}}{\rm w_1}\\
\Big[{\rm i}{\rm D_{\rm t}}+{\rm iu_p}+{\rm K_1}\Big]{\rm w_1}={\rm K_{2}^{'}}{\rm w_2}
\end{eqnarray}
and the equivalent second-order differential equations
\begin{equation} \label{Schrgb}
{\rm D_{\rm t}}^{2}{\rm w} _{i}+\Big[-{\rm i}\Delta {\rm K}\Big]{\rm D_{\rm t}}{\rm w}_i
+\Big[\pm {\rm D_{\rm t}}{\rm u}_{\rm p}+{\rm i} ({\rm K}_1+{\rm K_2}){\rm u_p}+({\rm K_1K_2-K_{1}^{'}K_{2}^{'}})
 -{\rm u_{p}^2}\Big]{\rm w} _{i}=0~,
\end{equation}
where the subindex $i=1,2$ and $\Delta {\rm K}={\rm K _1}-{\rm K _2}$.
Under the gauge transformation
\begin{equation} \label{mg1}
{\rm w} _{i}={\rm Z}_{i}\exp \left(-\frac{1}{2}\int ^{\rm
t}\Big[-{\rm i}\Delta {\rm K}\Big] d\tau\right)={\rm Z}_i({\rm
t})\textrm{e}^{\frac{1}{2}{\rm i\, t} \Delta {\rm K}}~,
\end{equation}
one gets
\begin{equation} \label{schz}
{\rm D_{\rm t}}^{2}{\rm Z}_i+Q_i({\rm t}){\rm Z}_i=0,
\end{equation}
where the `potentials' have the form
\begin{equation} \label{Q}
Q_{i} ({\rm t})=\Big[\pm {\rm D_{\rm t}}{\rm u}_{\rm p}+{\rm i}({\rm K_1}+{\rm K_2}){\rm u_p}+({\rm K_1K_2-K_{1}^{'}K_{2}^{'}})
 -{\rm u_{p}}^{2}\Big]-
\frac{1}{4}\Big[-{\rm i}\Delta {\rm K}\Big]^2
\end{equation}
$Q_{1,2}$ are functions that differ from the nonoperatorial parts
in Eqs.~(\ref{comp1})-(\ref{comp2b}) only by constant terms.
Indeed, one can obtain easily the following equations.

For the fermionic spinor component one gets
\begin{equation} \label{comp1}
{\rm D^{2}_{t}Z_1^{+}-\omega _0^2\Big[1 + 2\tan ^2 \omega _0
t-\frac{(\Delta K)^2}{4\omega _0^2}
-\frac{K_1K_2-K_{1}^{'}K_{2}^{'}}{\omega _0^2} +
i\frac{K_1+K_2}{\omega _0}\tan \omega _0 t\Big] Z_1^{+}=0}
\end{equation}
for $\kappa =1$, and
\begin{equation} \label{comp1b}
{\rm D^{2}_{t}Z_1^{-}+\omega _0^2\Big[1-2{\rm coth}^2 \omega _0 t
+\frac{(\Delta K)^2}{4\omega _0^2}
+\frac{K_1K_2-K_{1}^{'}K_{2}^{'}}{\omega_0^2}+i\frac{K_1+K_2}{\omega
_0}{\rm coth} \,\omega _0 t\Big] Z_1^{-}=0}
\end{equation}
for $\kappa =-1$.

The bosonic component fulfills
\begin{equation} \label{comp2}
{\rm D^{2}_{t}Z_2^{+}+\omega _0^2\Big[1 +\frac{(\Delta
K)^2}{4\omega _0^2} +\frac{K_1K_2-K_{1}^{'}K_{2}^{'}}{\omega _0^2}
-i\frac{K_1 +K_2}{\omega _0}\tan \omega _0t\Big] Z_2^{+}=0}~,
\end{equation}
for $\kappa =1$, and
\begin{equation} \label{comp2b}
{\rm D^{2}_{t}Z_2^{-}-\omega _0^2\Big[1 -\frac{(\Delta
K)^2}{4\omega _0^2} -\frac{K_1K_2-K_{1}^{'}K_{2}^{'}}{\omega _0^2}
-i\frac{K_1 + K_2}{\omega _0}{\rm coth} \,\omega _0t\Big]
Z_2^{-}=0}~,
\end{equation}
 for $\kappa =-1$. When ${\rm K_1=K_2=K}$ one gets the particular case studied in full detail above.

The more general bosonic modes have the form:
\begin{eqnarray}
{\rm Z_{2}^{+}(t;\alpha_{+},\beta_{+})=\alpha_{+}[\tan(\omega _0t) -{\rm i}]^{\frac{\Omega _1}{4\omega _0}}
[\tan(\omega _0t) +{\rm i}]^{\frac{\Omega _2}{4\omega _0}}}\, \nonumber\\
\times\, _{2} {\rm F}_{1}\left[\frac{\Omega _1 +\Omega _2}{4\omega
_0},\frac{\Omega _1 +\Omega _2}{4\omega _0}+1\,,
1+\frac{\Omega _1}{2\omega _0};\frac{1}{2}(\tan (\omega _0 {\rm t})-{\rm i})\right]\nonumber\\
 {\rm +\beta_{+}(-1)^{-\frac{\Omega _1}{2\omega _0}}[\tan(\omega _0t) -{\rm i}]^{-\frac{\Omega _1}{4\omega _0}}[\tan(\omega _0t) +{\rm i}]^{\frac{\Omega _2}{4\omega _0}}}\,\nonumber\\
\times\, _{2} {\rm F}_{1}\left[\frac{\Omega _2 -\Omega _1}{4\omega
_0},\frac{\Omega _2 -\Omega _1}{4\omega _0}+1\,, 1-\frac{\Omega
_1}{2\omega _0};\frac{1}{2}(\tan (\omega _0 {\rm t})-{\rm
i})\right]
\end{eqnarray}
and
\begin{eqnarray}
{\rm Z}_{2}^{-}({\rm t};\alpha_{-},\beta_{-}) =
\alpha_{-}[{\rm coth}(\omega _0{\rm t}) -1]^{\frac{\Omega _3}{4\omega _0}}
[\coth(\omega _0{\rm t}) +1]^{\frac{\Omega _4}{4\omega _0}}\, \nonumber\\
\times\, _{2} {\rm F}_{1}\left[\frac{\Omega _3 +\Omega _4}{4\omega
_0}+1,\,\frac{\Omega _3 +\Omega _4}{4\omega _0},
1+\frac{\Omega _3}{2\omega _0};-\frac{1}{2}({\rm coth} (\omega _0 {\rm t})-1)\right]\nonumber\\
 +{\rm \beta_{-}(-1)^{-\frac{\Omega _3}{2\omega _0}}4^{\frac{\Omega _3}{2\omega _0}}[{\rm coth}(\omega _0{\rm t}) -1]^{-\frac{\Omega _3}{4\omega _0}}[{\rm coth}(\omega _0{\rm t}) +1]^{\frac{\Omega _4}{4\omega _0}}}\,\nonumber\\
\times\, _{2} {\rm F}_{1}\left[\frac{\Omega _3 -\Omega _4}{4\omega
_0},\frac{\Omega _4 -\Omega _3}{4\omega _0}+1\,, 1-\frac{\Omega
_3}{2\omega _0};-\frac{1}{2}({\rm coth} (\omega _0 {\rm
t})-1)\right]~,
\end{eqnarray}

where
$$
{\rm \Omega _1=
\left(4\omega _0^2+(K_1+K_2)^2+4[(K_1+K_2)\omega _0-K_{1}^{'}K_{2}^{'}]\right)^{1/2}}~,
$$
$$
{\rm \Omega _2=
\left(4\omega _0^2+(K_1+K_2)^2-4[(K_1+K_2)\omega _0+K_{1}^{'}K_{2}^{'}]\right)^{1/2}}~,
$$
$$
{\rm\Omega _3=
\left(4\omega _0^2-(K_1+K_2)^2-4[{\rm i}(K_1+K_2)\omega _0-K_{1}^{'}K_{2}^{'}]\right)^{1/2}}~,
$$
$$
{\rm\Omega _4=
\left(4\omega _0^2-(K_1+K_2)^2+4[{\rm i}(K_1+K_2)\omega _0+K_{1}^{'}K_{2}^{'}]\right)^{1/2}}~.
$$


\bigskip

\noindent
\section{Possible applications of the K-modes}

{\bf 5.6.1 Waveguides.}

\noindent In view of the correspondence between mechanics and
optics, one can also provide an interpretation in terms of the
Helmholtz optics for light propagation in waveguides of special
profiles. The supersymmetry of the Helmholtz equation has been
studied by Wolf and collaborators \cite{W2}. To get the waveguide
application, one should switch from the temporal independent
variable to a spatial variable ${\rm t\rightarrow x}$ along which
we consider the inhomogeneity of the fiber whereas the propagation
of beams is along another supplementary spatial coordinate $\rm
z$. Thus, we turn the equations (\ref{comp1})-(\ref{comp2}) into
Helmholtz waveguide equations of the type (we take $\rm c=1$)
 \begin{equation}\label{H1}
{\rm \large[\partial _z^2 +\partial _x^2+\omega _0
^2n^2(x)\large]\varphi(x,z)=0}~,
\end{equation}
where the modes $\varphi(x,z)$ can be written in the form $\rm
w_{1,2}(x) e^{-ik_z z}$ for a fixed wavenumber $\rm k_z$ in the
propagating coordinate that is common to both wave functions and
the index profiles correspond to two pairs of bosonic-fermionic
waveguides and are given by
\begin{equation} \label{H2}
{\rm n_{b}^2(x) \sim 1- i\frac{2K}{k _0}\tan(k _0x)}~,
\quad {\rm  n_{f}^2(x)\sim -(1+2\tan ^2(\omega _0 x))- i\frac{2K}{k _0}\tan(k _0x)}~,
\end{equation}
and
\begin{equation} \label{H3}
 {\rm  n_b^2(x)\sim -1- i\frac{2K}{k _0}{\rm coth}(k _0x)~, \quad  n_f^2(x) \sim 1-2coth ^2(k _0 x)+ i\frac{2K}{k _0}{\rm coth}(k _0x)}~,
\end{equation}
respectively. In our units $\rm k_0 = \omega _0$. Eqs.~(\ref{H2}),
(\ref{H3}) can be obtained from Riccati equations of the type
($\rm c\neq 1$)
\begin{equation}\label{H4}
\rm \omega _0^2n_{\rm f,b}^2(x)/c^2=k^2\mp R_x-R^2~,
\end{equation}
where $\rm R(x)$ are Riccati solutions directly related to the Riccati solutions discussed in the previous sections.

\medskip

According to Chumakov and Wolf \cite{W2} a second waveguide
interpretation is possible describing two different Gaussian
beams, bosonic and fermionic, whose small difference in frequency
is given in terms of a small parameter $\epsilon$ (wavelength/beam
width), propagating in the {\em same} waveguide. In this
interpretation, the index profile is the same for both beams. For
illustration, let us take the normal oscillator Riccati solution
in the space variable $\rm x$, i.e., $\rm tan k_0 x$ that we
approximate to first order linear Taylor term $\rm  k_0 x$. Then,
the two beam interpretation leads to the following Riccati
equation (for details, see the paper of Chumakov and Wolf)
\begin{equation} \label{CW1}
{\rm \omega _{1,2}^2n^2(x)-\omega _{0}^2n^2(0)=\mp k_0-k_0^2x^2(1\mp \epsilon)}~.
\end{equation}
An almost exact, up to nonlinear corrections of order $\epsilon
^2$ and higher, supersymmetric pairing of the $\rm z$ wavenumbers
(propagating constants) occurs, except for the `ground state' one.
As noted by Chumakov and Wolf, supersymmetry connects in this case
light beams of different frequencies but having the same
wavelength in the propagation direction $\rm z$. This approach is
valid only in the paraxial approximation. Therefore, one should
know the small x behavior of the K-modes in order to hope to
detect them through stable interference patterns along the
waveguide axis.

\medskip

{\bf  5.6.2 Cavity physics.}

\noindent Another very interesting application of the K-modes in a
radial variable could be Schumann's resonances, i.e., the resonant
frequencies of the spherical cavity provided by the Earth's
surface and the ionosphere plasma layer \cite{CF04}. The Schumann
problem can be approached as a spherical Helmholtz equation
$[\nabla ^2_{\rm r} +k^2] \phi =0$ with Robin type (mixed)
boundary condition $\frac{\partial \phi}{\partial
n}|_{S}=C(\omega) \phi _S$, where $C(\omega)$ is expressed in
terms of the skin depth $\delta=\sqrt{2/(\mu _c \sigma \omega)}$
of the conducting wall, $\mu _c$ is its permeability and $\sigma$
is its conductivity. The eigenfrequencies fulfilling such boundary
conditions can be written as follows
\begin{equation} \label{S1}
\omega ^2 \approx \omega _0 ^2 [(1-I) +{\rm i}I]~,
\end{equation}
where $I$ is a complicated expression in terms of skin depths and
surface and volume integrals of Helmholtz solutions with 
Neumann boundary conditions $\frac{\partial \phi}{\partial
n}|_{S}=0$. It is worth noting the similarity between these
improved values of Schumann's eigenfrequencies and the
K-eigenfrequencies. Moreover, using the $Q$ parameter of the
cavity, one can write Eq.~(\ref{S1}) in the form
\begin{equation} \label{S2}
\omega ^2 \approx \omega _0 ^2 \Bigg[\left(1-\frac{1}{Q}\right) +{\rm i}\frac{1}{Q}\Bigg]~.
\end{equation}
This form shows that the modification of the real part of $\omega$
leads to a downward shift of the resonant frequencies, while the
contribution to the imaginary component changes the rate of decay
of the modes.

We point out that Jackson mentions in his textbook that the near
equality of the real and imaginary parts of the change in $\omega
^2$ is a consequence of the employed boundary condition, which is
appropriate for relatively good conductors. Thus, by changing the
form of $C(\omega)$ that could result from different surface
impedances, the relative magnitude of the real and imaginary parts
of the change in $\omega ^2$ can be made different. It is this
latter case that corresponds better to the K-modes.

\newpage

{\bf 5.6.3 Crystal models.}

\noindent There is also a strong mathematical similarity between
the K-modes and the solutions of Scarf's crystal model
\cite{Scarf58} based on the singular potential $V(x)=-V_0{\rm
cosec} ^2(\pi x/a)$, where $a$ is an arbitrary lattice parameter.
For this model the one-dimensional Schr\"odinger equation has the
form
\begin{equation}\label{scarf1}
\psi '' +(a/\pi)^2 \bigg[\lambda ^2 +\left(\frac{1}{4}-s^2\right){\rm cosec}^2(\pi x/a)\bigg]\psi =0~.
\end{equation}
For $0<x\leq a/2$, the general solution is
$$
\psi =[f(x)]^{\frac{1}{2}+s}\, _2{\rm F}_1\Big[\frac{1}{4}+\frac{1}{2}(s+\lambda), \frac{1}{4}+\frac{1}{2}(s-\lambda); 1+s; f ^2 (x)\Big]+
$$
\begin{equation}\label{scarf2}
 [f(x)]^{\frac{1}{2}-s}\, _2{\rm F}_1\Big[\frac{1}{4}-\frac{1}{2}(s-\lambda), \frac{1}{4}-\frac{1}{2}(s+\lambda); 1-s; f^2 (x)\Big]~,
\end{equation}
where $f(x)=\sin (\pi x/a)$ corresponds to the ${\rm z_i}(t)$
functions, and $s$ and $\lambda$ corresponding to -p and -q,
respectively, are related to the potential amplitude and energy
spectral parameter. Thus, by turning the K-oscillator equations
into corresponding Schr\"odinger equations, one could introduce
another analytical crystal model with possible applications in
photonics crystals.

\medskip

{\bf  5.6.4 Cosmology.}

\noindent Rosu and L\'opez-Sandoval apply the K-mode approach to
barotropic FRW cosmologies \cite{RLS04}. K- Hubble cosmological
parameters have been introduced and expressed as logarithmic
derivatives of the K-modes with respect to the conformal time. For
$\rm K \rightarrow 0$ the ordinary solutions of the common FRW
barotropic fluids have been obtained.

It is also worth noticing the analogy of the nonzero ${\rm K}$
oscillator case with the phenomenon of diffraction of atomic waves
in imaginary crystals of light (crossed laser beams)
\cite{crystL}. In fact, the ${\rm K}$ parameter is a counterpart
of the modulation parameter ${\rm Q}$ introduced by Berry and
O'Dell in their study of imaginary optical gratings. Roughly
speaking, the nonzero ${\rm K}$ modes could occur in an {\em
imaginary crystal of time} that could occur in some exotic
astrophysical conditions.

\medskip

\vskip 1ex \centerline{ \epsfxsize=230pt \epsfbox{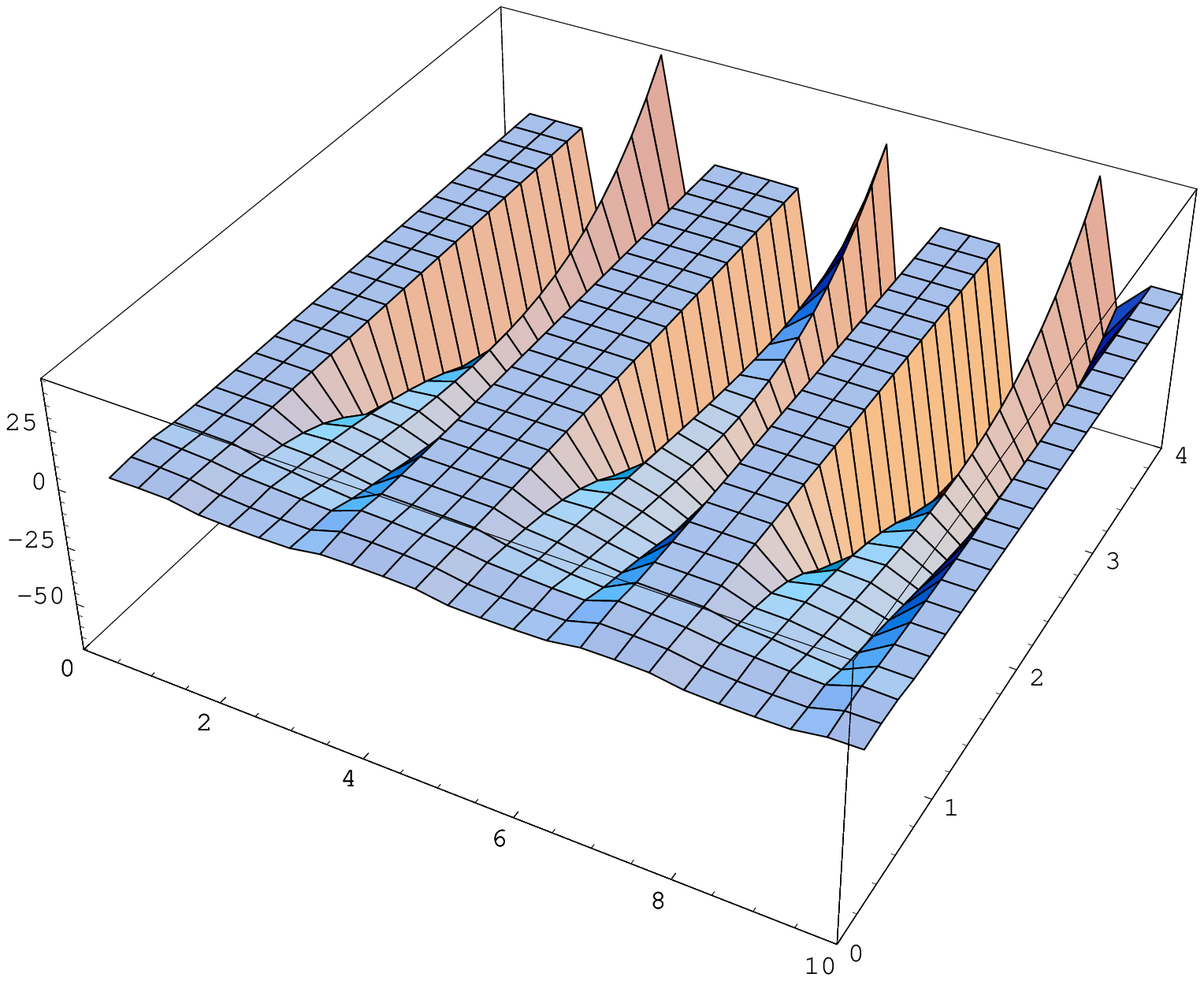}}
\vskip 3ex
\begin{center} {\small{Fig.
5.1:}$\quad$ The real part of the bosonic mode ${\rm
w_2^{+}(y;\frac{1}{2},\frac{1}{2}})$ for ${\rm t}\in [0,10]$ and
${\rm K\in[0,4]}$.}
\end{center}

\vskip 1ex \centerline{ \epsfxsize=230pt \epsfbox{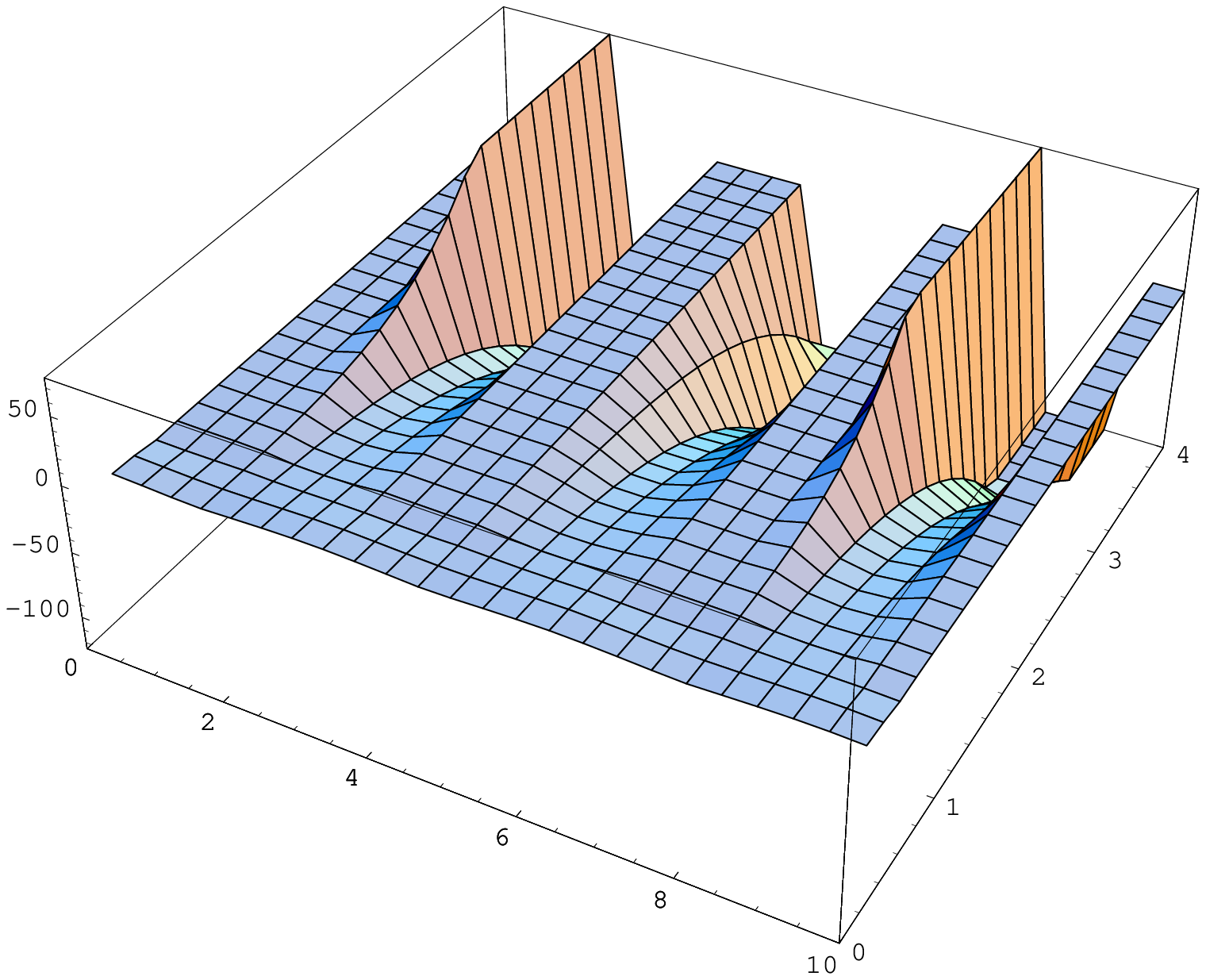}}
\vskip 3ex
\begin{center} {\small{Fig.
5.2:}$\quad$ The imaginary part of the bosonic mode ${\rm
w_2^{+}(y;\frac{1}{2},\frac{1}{2}})$ for ${\rm t}\in [0,10]$ and
${\rm K\in[0,4]}$.}
\end{center}

\vskip 1ex \centerline{ \epsfxsize=200pt \epsfbox{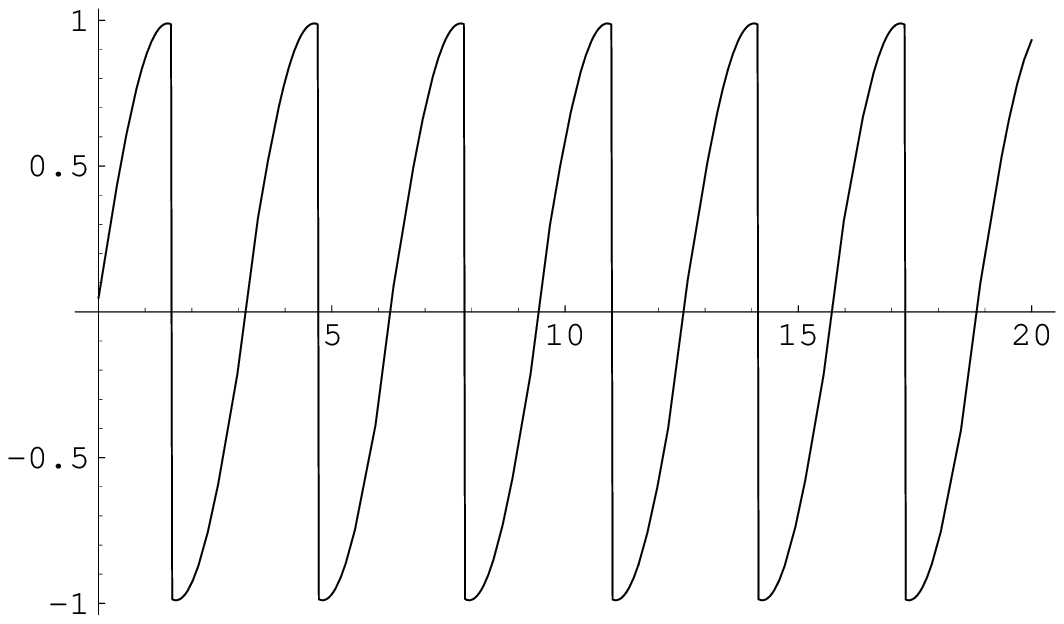}}
\vskip 3ex
\begin{center} {\small{Fig.
5.3:}$\quad$ The real part of the bosonic mode ${\rm
w_2^{+}(y;\frac{1}{2},\frac{1}{2}})$ for ${\rm t}\in [0,20]$ and
${\rm K}=0.01$.}
\end{center}

\vskip 1ex \centerline{ \epsfxsize=200pt \epsfbox{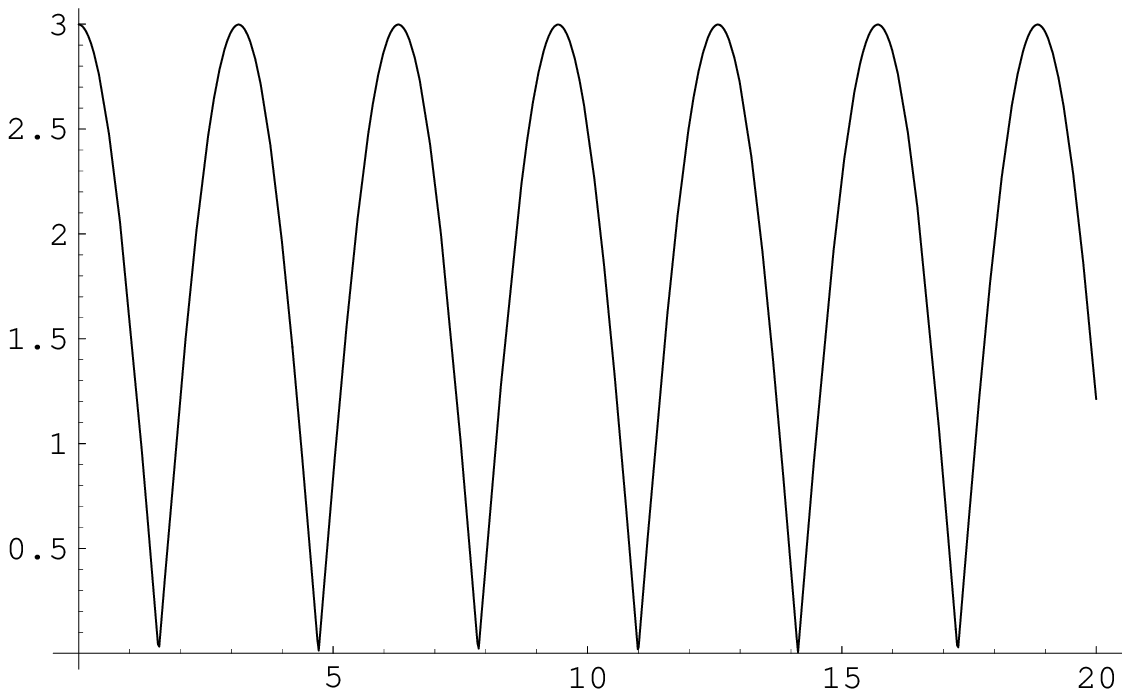}}
\vskip 3ex
\begin{center} {\small{Fig.
5.4:}$\quad$ The imaginary part of the bosonic mode ${\rm
w_2^{+}(y;\frac{1}{2},\frac{1}{2}})$ for ${\rm t}\in [0,20]$ and
${\rm K}=0.01$.}
\end{center}

\vskip 1ex \centerline{ \epsfxsize=200pt \epsfbox{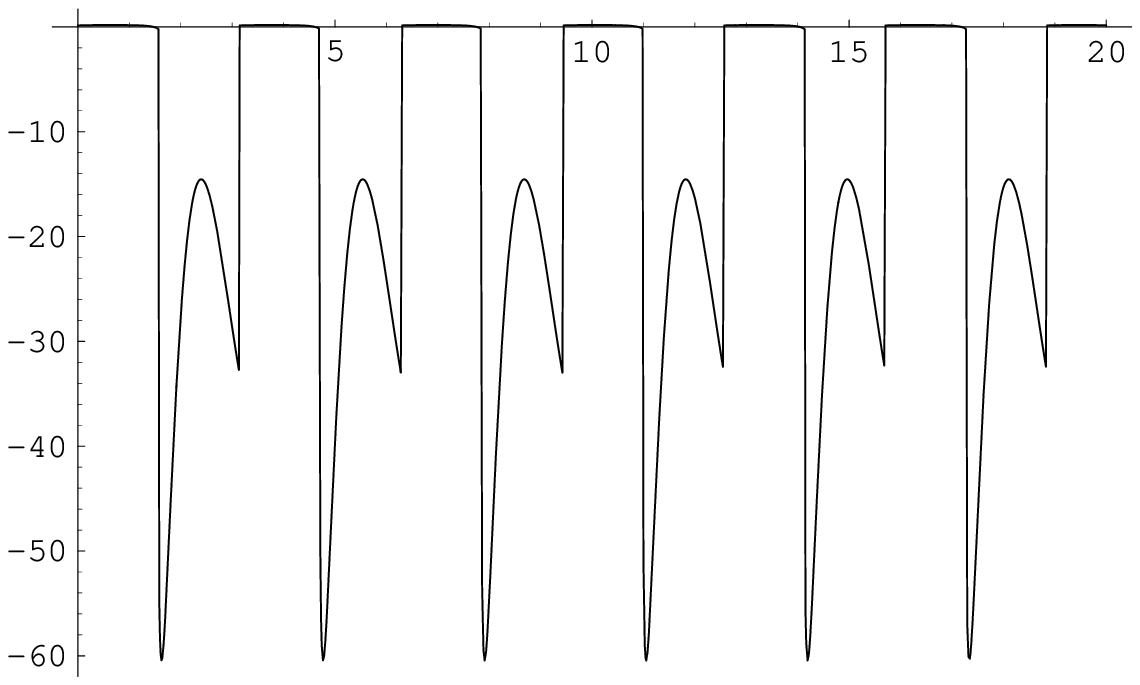}}
\vskip 3ex
\begin{center} {\small{Fig.
5.5:}$\quad$ The real part of the bosonic mode ${\rm
w_2^{+}(y;\frac{1}{2},\frac{1}{2}})$ for ${\rm t}\in [0,20]$ and
${\rm K}=2$.}
\end{center}

\vskip 1ex \centerline{ \epsfxsize=200pt \epsfbox{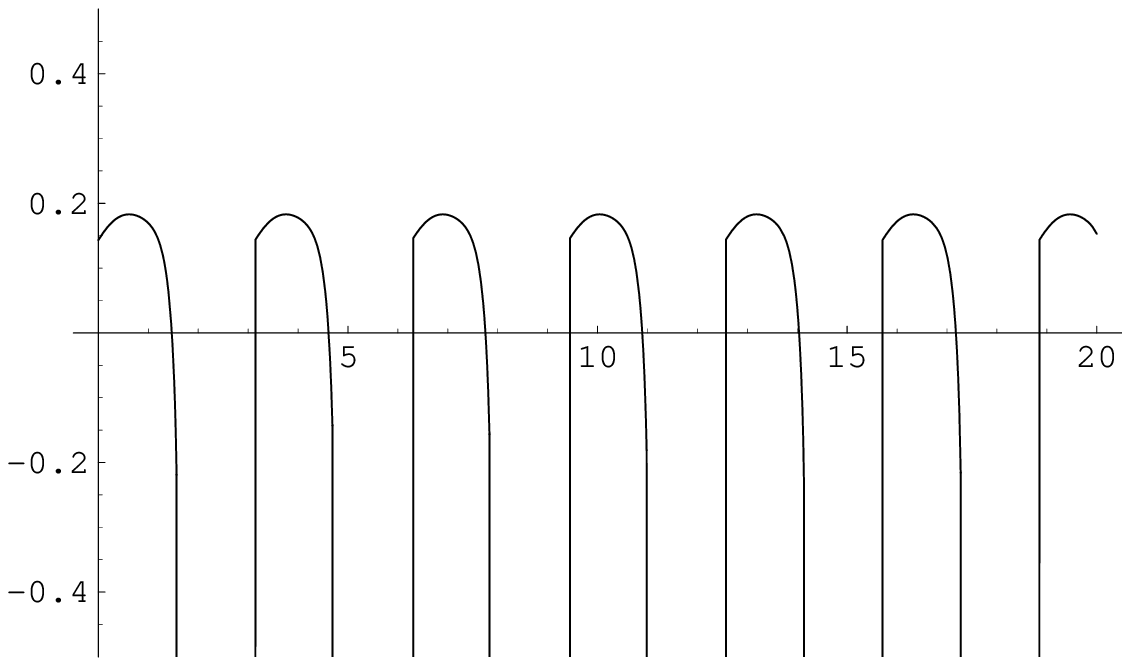}}
\vskip 3ex
\begin{center} {\small{Fig.
5.6:}$\quad$ The real part of the bosonic mode ${\rm
w_2^{+}(y;\frac{1}{2},\frac{1}{2}})$ for ${\rm t}\in [0,20]$ and
${\rm K}=2$ in the vertical strip [-0.5, 0.5].}
\end{center}

\vskip 1ex \centerline{ \epsfxsize=200pt \epsfbox{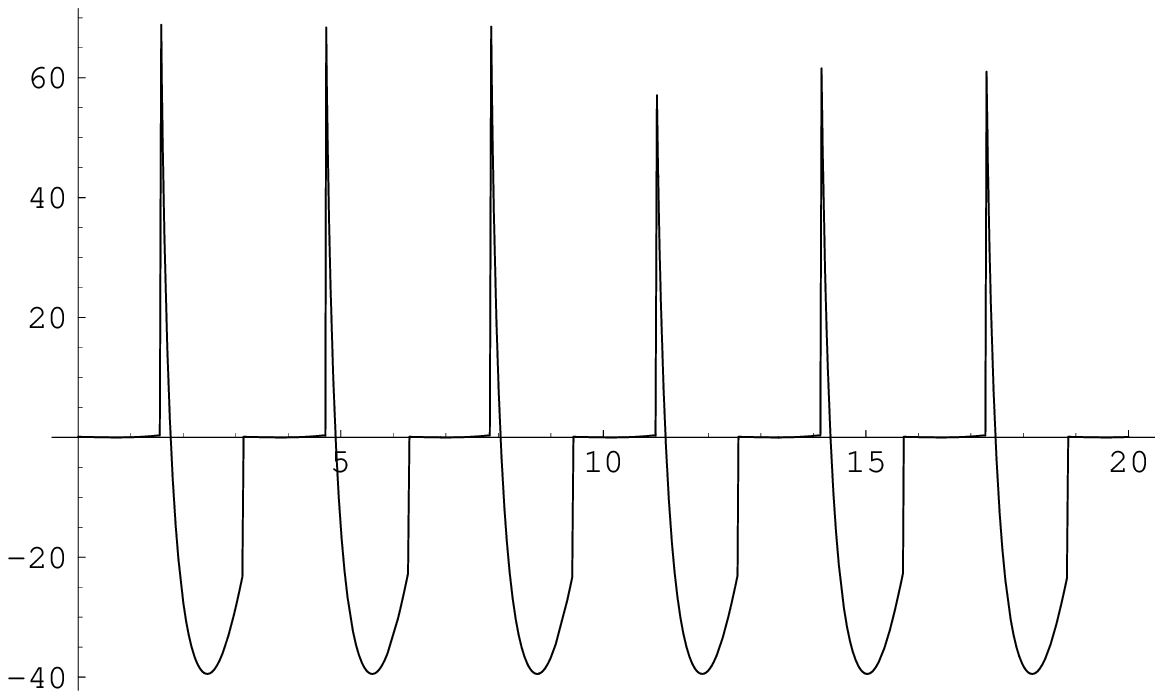}}
\vskip 3ex
\begin{center} {\small{Fig.
5.7:}$\quad$ The imaginary part of the bosonic mode ${\rm
w_2^{+}(y;\frac{1}{2},\frac{1}{2}})$ for ${\rm t}\in [0,20]$ and
${\rm K}=2$.}
\end{center}

\vskip 1ex \centerline{ \epsfxsize=220pt \epsfbox{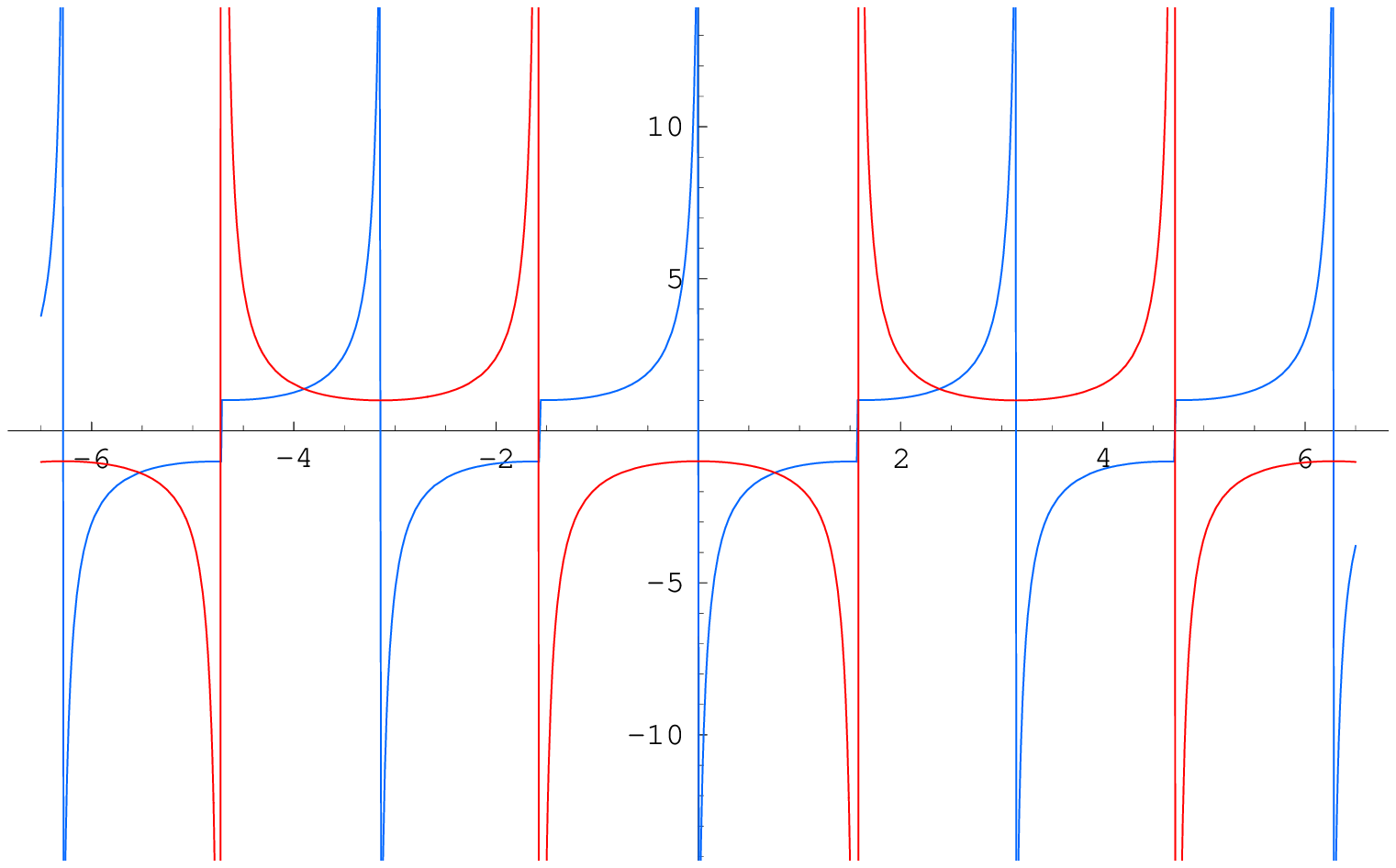}}
\vskip 3ex
\begin{center} {\small{Fig.
5.8:}$\quad$ The fermionic zero mode $-1/\cos{\rm  t}$, (red
curve), and the real part of $-1/{\rm w_2^{+}}$, (blue curve), for
${\rm K}=0.01$.}
\end{center}

\bigskip
\bigskip
\vskip 1ex \centerline{ \epsfxsize=220pt \epsfbox{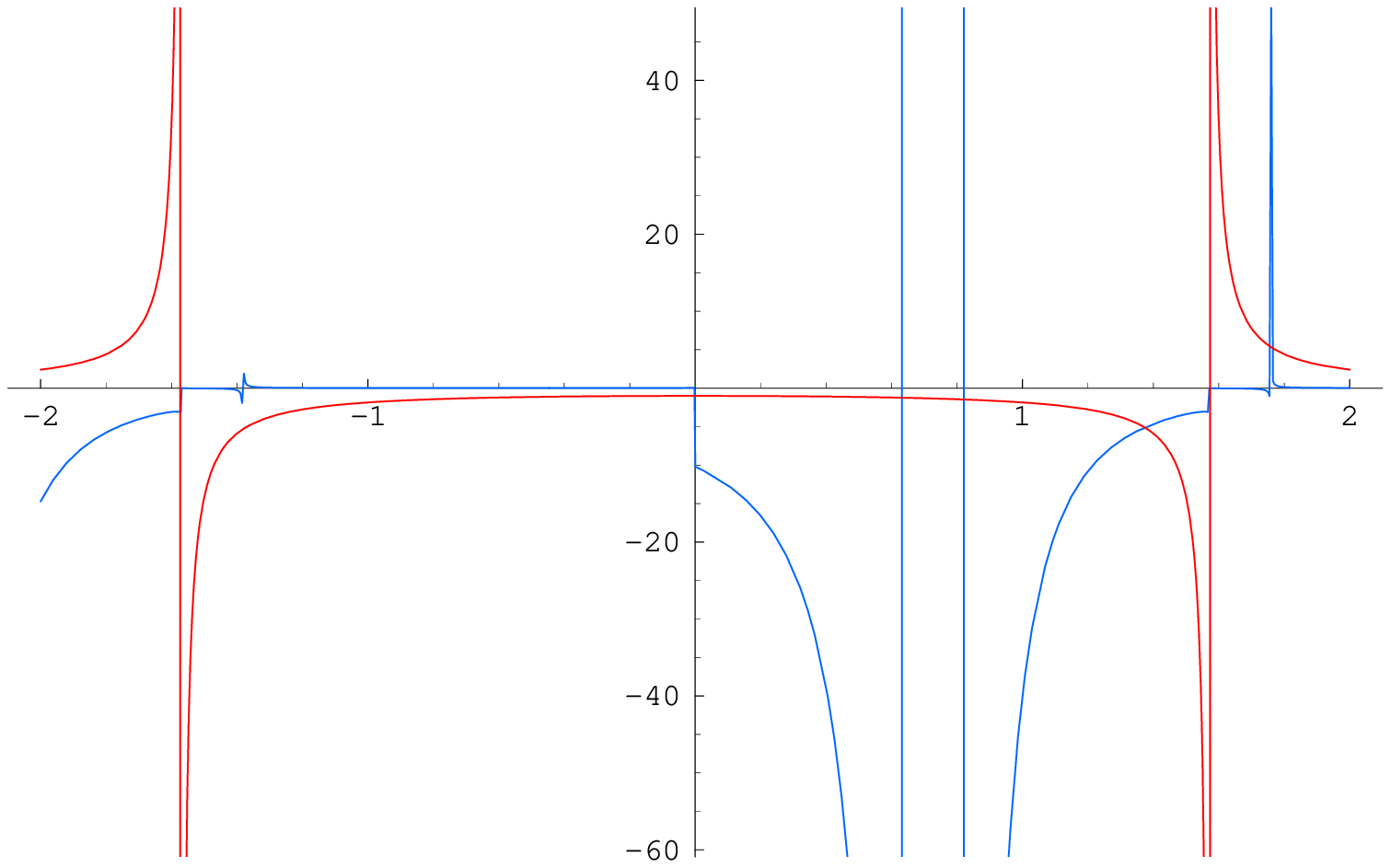}}
\vskip 3ex
\begin{center} {\small{Fig.
5.9:}$\quad$ The fermionic zero mode $-1/\cos{\rm  t}$, (red
curve), and the imaginary part of $-1/{\rm w_2^{+}}$, (blue
curve), for ${\rm K}=2$.}
\end{center}

\bigskip

\section{Quantum mechanics with Riccati nonhermiticity}

\noindent We have elaborated in the previous sections on an
interesting way of introducing imaginary parts (nonhermiticities)
in second order differential equations starting from a Dirac-like
matrix equation \cite{RLS04,A1}. The procedure is a complex
extension of the known supersymmetric connection between the Dirac
matrix equation and the Schr\"odinger equation. A detailed
discussion of the Dirac equation in the supersymmetric approach
has been provided by Cooper et al. \cite{cooper1,literature} in
1988, who showed that the Dirac equation with a Lorentz scalar
potential is associated with a susy pair of Schr\"odinger
Hamiltonians.
In the supersymmetric approach one uses the fact that the Dirac
potential, that we denote by $S$, is the solution of a Riccati
equation with the free term related to the potential function $U$
in the second order linear differential equations of the
Schr\"odinger type.

Indeed, writing the one-dimensional Dirac equation in the form
\begin{equation}\label{Dparticulas}
[\alpha p +\beta m +\beta R(x)] \psi (x) = E \psi (x)
\end{equation}
where $c=\hbar =1$, $p=-id/dx$, $m$ ($>0$) is the fermion mass,
and $R(x)$ is a Lorentz scalar. The wave function $\psi$ is a
two-component spinor $\big(\begin{array}{c} \psi _1\\ \psi _2
\end{array}\big)$ and the Pauli matrices $\alpha$ and $\beta$ are
the following
$$
\sigma _y=
\left( \begin{array}{cc}
0 & -{\rm i }\\
{\rm i} & 0\end{array} \right ) \qquad
{\rm  and}  \qquad 
\sigma _x=\left( \begin{array}{cc}
0 & 1\\
1 & 0 \end{array} \right )
$$
Writing the matrix Dirac equation in coupled system form leads to
\begin{equation}\label{cs1}
\big[D_x+m+R\big]\psi _1=E\psi _2
\end{equation}
\begin{equation}\label{cs2}
\big[-D_x+m+R\big]\psi _2=E\psi _1
\end{equation}
By decoupling one gets two Schr\"odinger equations for each spinor
component, respectively
\begin{equation}\label{cs3}
H_i\psi _i\equiv\big[-D_{x}^{2}+U_i\big]\psi _i =\epsilon \psi
_i~, \qquad \epsilon =E^2-m^2~,
\end{equation}
where $i=1, 2$, and
$$
U_i(x)=\left(R^2+2mR\mp dR/dx\right)~.
$$
One can also write factorizing operators for Eqs.~(\ref{cs3})
\begin{equation}\label{cs4}
A^{\pm} =\pm D_x +m+R
\end{equation}
such that
\begin{equation}\label{cs5}
H_1=A^{-}A^{+}-\frac{m}{2}~, \quad H_2=A^{+}A^{-}-\frac{m}{2}
\end{equation}

However, we have employed the method for the case of the classical
harmonic oscillator, which is the very specific situation in which
the Dirac mass parameter that we denoted by K was  treated as a
free parameter {\em equal} to the Dirac eigenvalue parameter $E$.
This is equivalent to Schr\"odinger equations at zero energy,
$\epsilon =0$. On the other hand, it is interesting to see how the
method works for negative energies, i.e., for a bound spectrum in
quantum mechanics. Here we briefly describe the method and next
apply it to the case of Morse potential.




\section{Complex extension with a single {\rm K} parameter}
We consider the slightly different Dirac-like equation with
respect to Eq.~(\ref{Dparticulas})
\begin{equation} \label{HDM}
\hat{\cal D}_{{\rm K}}W\equiv {\rm [\sigma _y D_{x}+\sigma _x (iR
+K)]W=KW}~,
\end{equation}
where
K is a (not necessarily positive) real constant. In the left hand
side of the equation, $\rm K$ stands as a mass parameter of the
Dirac spinor, whereas on the right hand side it corresponds to the
energy parameter. $R$ is an arbitrary solution of the Riccati
equation of the Witten type \cite{w81}
\begin{equation}\label{ricric}
R'\pm R^2=u~,
\end{equation}
where $u$ is the real part of the nonhermitic potential in the
Schr\"odinger equations we get.
Thus, we have an equation equivalent to a Dirac equation for a
spinor $W=\left( {\rm \begin{array}{cc}
\phi _1\\
\phi _0\end{array}} \right )\equiv\left( {\rm \begin{array}{cc}
{\rm w_f}\\
{\rm w_b}\end{array}} \right ) $
of mass $\rm K$ at the fixed energy $E={\rm K}$ but in a purely imaginary potential (optical lattices). This equation  
can be written as the following system of coupled equations
\begin{equation}\label{D1}
{\rm iD_{x}\phi _1+(iR+K)\phi _1=K\phi _0}
\end{equation}
\begin{equation}\label{D2}
{\rm -iD_{x}\phi _0+(iR+K)\phi _0=K\phi  _1}~.
\end{equation}

The decoupling of these two equations can be achieved by applying
the operator in Eq.~(\ref{D2}) to Eq.~(\ref{D1}) . For the
fermionic spinor component one gets
\begin{equation} \label{comp1}
{\rm D^{2}_{x}\phi_1-\Big[R^2-D_x R-i\,2KR\Big] \phi_1=0}  
\end{equation}
whereas the bosonic component fulfills
\begin{equation} \label{comp2}
{\rm D^{2}_{x}\phi_0-\Big[R^2+D_x R-i\,2KR\Big] \phi _0=0  }~.
\end{equation}
This is a very simple mathematical scheme for introducing a
special type of nonhermiticity directly proportional to the
Riccati solution.



%

\section{Complex extension with parameters {\rm K} and {\rm K'}.}

\noindent
A more general case in this scheme is to consider the following
matrix Dirac-like equation
$$
\Bigg[\left( \begin{array}{cc}
0 & -{\rm i }\\
{\rm i} & 0\end{array} \right ){\rm D_{\rm x}}+\left(
\begin{array}{cc}
0 & 1\\
1 & 0 \end{array} \right )\left( \begin{array}{cc}
 {\rm iR +K}& 0\\
0 &{\rm  iR+K}\end{array} \right )\Bigg]\left( \begin{array}{cc}
{\rm w}_1\\
{\rm w}_2 \end{array} \right )=
$$
\begin{equation} \label{Dg}
\left( \begin{array}{cc}
{\rm K^{'}}& 0\\
0 &{\rm  K^{'}}\end{array} \right )\left( \begin{array}{cc}
{\rm w_1}\\
{\rm w_2} \end{array} \right )~.
\end{equation}
The system of coupled first-order differential equations will be
now
\begin{eqnarray}
\Big[-{\rm i}{\rm D_{\rm x}}+{\rm iR}+{\rm K}\Big]{\rm w_2}={\rm K^{'}}{\rm w_1}\\ 
\Big[{\rm i}{\rm D_{\rm x}}+{\rm iR}+{\rm K}\Big]{\rm w_1}={\rm K^{'}}{\rm w_2}      
\end{eqnarray}

and the equivalent second-order differential equations
\begin{equation} \label{Schrgb}
{\rm D_{\rm x}}^{2}{\rm w} _{i} +\Big[\pm {\rm D_{\rm x}}{\rm
R}+2{\rm i} {\rm K}{\rm R}+({\rm K^2-K^{'2}})
 -{\rm R}^2\Big]{\rm w} _{i}=0~,
\end{equation}
where the subindex $i=1,2$ refers to the fermionic and bosonic
components, respectively.

\medskip

\section{Application to the Morse potential}

This potential is frequently used in molecular physics in
connection with the disassociation of diatomic molecules. In this
case, the Riccati solution is of the type
\begin{equation}\label{RicMorse}
R=A-B\textrm{e}^{-ax}~,
\end{equation}

Therefore, the second-order fermionic differential equation will
be
\begin{eqnarray}
D^{2}_{x}w_{1}&+&
\left[-\left(\bar{B}\textrm{e}^{-2ax}-\bar{C}_{1}\textrm{e}^{-ax}\right)\right.
+ (K^{2}-K^{\prime 2}) - A^{2} \nonumber\\
&+& \left. 2iK(A-B\textrm{e}^{-ax}) \right]w_{1}=0
\end{eqnarray}
where $\bar{B}=B^{2}$, and $\bar{C}_{1}=B(2A+a)$.

The solution is expressed as a superposition of Whittaker
functions
\begin{equation}
w_{1}=\alpha_{1}\textrm{e}^{ax/2} M_{\kappa _1, \mu}
\left(\frac{2B}{a}\,\textrm{e}^{-ax}\right)
+\beta_{1}\textrm{e}^{ax/2} W_{\kappa _1, \mu}
\left(\frac{2B}{a}\,\textrm{e}^{-ax}\right)
\end{equation}
$\kappa _1= \frac{A}{2a}\left(2+\frac{a}{A}-i\frac{2K}{A}\right)$
and $\mu =
\frac{A}{a}\left(\frac{K^{'2}-K^2}{A^2}-i\frac{2K}{A}\right)^{1/2}$.

The bosonic equation reads
\begin{eqnarray}
D^{2}_{x}w_{2}&+&
\left[-\left(\bar{B}\textrm{e}^{-2ax}-\bar{C}_{2}\textrm{e}^{-ax}\right)\right.
+ (K^{2}-K^{\prime 2}) - A^{2} \nonumber\\
&+& \left. 2iK(A-B\textrm{e}^{-ax}) \right]w_{2}=0
\end{eqnarray}
where $\bar{B}=B^{2}$, and $\bar{C}_{2}=B(2A-a)$.

The solution is a superposition of the following Whittaker
functions
\begin{equation} \label{superposition}
w_{2}=\alpha_{2}\textrm{e}^{ax/2} M_{\kappa _2, \mu}
\left(\frac{2B}{a}\,\textrm{e}^{-ax}\right)
+\beta_{2}\textrm{e}^{ax/2} W_{\kappa _2 , \mu}
\left(\frac{2B}{a}\,\textrm{e}^{-ax}\right)
\end{equation}
where $\kappa
_2=\frac{A}{2a}\left(2-\frac{a}{A}-i\frac{2K}{A}\right)$ and the
$\mu$ subindex is unchanged.

If we now place ourselves within the quantum mechanical (hermitic)
Morse problem we should take $\beta _2=0$ and $K=0$ in order to
achieve the exact correspondence with the bound spectrum problem
and eliminate the nonhermiticity. Moreover, the following
well-known connection with the associated Laguerre polynomials
\begin{equation} \label{Laguerre1}
M_{\frac{p}{2}+n+\frac{1}{2}, \frac{p}{2}}(y)=
y^{\frac{p+1}{2}}\textrm{e}^{-y/2}L_{n}^{p}(y)~, \quad
y=\frac{2B}{a}\textrm{e}^{-ax}
\end{equation}
can be used in our case with the following identifications
$$
\frac{p}{2}=\frac{K'}{a}~, \qquad K'=(A-an)
$$
i.e.,
$$
p=2\left(\frac{A}{a}-n\right)~.
$$
Then we can write the solution of the hermitic bosonic problem in
the well-known form
\begin{equation}\label{wn}
w _{2,n}(y) =\alpha
_2\left(\frac{2B}{a}\right)^{\frac{1}{2}}y^{\frac{A}{a}-n}\textrm{e}^{-y/2}L_{n}^{2(\frac{A}{a}-n)}(y)~.
\end{equation}
If we want to approach the nonhermitic problem we define by
analogy with Eq.~(\ref{Laguerre1})
\begin{equation}\label{Laguerre2}
M_{\kappa _2,\mu}(y)= y^{\mu
+\frac{1}{2}}\textrm{e}^{-y/2}L_{\kappa _2-\mu
-\frac{1}{2}}^{2\mu}(y)~,
\end{equation}
where $\kappa _2$ and $\mu$ are the complex parameters mentioned
before and the symbol corresponding to the associated Laguerre
polynomial representing now a Laguerre-like function introduced by
definition through Eq.~(\ref{Laguerre2}). The wave function of the
nonhermitic problem can be written as follows
\begin{equation}\label{wnonherm}
w _{2,nonherm}(y) =\alpha
_2\left(\frac{2B}{a}\right)^{\frac{1}{2}}y^{\mu}\textrm{e}^{-y/2}L_{\kappa
_2-\mu -\frac{1}{2}}^{2\mu}(y)~.
\end{equation}
For the nonhermitic fermionic problem, the formulas are similar
with the replacement of $\kappa _2$ by $\kappa _1$.

\medskip


\vskip 1ex \centerline{ \epsfxsize=230pt
\epsfbox{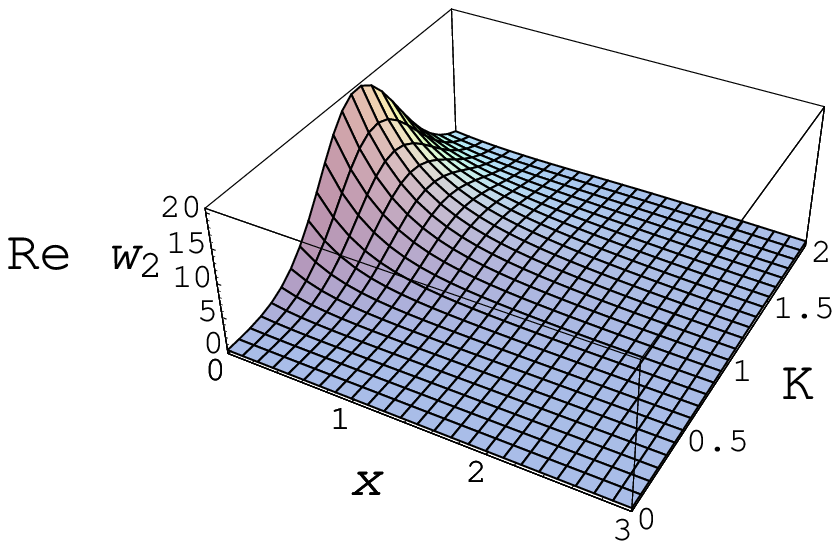}} \vskip 2ex
\begin{center}
{\small{Fig. 5.10:}$\quad$
Real part of the bosonic wave function $w_{2}$ in the range
$x\in[0,3]$ and ${\rm K\in[0,2]}$.}
\end{center}

\bigskip
\vskip 1ex \centerline{ \epsfxsize=230pt
\epsfbox{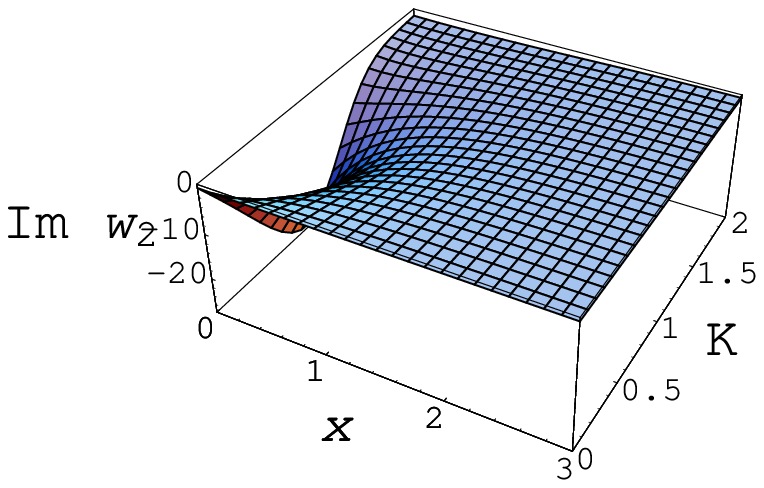}} \vskip 2ex
\begin{center}
{\small{Fig. 5.11:}$\quad$
Imaginary part of the bosonic wave function $w_{2}$ in the range
$x\in[0,3]$ and ${\rm K\in[0,2]}$.}
\end{center}

\bigskip
\vskip 1ex \centerline{ \epsfxsize=230pt
\epsfbox{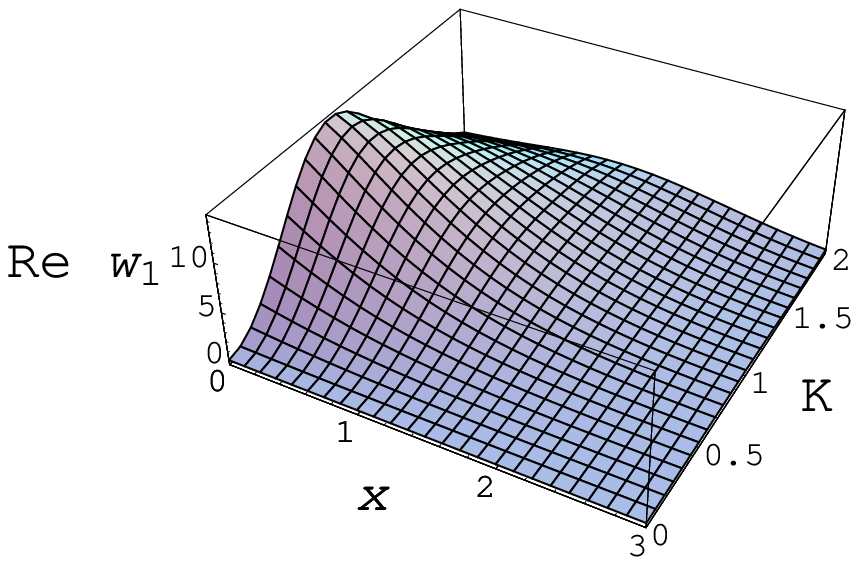}} \vskip 2ex
\begin{center}
{\small{Fig. 5.12:}$\quad$
Real part of the fermionic wave function $w_{1}$ in the range
$x\in[0,3]$ and ${\rm K\in[0,2]}$.}
\end{center}

\bigskip
\vskip 1ex \centerline{ \epsfxsize=230pt
\epsfbox{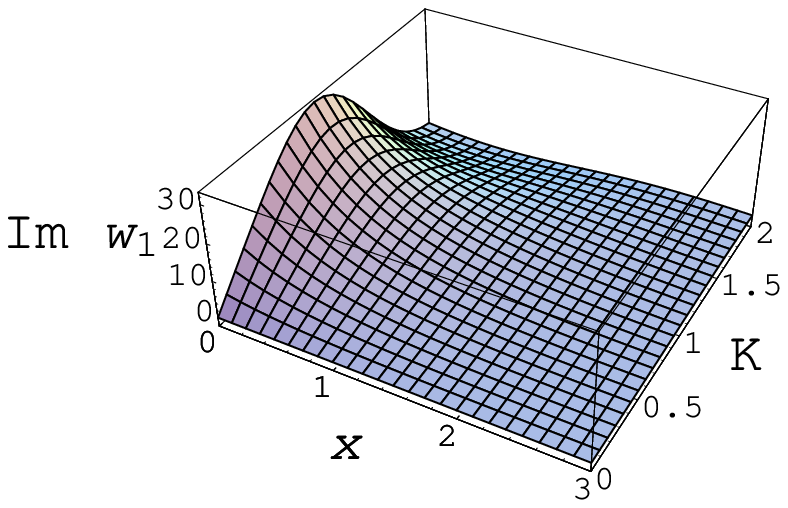}} \vskip 2ex
\begin{center}
{\small{Fig. 5.13:}$\quad$
Imaginary part of the fermionic wave function $w_{1}$ in the range
$x\in[0,3]$ and ${\rm K\in[0,2]}$.}
\end{center}

\bigskip

\section{Conclusion of the chapter}

By a procedure involving the factorization connection between the
Dirac-like equations and the simple second-order linear
differential equations of harmonic oscillator type, a class of
classical modes with a Dirac-like parameter describing their
damping and absorption (dissipation) has been introduced in this
chapter. While for zero values of the Dirac parameters the highly
singular fermionic modes are decoupled from their normal bosonic
harmonic modes, at nonzero values a coupling between the two types
of modes is introduced at the level of the matrix equation. These
interesting modes are given by the solutions of the
Eqs.~(\ref{comp1})-(\ref{comp2b}) and in a more general way by
Eqs.~(\ref{mg1}), (5.34)-(5.35) and are expressed in terms of
hypergeometric functions. Several possible applications in
different fields of physics are mentioned as well. Finally,
similar to the fact that the PT quantum mechanics can be
considered as a complex extension of standard quantum mechanics,
we notice that what we have done here is a particular type of
complex extension of the classical harmonic oscillator. The
complex supersymmetric extension introduced in the first sections
has been applied to exactly solvable quantum Morse problem. The
bosonic and fermionic wave functions have been obtained in
explicit form. This complex extension could have applications to
the diffraction of diatomic molecules on optical lattices (systems
of laser beams).






\part{SYNCHRONIZATION METHODS}

\chapter{Preliminary remarks on Part II}

\bigskip

The following remarks are pointed out in order to describe the
relationship between Part I and Part II of the thesis work, where
factorization methods for nonlinear ODE and synchronization of a
neuronal ensemble through feedback methods, respectively, have
been developed. The study of many biological systems from a
mathematical point of view is very stimulating because of the
possibilities to forecast the dynamic behavior of such biological
systems. An important example is the neuronal dynamics that
governs many living organisms. The neuronal dynamics performs the
processing of biological information by means of transmitted
signals. The pulse propagation along a nerve axon as described by
the FitzHugh-Nagumo equation that was solved in Section 2.4
through factorization methods is given by a travelling signal of
the kink type. The kink solution represents the transition between
two stable equilibrium states, and the "level change" travelling
waves are transmitted along the neuron axon \cite{scott}. It is a
very interesting challenge to study the synchronization dynamics
for a minimal ensemble of two neurons. However, the kink type
solutions that have been considered by us could not describe the
most realistic dynamical behavior for the neuronal ensemble. In
fact, the FitzHugh-Nagumo neuron model can be shown to be an
approximation (see, e.g. \cite{scott}) of the widely known
Hodgkin-Huxley (HH) neuron model \cite{hodgkin1}. Therefore, the
problem of neurons synchronization has been addressed in the more
realistic case of the HH neuron
model. 
Although theoretical attempts have been made to obtain analytical
solutions of the travelling wave type for the HH systems (see,
e.g. \cite{simpao,muratov}), it is an easier task to achieve
numerical results to study the synchronization dynamics. That is
why we do not pay attention to the HH travelling waves in the
second part. The nonlinear control theory is a very suitable way
to study and search for the intrinsic mechanisms underlying the
synchronization phenomena. In the next Chapters 7 and 8, the
problem of synchronizing a minimal ensemble of two HH neurons is
stated, and results on its synchronized dynamics are achieved by
implementing adequate feedback schemes.



\chapter{Synchronization of chaotic dynamics and neuronal systems}






\bigskip
\bigskip

{\small {\bf Abstract}. In this chapter, the synchronization
phenomena and the concept of chaos, their importance in natural
process and engineering systems are reviewed. A brief overview of
synchronization methods for the control of chaos and its
applications in biological systems is presented. Neuronal
synchronization activity and its role on brain dynamics is also
discussed in order to state the problem of neuronal
synchronization employing nonlinear control theory tools. In
addition, the dynamical model for the Hodgkin-Huxley (HH) neurons
is described.}

{


\bigskip


\section{Introduction}

Synchronization phenomena are very important processes occurring
in nature and often produced as a desired behavior in engineering
systems. In very general terms, the synchronization of coupled
systems means "to share time or events". It refers to the way in
which networked elements, due to their dynamics, communicate and
exhibit collective behavior \cite{nigel}. Some examples are the
observed synchronized flashing of fireflies, synchronization of
cells in a beating heart, the quantum synchronization in
superfluidity and superconductivity, in the phenomena related to
Josephson tunneling. Other important examples are the generated
synchronization in computer chips, communication systems and
global positioning systems.

On the other hand, the disquieting question about the exact
forecast of the evolution in time of diverse systems produced the
discovery of the existence of \textit{chaos}. The concept of chaos
usually refers to the issue of whether or not it is possible to
make long-term predictions about the behavior of a system. There
exist several mathematical definitions of chaos, however, all of
them express the property of high sensitivity to the initial
conditions. Such a characteristic implies that two even
arbitrarily close trajectories, separate exponentially in the
course of time. A deterministic system is said to be chaotic if
necessary requirements of nonlinearity and dimensionality (of at
least three) are characteristic for that system \cite{boccaletti}.

During the last fifteen years there have been an increased
interest in studying chaotic systems, since the proved fact that
chaos "cannot be forecasted but it can be controlled"
\cite{andrievskii}. The issue of the control of chaos is of
interest for both theorist and control engineers. Interest arises
because of many observations show that the chaotic behavior is
common in nature, for instance, chaotic dynamics can be found in
meteorology, plasma physics, heart and brain of living organisms;
and experimental results dealing with the control of chaotic
systems lead us to practically an unlimited amount of technical
applications in mechanical and space engineering, electrical and
electronic systems, communication and information systems, etc.
\cite{andrievskii2}.

Two applications of the control of chaos have been widely studied
for the past few year: the control and use of chaos for
communication systems, and the synchronization (and suppression)
of chaotic dynamics for several communication schemes
\cite{boccaletti,femat-01}.

In the following section a review of synchronization methods
concerning the control of chaotic dynamics and its applications in
biological systems is presented. Then we focus on synchronization
of a neuronal system comprised of two isolated HH neurons, whose
dynamical model is also described in final section. In Chapter 8,
the obtained results employing the mathematical tools provided by
the nonlinear control theory \cite{isidori1}, that can be applied
in order to show the way the HH neurons can synchronize are
presented. We find that isolated neurons unidirectionally couple
and synchronize through feedback action. In addition, robust
synchronization dynamics is obtained by implementing a dynamic
compensator.

\section{Synchronization methods for the control of chaos}

Many approaches have been proposed to control chaotic dynamics of
systems. Open loop or non feedback methods and closed loop or
feedback methods \cite{boccaletti,andrievskii,andrievskii2} have
been developed in order to produce the desired behavior in chaotic
systems. Two basic problems concerning the control of chaotic
dynamics through feedback methods are identified: synchronization
and suppression of chaos. Suppression of chaos consists in
stabilization of the system around regular orbits or equilibrium
points. The chaos synchronization problem has the characteristic
that the receiver (slave) system must track in some sense the
trajectories of the sender (master) system \cite{fematgual}.
Because of uncertainties may appear in the chaos control problem,
adaptive schemes are implemented in order to achieve robust
synchronization dynamics. Synchronization can be achieved for
identical chaotic systems with (obviously) different initial
conditions \cite{pecora}, however, researchers have found that non
identical systems synchronize when adequate developed feedback
schemes are applied \cite{fematgual,femat,femat-97}.

From the standpoint of the geometrical control theory, the
synchronization problem can be seen as a stabilization problem.
For a defined synchronization error $x_e=x_{e,M}-x_{e,S}$, where
$x_e$ represents the difference between the master and slave
system states, and $x \in R^d,$ $e=1,2,...,d$, there exists a
synchronization error system \cite{femat} whose trajectories
exponentially converge to zero under a feedback control action;
consequently, the master and slave systems unidirectionally
couple.

In the following chapter, the statement of the synchronization
problem and its solution using the geometrical control theory for
a proposed chaotic neuronal system is explained in detail. Also,
because of appearing uncertain system states, i.e., non accurately
measured states, an adaptive scheme is implemented via
construction of a state observer or uncertainty estimator that
guarantees robust synchronization.

\section{Applications of synchronization methods in biological systems}

Synchronization methods for the control of chaos where feedback
action is implemented have a wide variety of technical and
scientific applications. Among the main technical applications of
chaos synchronization can be found a diversity of schemes for
communication systems. The scientific applications are directed to
study properties, regularities, and mechanisms of the behavior of
physical, chemical and biological systems. Moreover, very
interesting results are obtained when the control methods are
applied to experimental systems. Examples of suppression and
synchronization of chaos can be found in biomechanical systems,
medicine, biology and ecology: design of feedback pacemakers,
suppression of oscillating epileptiform activity in neural
networks, control of population dynamics in plankton and other
biological species, etc.

The main interest in Part II is to apply feedback synchronization
methods to a chaotic neuronal system. Two HH neurons are regarded
as a system of two dynamical subsystems, with the aim to show that
synchronized dynamics is achieved through feedback strategies.
Synchronization and suppression of chaos, using the tools of
control theory, could provide insights to understand and show
their relevance in the processing of biological information in
neural networks.

\section{Synchronized dynamics of neurons}

Synchronization of neuronal activity patterns (action potentials)
is a fundamental topic in the modern research of brain dynamics.
Experimental evidence reveals that synchronization phenomena are
basic for the processing of biological information. It has been
demonstrated \cite{gray1} that large ensembles of neurons whose
functionality is related to visual perception synchronize their
oscillatory activity (in the gamma frequency range, 40-60 Hz) when
stimulated. Some researchers consider that neuronal
synchronization allows the brain to solve the so-called binding
problem \cite{haken}. Take for example a car: it may be
characterized by its shape, color, emitted noise, and so on. All
these features are processed in different parts of the brain,
however we conceive the car as a single entity because of still
unknown binding procedures. More recent studies suggest that
synchronization is a basic mechanism for consciousness
\cite{wakefield,edelman,rodriguez}. Also, Parkinsonian tremor and
epileptic seizures are closely related with this mechanism
\cite{haken,netoff}.

Recently, it was shown that two coupled living neurons synchronize
their activity patterns when depolarized by an external current
\cite{elson1}. However, the whole underlying mechanisms are not
completely understood. From a theoretical point of view, neurons
are considered as nonlinear oscillators. A lot of theoretical
studies have been carried out to investigate the dynamics of
single neurons and neural networks. The most employed and
realistic neuron models are the Hodgkin-Huxley and the
FitzHugh-Nagumo systems \cite{haken}, that have been used in
detailed studies of neuron behavior under external forcing.
Moreover, within these models one can consider the change of
dynamical parameters and implications on the neuronal activity,
for instance, the strength of synaptic conductance, and intrinsic
or added noise. The neuronal synchronization problem is addressed
regarding diffusive coupling, or modelling unidirectionally
coupled master-slave systems. Many people believe that control
theory could be very useful to address the problem of
synchronization in ensembles of neurons. The mathematical tools
provided by the nonlinear control theory, can suitably be applied
to the mathematical model of a system comprised of two noiseless
HH neurons.

\bigskip
\bigskip



\section{The Hodgkin-Huxley model of the neuron}

The brain is the most complex system known to us. Understanding
the way it works and its structure has been a very interesting
research issue during many decades. The basic units which
integrate the brain tissue and, in general, any nervous system,
are the nerve cells or neurons. The transmission of signals and
the processing of biological information are carried out through
complicated interactions between large ensembles of neurons.
External and internal stimuli generate the biological information
which propagates from the sensory sites to specific areas of the
brain, for instance, the visual, olfactory and auditive
perception, and the conscious sensory-motor activity.

Most of the nerve cells generate a series of voltage spiking
sequences called the membrane action potential, in response to
external stimuli performing the information processing. These
pulses of the action potential originate at the cell body and
propagate down the axon at constant amplitude and velocity. They
can be transmitted via synaptic coupling to another nerve cell
which is stimulated by the corresponding current. The electric
behavior of the cell axon membrane is described by the net ion
flux through a great amount of potassium and sodium ionic
channels; each ion passing from the inner (outer) to the outer
(inner) side of the cell. It is known that potassium and sodium
channels are composed of four independent \textit{gates}, which
can be in a permissive or non permissive state. The potassium ions
cross the membrane only through channels that are specific for
potassium. If the four gates of a potassium channel are in
permissive state, then the channel is open and potassium ions flow
through it. The sodium ions cross the membrane only through
channels that are specific for sodium. Three of the gates for a
sodium channel are activation gates, and one is inactivation gate.
All of them must be in permissive state to allow sodium ions to
cross the sodium channel.

Several neuron models have been proposed to describe the dynamics
of the action potentials. However, the most widely used is still
the realistic HH neuron model \cite{hodgkin1}. In the early
1950's, Hodgkin and Huxley developed and published a series of
investigations where they studied the electrophysiology of the
squid giant axon. Their results allow them to calculate the total
membrane current as the sum of the potassium and calcium ionic
channels currents and the capacitive current,
\begin{equation}
I_{m}(t)=I_{ionic}(t)+C_{m}\frac{dV(t)}{dt}.\label{ho1}
\end{equation}
In addition, based on the large number of realized experiments
they postulated a phenomenological model that turned itself into a
paradigm for the generation of the action potential in the squid
axon. We follow next the recent discussion in the book of C. Koch
\cite{koch1} for a compact presentation of the main statements of
the HH model:\\

1. The action potential involves two major voltage-dependent ionic
conductances, a sodium conductance $g_{Na}$ and a potassium
conductance $g_{K}$. They are independent from each other. A
third, smaller "leak" conductance $g_{l}$ does not depend on the
membrane potential. The total ionic current flowing is given by
the following equation
\begin{equation}
I_{ionic}(t)= I_{Na}+I_{K}+I_{leak}. \label{ho2}
\end{equation}

2. The individual ionic currents $I_{i}(t)$ are linearly related
to driving potential via Ohm's law,
\begin{equation}
I_{i}(t)= g_{i}(V(t),t)(V(t)-E_{i}) \label{ho3}
\end{equation}
where the ionic reversal potential $E_{i}$ is given by Nernst's
equation for the appropriate ionic species. Depending on the
balance between the concentration difference of the ions and the
electrical field across the membrane separating the intracellular
cytoplasm from the extracellular milieu, each ionic species has an
associated "ionic battery". Conceptually, there exist an
equivalent electrical circuit to describe the axonal membrane.

3. Each of the two ionic conductances is expressed as a maximum
conductance, $g_{Na}$ and $g_{K}$, multiplied by a numerical
coefficient representing the fraction of the maximum conductance
actually open. These numbers are functions of one or more fictive
gating particles Hodgkin and Huxley introduced to describe the
dynamics of the conductances. In their original model, they talked
about activating and inactivating gating particles. Each gating
particle can be in one of two possible states, open or close,
depending on time and on the membrane potential. In order for the
conductance to open, all of these gating particles must be open
simultaneously. The entire kinetic properties of their model are
contained in these variables.

The gating particles, also known as gating variables, Hodgkin and
Huxley presented are usually denoted by $n$, $m$ and $h$. They are
the same as the currently known ion channel gates.

The following set of four coupled nonlinear differential equations
represents the complete HH neuron dynamical model \cite{hodgkin1}:
\begin{eqnarray}
C_{m} \frac{dV}{dt} &=& I_{ext} - g_{K}n^4 \left( V-V_{K} \right)
- g_{Na} m^3 h \left( V-V_{Na} \right) - g_{l}\left( V-V_{Na}
\right),
\label{ho4}\\
 \frac{dn}{dt} &=& \alpha_{n}\left( V \right)\left( 1-n \right) -
 \beta_{n}\left( V \right) n,
 \label{ho5}\\
 \frac{dm}{dt} &=& \alpha_{m}\left( V \right)\left( 1-m \right) -
 \beta_{m}\left( V \right) m,
 \label{ho6}\\
 \frac{dh}{dt} &=& \alpha_{h}\left( V \right)\left( 1-h \right) -
 \beta_{h}\left( V \right) h,
 \label{ho7}
\end{eqnarray}
where $V$ represents the membrane potential, $n$ is the
probability of any given potassium channel gate being in the
permissive state (activation of the potassium flow current), $m$
is the probability of any given activation sodium channel gate
being in the permissive state (activation of the sodium flow
current), and $h$ is the probability of any given inactivation
sodium channel gate being in the permissive state (inactivation of
the sodium flow current). Gating variables are dimensionless and
within the range $[0,1]$. $C_{m}$ is the membrane capacitance,
$g_{K}$, $g_{Na}$ and $g_{l}$ are the maximum ionic and leak
conductances, while $V_{K}$, $V_{Na}$ and $V_{l}$ stand for the
ionic and leak reversal potentials. The external stimulus current
can be modelled by the term $I_{ext}$, usually a tonic or periodic
forcing. The respective differential equations for $n$, $m$ and
$h$, describe the transition from open to closed states for the
gating variables. The explicit form of the functions
$\alpha_{j}(V)$ and $\beta_{j}(V)$ $(j=n, m, h)$ in Eqs.
(\ref{ho5})-(\ref{ho7}) is given as follows \cite{hodgkin1,koch1},
\begin{eqnarray}
\alpha_{n}=\frac{0.01(V+10)}{\{ \textrm{exp} [(V+10)/10] -1\}},
\quad
\beta_{n}=0.125\textrm{exp}(V/80);\label{ho8}\\
\alpha_{m}=\frac{0.1(V+25)}{\{ \textrm{exp} [(V+25)/10]
-1\}},\quad
\beta_{m}=4\textrm{exp}(V/18);\label{ho9}\\
\alpha_{h}=0.07\textrm{exp}(V/20),\quad \beta_{h}=\frac{1}{\{
\textrm{exp} [(V+30)/10] +1\}}.\label{ho10}
\end{eqnarray}
Also, nominal values for the system parameters can be found in
\cite{hodgkin1,koch1}.

\chapter{Unidirectional synchronization of Hodgkin-Huxley neurons}






\bigskip
\medskip

{\small {\bf Abstract}. Synchronization dynamics of two noiseless
Hodgkin-Huxley (HH) neurons under the action of feedback control
is studied. The spiking patterns of the action potentials evoked
by periodic external modulations attain synchronization states
under the feedback action. Numerical simulations for the
synchronization dynamics of regular-irregular desynchronized
spiking sequences are displayed. The results are discussed in
context of generalized synchronization. It is also shown that the
HH neurons can be synchronized in face of unmeasured states.}


\noindent







 \section{Introduction}

 For several decades many attempts have been addressed to
 understand the processing of biological information in single
 neurons and neural networks. Experimental reports
 \cite{gray,meister,kreiter} suggest that the synchronization
 plays a very important role in the processing of information by
 large ensembles of neurons. Recently, it has been demonstrated
 that a minimal ensemble of two coupled living neurons fire
 synchronized spiking activity when depolarized by an external DC
 current \cite{elson}. However, total neural mechanisms underlying
 synchronization are not well understood yet. The Hodgkin-Huxley
 neurons are usually used as realistic models of neuronal systems,
 for studying neuronal synchronization. Some theoretical
 approaches investigate the synchronization phenomena
 considering diffusive coupling and the influence of intrinsic
 noise as a promoter of neuronal activity \cite{casado,casado1},
 and studying the synchronization dynamics related to the rhythmic
 oscillations phenomena (theta and gamma frequency rhythms) in
 neurons of localized areas of the brain \cite{wang1,acker,ermentrout}.
 In addition, the forcing of HH neurons by external stimulus has been widely studied
 \cite{yu,parmananda,wang,lee,tanabe} for tonic or periodic
 currents that trigger the action potential displaying spike
 activity and refractory dynamics.

     On the other hand, synchronization of chaotic systems
 is a relatively recent phenomena \cite{pecora} which can be understood
 from nonlinear geometrical control theory \cite{fematgual,femat}. In this chapter we
 study the synchronized behavior of two silent (i.e., the
 autonomous HH systems exhibit fixed point dynamics) \cite{parmananda} HH
 neurons, proposing an unidirectionally coupled synchronization
 system. In the chaos synchronization problem the trajectories of a slave system
 must track, in some sense, the trajectories of a master system even
 though slave and master systems may be different.
 The obtained results contribute in the theoretical framework of neurons synchronization, and
 relates the phenomena to the well-posed concept of generalized synchronization
 (GS) \cite{parlitz}. Because of uncertain states cannot be accurately measured in practice,
 they are not available to do control, for instance, the ionic channels activation. Then, an
 approach for robust synchronization via construction of an uncertainty estimator, is implemented.

 The organization of this chapter is as follows. In Section 8.2 the HH neuronal systems are
 described. The statement of the problem for unidirectionally coupling
 synchronization of HH neurons is given in Section 8.3. The HH neuronal
 synchronization dynamics obtained through a stabilizing control law is
 studied in Section 8.4. In Section 8.5 the generalized and robust synchronization
 are discussed, and a final conclusion section is given.
\bigskip

\section{The Hodgkin-Huxley system redefined}


We redefine the HH system of equations in order to state the
synchronization problem of two HH neurons. Let $x_{i,M}$ and
$x_{i,S}$ ($i=1,2,3,4$ and subscripts $M,S$ stand for the master
and slave system, respectively) be the four variables $V$, $n$,
$m$ and $h$ in each system. As the meaning of synchronous behavior
is ``to share time or events'', we shall consider two HH neurons
modelled by Eqs. (\ref{ho4})-(\ref{ho7}). Thus, the master system
is represented by the following set of equations:
\begin{eqnarray}
\dot x_{1,M} &=& 1/C_{m_{M}} \left[ I_{ext_{M}}(t) \right.
 - g_{K_{M}}x^4_{2,M} \left( x_{1,M}-V_{K_{M}} \right)  \nonumber \\
&&- g_{Na_{M}} x_{3,M}^3 x_{4,M} \left( x_{1,M} -V_{Na_{M}}
\right) \left. - g_{l_{M}}\left( x_{1,M}-V_{l_{M}} \right)
\right],
\label{eq5}\\
\dot x_{2,M} &=& \alpha_{n}(x_{1,M})\left( 1-x_{2,M} \right) -
 \beta_{n}(x_{1,M})x_{2,M}\, ,
 \label{eq6}\\
\dot x_{3,M} &=& \alpha_{m}(x_{1,M})\left( 1-x_{3,M} \right) -
 \beta_{m}(x_{1,M})x_{3,M}\, ,
 \label{eq7}\\
\dot x_{4,M} &=& \alpha_{h}(x_{1,M})\left( 1-x_{4,M} \right) -
 \beta_{h}(x_{1,M})x_{4,M} \, ,
 \label{eq8}
\end{eqnarray}
and the slave system is proposed to be governed by the equations:
\begin{eqnarray}
\dot x_{1,S} &=& 1/C_{m_{S}} \left[ I_{ext_{S}}(t) \right.
 - g_{K_{S}}x^4_{2,S} \left( x_{1,S}-V_{K_{S}} \right)  \nonumber \\
&&- g_{Na_{S}} x_{3,S}^3 x_{4,S} \left( x_{1,S} -V_{Na_{S}}
\right) \left. - g_{l_{S}}\left( x_{1,S}-V_{l_{S}}
\right)\right]+u ,
\label{eq9}\\
\dot x_{2,S} &=& \alpha_{n}(x_{1,S})\left( 1-x_{2,S} \right) -
 \beta_{n}(x_{1,S})x_{2,S}\, ,
 \label{eq10}\\
\dot x_{3,S} &=& \alpha_{m}(x_{1,S})\left( 1-x_{3,S} \right) -
 \beta_{m}(x_{1,S})x_{3,S}\, ,
 \label{eq11}\\
\dot x_{4,S} &=& \alpha_{h}(x_{1,S})\left( 1-x_{4,S} \right) -
 \beta_{h}(x_{1,S})x_{4,S} \, ,
 \label{eq12}
\end{eqnarray}
where the added term $u$ in Eq. (\ref{eq9}) represents a feedback
synchronization force. Nominal values are considered for
parameters of the master system: $C_{m_{M}}=1$ $\mu F/cm^2$,
$g_{K_{M}}=36$ $m\Omega^{-1}/cm^2$, $g_{Na_{M}}=120$
$m\Omega^{-1}/cm^2$, $g_{l_{M}}=0.3$ $m\Omega^{-1}/cm^2$,
$V_{K_{M}}=12$ $mV$, $V_{Na_{M}}=-115$ $ mV$ and
$V_{l_{M}}=-10.613$ $mV$, and parameters for the slave system are
chosen with a difference of $10$ $\%$ from the nominal values:
$C_{m_{S}}=0.9$ $\mu F/cm^2$, $g_{K_{S}}=32.4$
$m\Omega^{-1}/cm^2$, $g_{Na_{S}}=108$ $m\Omega^{-1}/cm^2$,
$g_{l_{S}}=0.27$ $m\Omega^{-1}/cm^2$, $V_{K_{S}}=10.8$ $mV$,
$V_{Na_{S}}=-103.5$ $mV$ and $V_{l_{S}}=-9.5517$ $mV$. Under such
a parameters both neurons shall not "share time solutions", then
they cannot be synchronous.
\bigskip

\section{Synchronization problem statement}

In the nonlinear control theory a synchronization problem can be
stated as a feedback stabilization one \cite{femat,femat-97}. It
has been established that for a defined synchronization error,
$x_{e} = x_{e,M}-x_{e,S}$ (where $x \in R^d,$ $e=1,2,...,d$),
there exists a synchronization error (dynamical) system
\cite{femat} whose trajectories exponentially converge to zero
under a feedback control $u$. Hence, it can be said that the
master and slave systems (unidirectionally) couple and attain a
synchronization dynamical state under the action of the control
command $u$. The following definition of Exact Synchronization is
presented in \cite{fematgual}:

{\bf Definition 1.} It is said that two chaotic systems are
exactly synchronized if the synchronization error, $x_{e} =
x_{e,M}-x_{e,S}$, exponentially converges to the origin. This
implies that at a finite time $x_{e,S}=x_{e,M}$.

In this section, a proof for the Exact Synchronization of
two HH neurons is provided.\\

{\bf Lemma 1.} \begin{em}Consider the two HH neuronal systems
represented in Eqs. (\ref{eq5})-(\ref{eq12}). Such two silent HH
neurons attain dynamical states of Exact Synchronization for all
$t>t_{0}\ge 0$ under the action of a nonlinear feedback control
despite parametric differences for any initial condition $x_{i}(0)
= x_{i,M}(0)-x_{i,S}(0)$ in the domain physically realizable.
\end{em}\\

{\bf Proof.} Let us define the following synchronization error
system for the HH neuronal systems:
\begin{equation}
 \dot x = F(x,t)+ G(x)u = \Delta f(x) + \Delta I(t) - Bu, \quad y
= h(x) = x_{1} \, ,\label{eq13}
\end{equation}
where
\begin{eqnarray}
\Delta f(x) &=& \left({\rm \begin{array} {cc}
 \{-1/C_{m_{M}}\left[g_{K_{M}}x^4_{2,M}(x_{1,M}-V_{K_{M}})\right.
+g_{Na_{M}} x_{3,M}^3x_{4,M}(x_{1,M} \\
-V_{Na_{M}}) \left.+ g_{l_{M}}(x_{1,M}-V_{l_{M}}) \right]
+1/C_{m_{S}} \left[g_{K_{S}}x^4_{2,S} (x_{1,S}-V_{K_{S}})\right. \\
+ g_{Na_{S}} x_{3,S}^3 x_{4,S} (x_{1,S} -V_{Na_{S}})
\left.+g_{l_{S}}(x_{1,S}-V_{l_{S}})\right]\}\\
\{ \alpha_{n}(x_{1,M})(1-x_{2,M})-\beta_{n}(x_{1,M})x_{2,M}\\
-\alpha_{n}(x_{1,S})(1-x_{2,S})+\beta_{n}(x_{1,S})x_{2,S} \}\\
\{ \alpha_{m}(x_{1,M})(1-x_{3,M})-\beta_{m}(x_{1,M})x_{3,M}\\
-\alpha_{m}(x_{1,S})(1-x_{3,S})+\beta_{m}(x_{1,S})x_{3,S} \}\\
\{ \alpha_{h}(x_{1,M})(1-x_{4,M})-\beta_{h}(x_{1,M})x_{4,M}\\
-\alpha_{h}(x_{1,S})(1-x_{4,S})+\beta_{h}(x_{1,S})x_{4,S} \}
\end{array}} \right) \, , \nonumber\\
\Delta I(t)&=&\left(\frac{I_{ext_{M}}(t)}{C_{m_{M}}}\right.
\left.-\frac{I_{ext_{S}}(t)}{C_{m_{S}}}, 0, 0, 0 \right)^T, \quad
B=\left(1, 0,0,0 \right)^T, \nonumber
\end{eqnarray}
$\Delta f(x)$ is a smooth vector field, $\Delta I(t)$ is the
difference between the external exciting forces and $y \in R$
represents the measured state of the system. The relative degree
of a system is defined as the number of times one has to
differentiate the output $y(t)$ before the control action $u$
explicitly appears \cite{isidori1}. It can be easily shown that
the relative degree for the system (\ref{eq13}) is $\rho=1$. Let
us consider now the \textbf{Proposition 4.4.2} given in
\cite{isidori1}. The following stabilizing control law is
proposed,
\begin{equation}
u=\frac{1}{L_{g}L_{f}^{\rho-1}h(x)}\left(-L_{f}^{\rho}h(x)
-c_{0}h(x)-c_{1}L_{f}h(x)-...-c_{\rho-1}L_{f}^{\rho-1}h(x)\right)\label{eq-u1}
\end{equation}
where $L_{f}h(x)$ stands for the Lie derivative of the function
$h(x)$ along the vector field $f$, and the constant parameters
$c_{0},c_{1},...,c_{\rho-1}$, belong to the polynomial
$p(s)=c_{0}+c_{1}s+...+c_{\rho-1}s^{\rho-1}+s^{\rho}$ with all its
eigenvalues having negative real part. Then, the synchronization
error system (\ref{eq13}) is stabilized by the control law
$u=[-L_{F}h(x)-C_{0}h(x)]/L_{G}h(x)$, and $C_{0}$ is a positive
real constant that represents the convergence rate. Under the
control action given by
\begin{eqnarray}
\label{eq-u}
 u &=& 1/C_{m_{M}} \left[I_{ext_{M}}(t)-g_{K_{M}}x^4_{2,M} (x_{1,M}-V_{K_{M}})\right.
-g_{Na_{M}}x_{3,M}^3x_{4,M}(x_{1,M}-V_{Na_{M}})\nonumber\\
 &&- g_{l_{M}}(x_{1,M}-V_{l_{M}})\left. \right]
-1/C_{m_{S}} \left[ I_{ext_{S}}(t) \right.
 - g_{K_{S}}x^4_{2,S} \left( x_{1,S}-V_{K_{S}} \right)  \nonumber \\
&&- g_{Na_{S}} x_{3,S}^3 x_{4,S} \left( x_{1,S} -V_{Na_{S}}
\right) \left. - g_{l_{S}}\left( x_{1,S}-V_{l_{S}} \right)\right]
 + C_{0}\left(x_{1,M}-x_{1,S} \right)
\end{eqnarray}
\noindent Eq. (\ref{eq13}) becomes
\begin{eqnarray}
\dot x_{1} &=& -C_{0}x_{1}, \label{eqerr1}\\
\dot x_{2} &=& \alpha_{n}(x_{1,M})- \alpha_{n}(x_{1,M},x_{1})-
 [\alpha_{n}(x_{1,M})+\beta_{n}(x_{1,M})-\alpha_{n}(x_{1,M},x_{1})\nonumber \\
 &&-\beta_{n}(x_{1,M},x_{1})]x_{2,M}-[\alpha_{n}(x_{1,M},x_{1})+\beta_{n}(x_{1,M},x_{1})]x_{2},\label{eqerr2}\\
\dot x_{3} &=& \alpha_{m}(x_{1,M})- \alpha_{m}(x_{1,M},x_{1})-
 [\alpha_{m}(x_{1,M})+\beta_{m}(x_{1,M})-\alpha_{m}(x_{1,M},x_{1})\nonumber \\
 &&-\beta_{m}(x_{1,M},x_{1})]x_{3,M}-[\alpha_{m}(x_{1,M},x_{1})+\beta_{m}(x_{1,M},x_{1})]x_{3},\label{eqerr3}\\
\dot x_{4} &=& \alpha_{h}(x_{1,M})- \alpha_{h}(x_{1,M},x_{1})-
 [\alpha_{h}(x_{1,M})+\beta_{h}(x_{1,M})-\alpha_{h}(x_{1,M},x_{1})\nonumber\\
 &&-\beta_{h}(x_{1,M},x_{1})]x_{4,M}-[\alpha_{h}(x_{1,M},x_{1})+\beta_{h}(x_{1,M},x_{1})]x_{4}\label{eqerr4}.
\end{eqnarray}
\noindent Because of the system (\ref{eqerr1})-(\ref{eqerr4}) is
minimum-phase (see Appendix A at the end of the chapter), the
obtained control law $u$ leads the trajectories of the
synchronization error system to asymptotically converge to zero in
a finite time. $\Box$
\\

{\bf Remark 1.} It should be noted that, by definition, Exact
Synchronization implies that $x_{M}\equiv x_{S}$ for any time
$t>t_{0}\ge0$ and $x(0)=x_{M}(0)-x_{S}(0)$ in a given domain. Now,
also by definition, GS corresponds to the state where the states
of slave systems can be written as a function of the master states
(i.e., $x_{S}=\psi(x_{M})$). Thus, as we shall see below, the
unidirectional synchronization of noiseless HH neurons is exact
and $x_{S}=Ix_{M}$, where $I$ stands for the identity matrix.\\

In order to establish the GS between the neurons
(\ref{eq5})-(\ref{eq12}), the following results state the
conditions to derive an expression for $x_S=\psi(x_M)$. Thus, we
depart from the Fact 1.

The following definitions are useful concepts presented in
\cite{isidori1}. For two vector fields $f$ and $g$, both defined
on an open subset $U$ of $R^{n}$ (i.e., $U\subset R^n$), the
\textit{Lie bracket} $[f,g]$ is a third vector field defined by
$[f,g](x)=\frac{\partial g}{\partial x}f(x)-\frac{\partial
f}{\partial x}g(x)$, where $\frac{\partial g}{\partial x}$ and
$\frac{\partial f}{\partial x}$ are Jacobian matrices.

For a given set of $d$ vector fields $f_{1}$,...,$f_{d}$, all
defined on the same open set $U$, let
$\Delta(x)=span\{f_{1}(x),...,f_{d}(x)\}$ be a subspace of $R^n$
spanned by the vectors $f_{1}$,...,$f_{d}$, at any fixed point $x$
in $U$. The subspace $\Delta(x)$ of $R^n$, for $x\in U$, is called
a \textit{distribution}. A distribution $\Delta$ is
\textit{involutive} if $\tau_{1}\in\Delta$, $\tau_{2}\in\Delta$
$\Rightarrow$ $[\tau_{1},\tau_{2}]\in\Delta$, where $\tau_{1}$ and
$\tau_{2}$ are any pair of vector fields belonging to $\Delta$.\\

{\bf Fact 1 \cite{isidori1}.} Consider an affine nonlinear system
$\dot x=f(x)+g(x)u$; where $x\in \Omega \subseteq R^n$, $u\in R$,
$g,f:R^n\rightarrow R^n$ are smooth vector fields. Besides, let us
consider that $y=h(x)$ for any smooth function $h(x)$. If
involutivity condition is satisfied, then the mappings
$\Phi_1:R^n\rightarrow R^\rho$, $x\mapsto z$ and
$\Phi_2:R^n\rightarrow R^{n-\rho}$, $x\mapsto (z,\nu)$ are such
that the affine nonlinear system can be written in the canonical
form
\begin{eqnarray}
\dot z_i&=&z_{i+1}, \quad i=1,2,...,\rho-1,\nonumber\\
\dot z_\rho&=&\alpha(z,\nu)+\beta(z,\nu)u,\label{faeq1}\\
\dot \nu&=&\zeta(z,\nu),\nonumber
\end{eqnarray}
and can be derived from Lie derivatives of the output function
$h(x)$ along the vector fields $f(x)$ and $g(x)$ as follows
\begin{eqnarray}
z=\Phi_{1}(x)=\left({\rm \begin{array} {cc}
h(x)\\
L_{f}h(x)\\
...\\
L_{f}^{\rho-1}h(x)
\end{array}} \right) \label{faeq2}
\end{eqnarray}
and
\begin{eqnarray}
\nu=\Phi_{2}(x)=\left({\rm \begin{array} {cc}
\phi_{\rho+1}(x)\\
\phi_{\rho+2}(x)\\
...\\
\phi_{n}(x)
\end{array}} \right), \label{faeq3}
\end{eqnarray}
moreover, it is always possible to chose $\phi_{\rho+1},...,
\phi_{n}$ in such a way that
\begin{equation}
L_{g}\phi_{j}(x)=0, \quad \rho+1\leq j\leq n. \label{faeq4}
\end{equation}
$\Box$

The Fact 1 is well known in nonlinear control theory. Here it is
included for clarity in presentation and exploited in neuronal
synchronization towards robust feedback synchronization of HH
neurons.

{\bf Fact 2 \cite{isidori1}.} If exists the map
$\Phi=(\Phi_{1},\Phi_{2}):R^{n}\rightarrow R^{n}$, $x\mapsto
(z,\nu)$ derived from (\ref{faeq2}) and (\ref{faeq3}), then there
exists the inverse $\Phi^{-1}(\Phi(x))=x\in \Omega\subset R^{n}$.
This fact is proved since $h(x)$, $L_{f}h(x)$, ...,
$L_{f}^{\rho-1}h(x)$ and $\phi_{\rho+1}(x)$, ..., $\phi_{n}(x)$
are linearly independent at any $x$ in the neighborhood
$U\subset\Omega\subseteq R^{n}$ of the point $x^{0}$ in $\Omega$.

\bigskip

\section{Synchronizing the Hodgkin-Huxley neurons}

Once obtained the control action (\ref{eq-u}), it can be directly
implemented in Eq. (\ref{eq9}) for the slave system leading to the
following set of coupled nonlinear differential equations,
\begin{eqnarray}
\dot x_{1,M} &=& 1/C_{m_{M}} \left[ I_{ext_{M}}(t) \right.
 - g_{K_{M}}x^4_{2,M} \left( x_{1,M}-V_{K_{M}} \right)  \nonumber \\
&&- g_{Na_{M}} x_{3,M}^3 x_{4,M} \left( x_{1,M} -V_{Na_{M}}
\right) \left. - g_{l_{M}}\left( x_{1,M}-V_{l_{M}} \right)
\right],
\label{eq16}\\
\dot x_{2,M} &=& \alpha_{n}(x_{1,M})\left( 1-x_{2,M} \right) -
 \beta_{n}(x_{1,M})x_{2,M}\, ,
 \label{eq17}\\
\dot x_{3,M} &=& \alpha_{m}(x_{1,M})\left( 1-x_{3,M} \right) -
 \beta_{m}(x_{1,M})x_{3,M}\, ,
 \label{eq18}\\
\dot x_{4,M} &=& \alpha_{h}(x_{1,M})\left( 1-x_{4,M} \right) -
 \beta_{h}(x_{1,M})x_{4,M} \, ,
 \label{eq19}\\
 \dot x_{1,S} &=& 1/C_{m_{M}} \left[ I_{ext_{M}}(t) \right.
 - g_{K_{M}}x^4_{2,M} \left( x_{1,M}-V_{K_{M}} \right)  \nonumber \\
&&- g_{Na_{M}} x_{3,M}^3 x_{4,M} \left( x_{1,M} -V_{Na_{M}}
\right) \left. - g_{l_{M}}\left( x_{1,M}-V_{l_{M}} \right)
\right]\nonumber\\
 &&+C_{0}\left(x_{1,M}-x_{1,S} \right) ,
\label{eq20}\\
\dot x_{2,S} &=& \alpha_{n}(x_{1,S})\left( 1-x_{2,S} \right) -
\beta_{n}(x_{1,S})x_{2,S}\, ,
\label{eq21}\\
\dot x_{3,S} &=& \alpha_{m}(x_{1,S})\left( 1-x_{3,S} \right) -
\beta_{m}(x_{1,S})x_{3,S}\, ,
\label{eq22}\\
\dot x_{4,S} &=& \alpha_{h}(x_{1,S})\left( 1-x_{4,S} \right) -
\beta_{h}(x_{1,S})x_{4,S} \, . \label{eq23}
\end{eqnarray}

The above set of equations represents the dynamics of the HH
neuronal synchronization when the control action $u$ is
implemented. The right-hand side of Eq. (\ref{eq20}) describes the
new induced dynamics of the slave system. The master and slave
systems unidirectionally couple through $u$; also parametric
differences have been subtracted. The term containing the
convergence rate $C_{0}$ can be interpreted as a synaptic-like
control current (divided by a constant capacitance) being
$C_{0}C_{m_{M}}$ a constant synaptic conductance. In the framework
of geometrical control and its applications on communicating
systems, the synchronization of chaotic dynamics for the HH
neurons could be understood as synchronization of a transmitter
(master)-receiver (slave) system. Interpretation of HH neurons as
chaotic systems in context of communicating systems, which
transmit information, could provide insight to understand the way
the biological information is processed in neuronal ensembles.

Numerical simulations were carried out for the HH neuronal
synchronization system. Sinusoidal exciting modulations are
considered, and the amplitud and frequency parameters are chosen
within the \textit{U}-shaped curve shown in Fig. 1.(b) of
reference \cite{parmananda}. This curve encapsulates the region of
parameter space (in amplitude and frequency domain) where the
exciting modulations trigger spike trains of the action potential
in the model system of single silent HH neurons.

Fig. 8.1 shows desynchronized regular (master system) and
irregular (slave system) spiking patterns and the transition to a
regular synchronized state of the action potentials. The applied
forcing functions are $I_{ext_{M}}(t)=-2.58\textrm{sin}(.245t)$
and $I_{ext_{S}}(t)=-3.15\textrm{sin}(.715t)$. Initial conditions
were chosen as $x_{i,M}(0)=(10\,mV,0,0,$ 0) and
$x_{i,S}(0)=(0,0,0,0)$, and the control action was implemented at
time $t_{0}=180$ $ms$. A choice for $C_{0}=0.3$ leads to rapid
synchronization convergence for the refractory period. The
activation and inactivation dynamics for the ionic channels also
attains synchronization state. Fig. 8.2 shows the evolution in
time of voltage per second which is supplied to the neuronal
system to achieve the synchronization. Fig. 8.3 shows the phase
locking of the action potentials in synchronized state of Fig.
8.1.
\bigskip
\vskip 1ex \centerline{ \epsfxsize=210pt \epsfbox{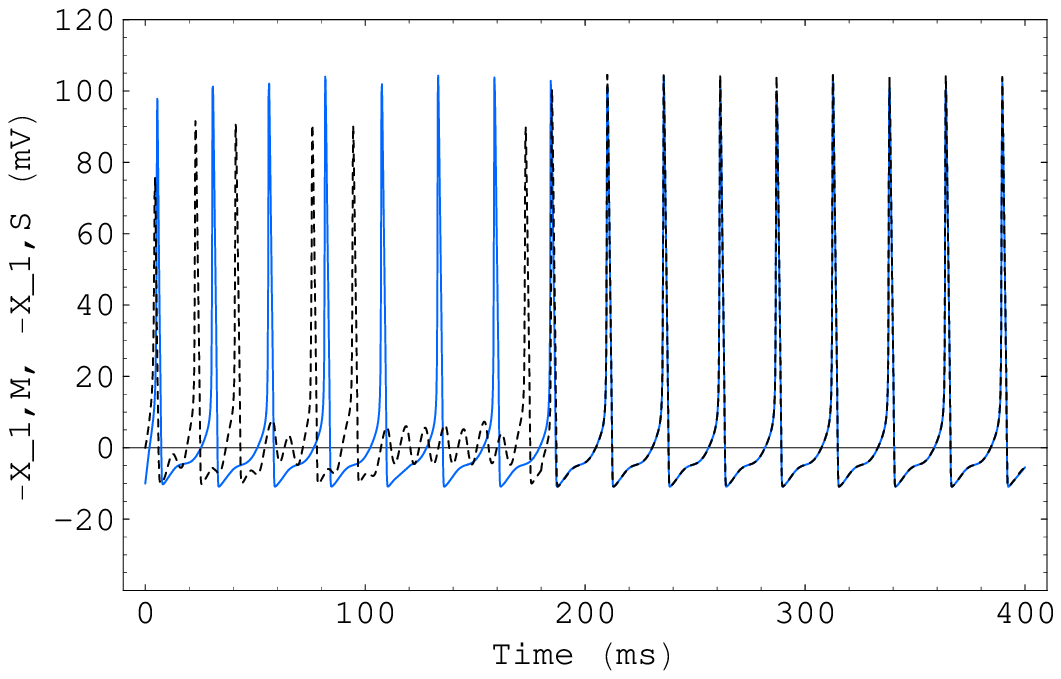}}
\vskip 3ex
\begin{center} {\small{Fig.
8.1:}$\quad$ Spiking patterns of the master (solid line) and slave
(dashed line) systems for the action potentials in desynchronized
and synchronized states. The forcing functions amplitud and
frequency parameters as specified in the text:
$I_{ext_{M}}(t)=-2.58\textrm{sin}(.245t)$,
$I_{ext_{S}}(t)=-3.15\textrm{sin}(.715t)$.}
\end{center}

\vskip 1ex \centerline{ \epsfxsize=200pt \epsfbox{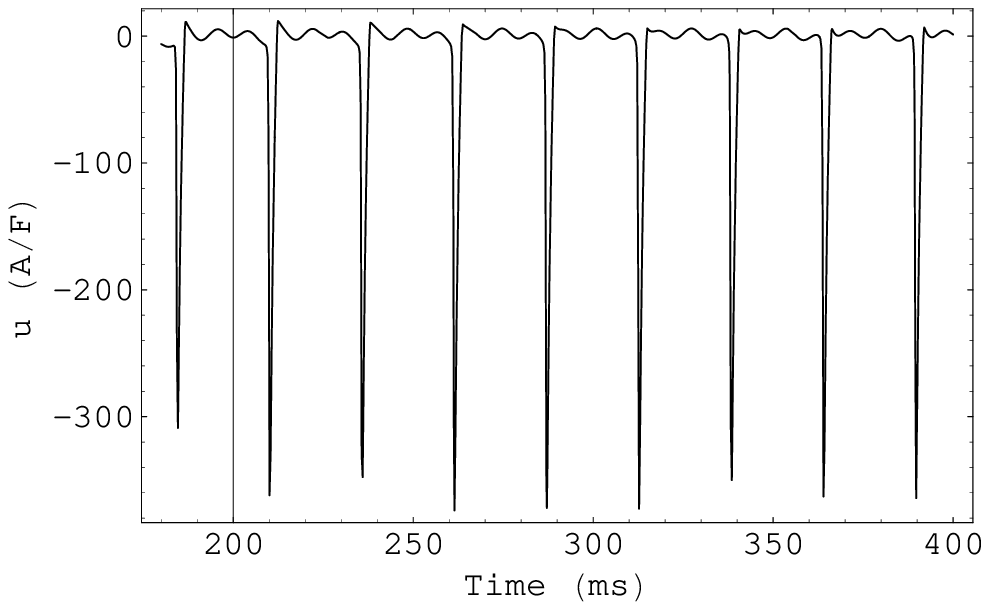}}
\vskip 3ex
\begin{center} {\small{Fig.
8.2:}$\quad$ Dynamical response of the implemented control action
of Fig 8.1.}
\end{center}
\bigskip
\bigskip

\vskip 1ex \centerline{ \epsfxsize=200pt \epsfbox{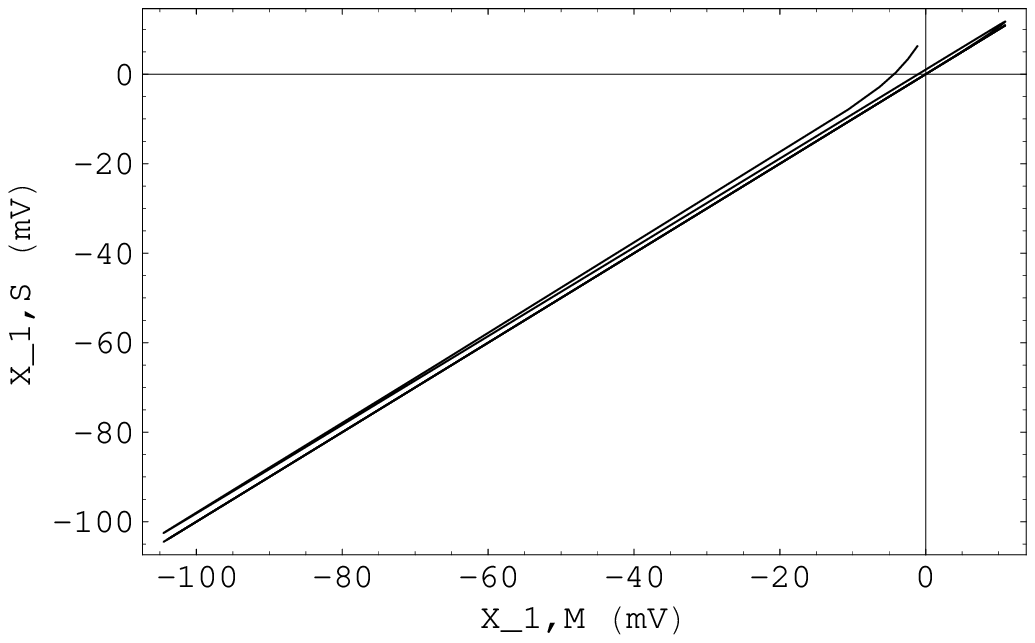}}
\vskip 3ex
\begin{center} {\small{Fig.
8.3:}$\quad$ Phase locking of the synchronized action potentials
of Fig 8.1.}
\end{center}
\bigskip

Note that the nonlinear feedback (\ref{eq-u}) requires information
about currents flowing through the membrane and the ionic channels
of the master and slave neurons. Such a coupling cannot be
implemented in practice on real neurons. However, as we shall see
below, the controller (\ref{eq-u}) allows to discuss the robust
synchronization in context of GS, which is the most significant
phenomenon in chaotic synchronization. Thus, once neuron
synchronization is discussed in terms of GS, the robust
synchronization is proposed by relaxing the nonlinear controller
(\ref{eq-u}) towards a linear approach. This linear approach is
robust in the sense that synchronization is induced in face of
parameter mismatches and differences between amplitud and
frequency parameters in external current entering into master and
slave neurons.
\bigskip

\newpage
\section{Generalized and robust synchronization}

\medskip

{\bf 8.5.1 Generalized synchronization}

In this subsection, we derive the mappings $\Phi_1:R^n\rightarrow
R^\rho$, $x\mapsto z$ and $\Phi_2:R^n\rightarrow R^{n-\rho}$,
$x\mapsto (z,\nu)$, and write the HH model of neurons in canonical
form (\ref{faeq1}) towards generalized synchronization (GS).

Dynamical models for each HH neuron can be written in nonlinear
affine form $\dot x=f(x)+g(x)u$, with
\begin{eqnarray}
f(x)=\left({\rm \begin{array} {c} \{1/C_{m}[I_{ext} - g_{K}x_{2}^4
\left(
x_{1}-V_{K} \right) \\
- g_{Na} x_{3}^3 x_{4} \left( x_{1}-V_{Na} \right) - g_{l}\left(
x_{1}-V_{Na}
\right)]\}\\
\alpha_{n}(x_{1})(1-x_{2}) -
 \beta_{n}(x_{1})x_{2}\\
\alpha_{m}(x_{1})(1-x_{3}) -
 \beta_{m}(x_{1})x_{3}\\
\alpha_{h}(x_{1})(1-x_{4}) -
 \beta_{h}(x_{1})x_{4}
\end{array}} \right),\quad g(x)=\left({\rm \begin{array} {c}
1\\
0\\
0\\
0
\end{array}} \right),\label{eq24}
\end{eqnarray}
and the output $y=h(x)$ is given by the membrane potential, i.e.,
$h(x)=x_{1}$. By computing the Lie derivatives of the output
function along the vector fields (\ref{eq24}), we obtain
\begin{equation}
z=\Phi_{1}(x)=h(x)=x_{1}
\end{equation}
and
\begin{eqnarray}
\nu=\Phi_{2}(x)=\left({\rm \begin{array} {c}
\phi_{1}(x)\\
\phi_{2}(x)\\
\phi_{3}(x)
\end{array}} \right),
\end{eqnarray}
then according to Eqs. (\ref{faeq3}) and (\ref{faeq4}) $\nu$ can
be chosen as
\begin{eqnarray}
\nu=\left({\rm \begin{array} {c}
\nu_{1}\\
\nu_{2}\\
\nu_{3}
\end{array}} \right)
=\left({\rm \begin{array} {c}
x_{2}\\
x_{3}\\
x_{4}
\end{array}} \right).
\end{eqnarray}
Consequently, the previous HH model (\ref{ho4})-(\ref{ho7}) is
transformed into
\begin{eqnarray}
\dot z_{1} &=& 1/C_m[I_{ext} - g_{K}\nu_{1}^4 \left( z_{1}-V_{K}
\right) -
g_{Na} \nu_{2}^3 \nu_{4} \left( z_{1}-V_{Na} \right)\nonumber\\
&& - g_{l}\left( z_{1}-V_{Na}\right)] + u\, ,\\
\dot \nu_{1}&=&\alpha_{n}\left( z_{1} \right)\left( 1-\nu_{1}
\right) -
 \beta_{n}\left( z_{1} \right) \nu_{1}\, ,\\
\dot \nu_{2}&=&\alpha_{m}\left( z_{1} \right)\left( 1-\nu_{2}
\right) -
 \beta_{m}\left( z_{1} \right) \nu_{2}\, ,\\
\dot \nu_{3}&=&\alpha_{h}\left( z_{1} \right)\left( 1-\nu_{3}
\right) -
 \beta_{h}\left( z_{1} \right) \nu_{3}\, .
\end{eqnarray}

In what follows we show how the GS can be studied in HH neurons by
departing from Lemma 1 and Fact 1. To this end, we can separately
transform both master and slave neurons. In this manner, we shall
derive the maps $\Phi_{M}(x_{M})$ and $\Phi_{S}(x_{S})$ to get
\begin{eqnarray}
\left({\rm \begin{array} {c}
z_{M}\\
\nu_{M}
\end{array}} \right)=\left({\rm \begin{array} {c}
\Phi_{1M}(x_{M})\\
\Phi_{2M}(x_{M})
\end{array}} \right)
\quad \textrm{and} \quad \left({\rm \begin{array} {c}
z_{S}\\
\nu_{S}
\end{array}} \right)=\left({\rm \begin{array} {c}
\Phi_{1S}(x_{S})\\
\Phi_{2S}(x_{S})
\end{array}} \right),\label{gs1}
\end{eqnarray}
from where each HH neuron can be transformed into (\ref{faeq1})
and driving signal (\ref{eq-u1}) induces the master behavior onto
slave neuron. Then, if stability holds and neurons are
minimum-phase systems, $(z_{S},\nu_{S})\rightarrow
(z_{M},\nu_{S}^{*})$ for $t>t_{0}\geq 0$, for any initial
conditions $(z(0),\nu(0))=(\Phi_{1}(x(0)),\Phi_{2}(x(0)))$ in
physical domain. Note that $\nu_{S}^{*}$ is a stable manifold
which can correspond to the stable manifold of the master neuron
$\nu_{M}^{*}$. In this case complete synchronization is achieved.
In case $\nu_{S}^{*}\neq \nu_{M}^{*}$, the partial state
synchronization is attained \cite{fematgual}. Anyway, the
composition $\Phi_{S}^{-1}(\Phi_{1}(x_{M});\nu_{S}^{*})=x_{S}\in
\Omega\subset R^{n}$, where $\Omega$ denotes the physical domain.
In particular, if $\nu_{S}^{*}\equiv \nu_{M}^{*}$ for all time
$t>t_{0}\geq 0$, where $t_{0}$ stands for time of turning on the
control, then
$x_{S}=\Phi_{S}^{-1}(\Phi_{1M}(x_{M}),\Phi_{2M}(x_{M}))$. Since HH
neurons (\ref{ho4})-(\ref{ho7}) are minimum-phase systems (see
Appendix A at the end of the chapter), the GS yields the following
relation
\begin{eqnarray}
x_{S}=\left({\rm \begin{array} {c}
h_{S}^{-1}(h_{M}(x_{M}))\\
x_{2S}^{*}\\
x_{3S}^{*}\\
x_{4S}^{*}
\end{array}} \right).\label{gs2}
\end{eqnarray}
Now, it should be pointed out that driving signal (\ref{eq-u1})
has full information about states of both master and slave
neurons. This situation cannot be physically realizable (for
example, currents due to the ionic channels activity cannot be
available for feedback). In next paragraphs, a robust approach is
taken from open literature to show how the HH neurons can be
synchronized.

\medskip

{\bf 8.5.2 Robust synchronization}

The nonlinear controller (\ref{eq-u}) allows to obtain states of
Exact Synchronization for the silent HH neurons represented by
systems (\ref{eq5})-(\ref{eq12}). However, because of measurements
for activation and inactivation of the ionic channels cannot be
physically carried out, implementation of the control action
(\ref{eq-u}) would be unpractical. Then, an adaptive scheme to
yield robust synchronization is realized by using a modified
feedback control law. The synchronization error system
(\ref{eq13}) can be represented in the following extended form
\cite{femat},
\begin{eqnarray}
\dot z_{1} = \eta + \beta_{E}(z)u,\quad \dot \eta =
\Gamma(z_{1},\eta,\nu,u),\quad \dot \nu = \zeta(z_{1},\nu),\quad
y=z_{1}, \label{eqex1}
\end{eqnarray}
where the invertible coordinates change $x=(z_{1},\nu)$ has been
developed; $\eta=\Delta f_{1}(z_{1},\nu)+ \Delta I_{1}(t)$,
represents an augmented state which lumps the uncertain terms (the
ionic channels activation and inactivation variables for the
master and slave systems) contained in $\Delta f_{1}$, $\nu$ is
the state vector for the internal dynamics and $\beta_{E}(z)=-1$.
In order to stabilize the synchronization error system, we
consider the nonlinear controller (\ref{eq-u}) and observe that it
can be written in the following (linearizing-like) form
\begin{equation}
u= [ \eta + kz_{1} ] \label{eqex-u1}
\end{equation}
where $k \in R_{+}$ represents a control gain value. However,
since the control law (\ref{eqex-u1}) depends on the uncertain
state $\eta$, it is not physically realizable. The problem of
estimating ($z_{1}$,$\eta$) is solved by using a high-gain
observer (dynamic compensator) \cite{femat},
\begin{eqnarray}
\dot{\hat{z}}_{1} &=& \hat{\eta} -u +
L_{0}\kappa_{1}^{*}(z_{1}-\hat{z}_{1}),\label{eqest1}\\
\dot{\hat{\eta}} &=&
L_{0}^{2}\kappa_{2}^{*}(z_{1}-\hat{z}_{1}),\label{eqest2}
\end{eqnarray}
where ($\hat{z}_{1}$,$\hat{\eta}$) are the estimated values of
($z_{1}$,$\eta$); $L_{0}$ is the unique tuning parameter, and
represents a high-gain estimation parameter that can be
interpreted as the uncertainties estimation rate. The parameters
$\kappa_{1,2}$ are chosen for the polynomial
$P(s)=s^2+\kappa_{2}s+\kappa_{1}=0$ with all its eigenvalues in
the left-half complex plane.

The linearizing control law with uncertainty estimation that,
together with the dynamic compensator
(\ref{eqest1})-(\ref{eqest2}), stabilizes the synchronization
error at the origin and, consequently, synchronizes the HH
neuronal systems now becomes
\begin{equation}
u = [ \hat{\eta} + k\hat{z}_{1} ]. \label{eqex-u2}
\end{equation}

A stability analysis for the closed loop system (\ref{eqex1}),
(\ref{eqest1})-(\ref{eqex-u2}) is provided in Appendix B at the
end of the chapter. A tuning algorithm for stability and duration
time is also provided in \cite{bowong}. Thus, controller
(\ref{eqest1})-(\ref{eqex-u2}) is a general approach to
synchronization of HH neurons despite it lacks knowledge about the
states of activation and/or inactivation of the potassium and
sodium ionic channels.

It is pointed out that the modified feedback control law
(\ref{eqest1})-(\ref{eqex-u2}) yields Complete Practical
Synchronization \cite{fematgual}, i.e., the trajectories of the
synchronization error system converge around the origin within a
ball of radius $L_{0}^{-1}$.

Once obtained the modified control law, it can be implemented in
systems (\ref{eq5})-(\ref{eq12}). Then, we are led to the
following extended system of differential equations that
guarantees the robust synchronization of HH neurons,
\begin{eqnarray}
\dot x_{1,M} &=& 1/C_{m_{M}} \left[ I_{ext_{M}}(t) \right.
 - g_{K_{M}}x^4_{2,M} \left( x_{1,M}-V_{K_{M}} \right)  \nonumber \\
&&- g_{Na_{M}} x_{3,M}^3 x_{4,M} \left( x_{1,M} -V_{Na_{M}}
\right) \left. - g_{l_{M}}\left( x_{1,M}-V_{l_{M}} \right)
\right],
\label{eqac1}\\
\dot x_{2,M} &=& \alpha_{n}(x_{1,M})\left( 1-x_{2,M} \right) -
 \beta_{n}(x_{1,M})x_{2,M}\, ,
 \label{eqac2}\\
\dot x_{3,M} &=& \alpha_{m}(x_{1,M})\left( 1-x_{3,M} \right) -
 \beta_{m}(x_{1,M})x_{3,M}\, ,
 \label{eqac3}\\
\dot x_{4,M} &=& \alpha_{h}(x_{1,M})\left( 1-x_{4,M} \right) -
 \beta_{h}(x_{1,M})x_{4,M} \, ,
 \label{eqac4}\\
\dot x_{1,S} &=& 1/C_{m_{S}} \left[ I_{ext_{S}}(t) \right.
 - g_{K_{S}}x^4_{2,S} \left( x_{1,S}-V_{K_{S}} \right)  \nonumber \\
&&- g_{Na_{S}} x_{3,S}^3 x_{4,S} \left( x_{1,S} -V_{Na_{S}}
\right) \left. - g_{l_{S}}\left( x_{1,S}-V_{l_{S}}
\right)\right]+(\hat{\eta} + k\hat{z}_{1}) ,
\label{eqac5}\\
\dot x_{2,S} &=& \alpha_{n}(x_{1,S})\left( 1-x_{2,S} \right) -
 \beta_{n}(x_{1,S})x_{2,S}\, ,
 \label{eqac6}\\
\dot x_{3,S} &=& \alpha_{m}(x_{1,S})\left( 1-x_{3,S} \right) -
 \beta_{m}(x_{1,S})x_{3,S}\, ,
 \label{eqac7}\\
\dot x_{4,S} &=& \alpha_{h}(x_{1,S})\left( 1-x_{4,S} \right) -
 \beta_{h}(x_{1,S})x_{4,S} \, ,
 \label{eqac8}\\
\dot{\hat{z}}_{1} &=&  -k\hat{z}_{1}+
L_{0}\kappa_{1}^{*}((x_{1,M}-x_{1,S})-\hat{z}_{1}),\label{eqac9}\\
\dot{\hat{\eta}} &=&
L_{0}^{2}\kappa_{2}^{*}((x_{1,M}-x_{1,S})-\hat{z}_{1}),\label{eqac10}
\end{eqnarray}

Fig. 8.4 shows the attained robust synchronization dynamics for
the master (solid line) and slave (dashed line) systems when the
modified feedback control law has been implemented. The control
gain value was chosen as $k=1$, the $\kappa_{1,2}$ parameters were
chosen for the polynomial $P(s)$ with its eigenvalues located at
$s=-20$, and the high-gain parameter is $L_{0}=50$. The applied
forcing functions are taken as in Section 8.4,
$I_{ext_{M}}(t)=-2.58\textrm{sin}(.245t)$ and
$I_{ext_{S}}(t)=-3.15\textrm{sin}(.715t)$. Initial conditions were
chosen as $x_{i,M}(0)=(10\,mV,0,0,$ 0) and $x_{i,S}(0)=(0,0,0,0)$,
and the modified control law was implemented at time $t_{0}=200$
$ms$. Dynamics of the ionic channels is also synchronized by the
modified control law.  Fig. 8.5 shows the evolution in time of
voltage per second which is supplied to the neuronal system to
achieve the robust synchronization. Fig. 8.6 shows the phase
locking of the action potentials for the robust synchronization
state of Fig. 8.4.

\bigskip
\vskip 1ex \centerline{ \epsfxsize=210pt
\epsfbox{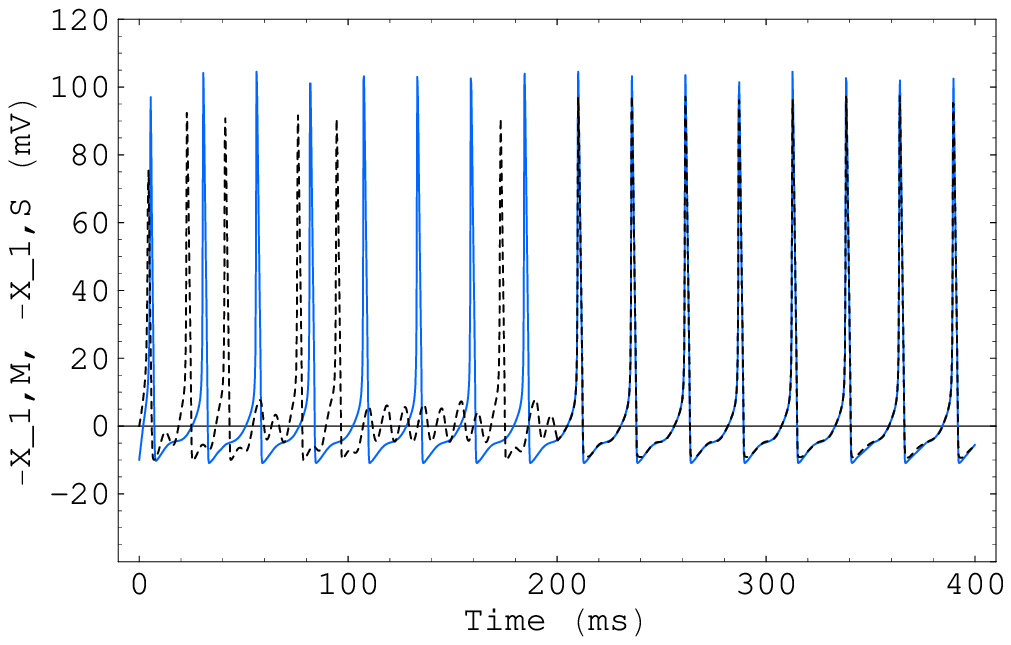}} \vskip 3ex
\begin{center} {\small{Fig.
8.4:}$\quad$ Spiking patterns of the master (solid line) and slave
(dashed line) systems for the action potentials in desynchronized
state and the transition to a robust synchronization state when
the modified feedback control law is implemented. The forcing
functions are $I_{ext_{M}}(t)=-2.58\textrm{sin}(.245t)$,
$I_{ext_{S}}(t)=-3.15\textrm{sin}(.715t)$.}
\end{center}

\bigskip
\bigskip

\vskip 1ex \centerline{ \epsfxsize=200pt
\epsfbox{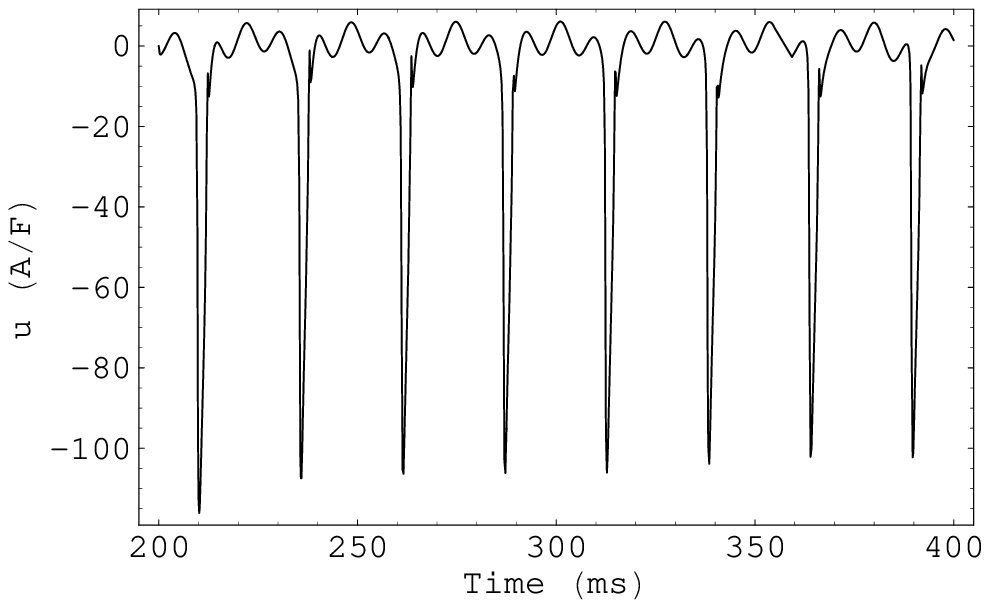}} \vskip 3ex
\begin{center} {\small{Fig.
8.5:}$\quad$ Dynamical response of the implemented modified
control law of Fig 8.4.}
\end{center}
\bigskip
\bigskip

\vskip 1ex \centerline{ \epsfxsize=200pt
\epsfbox{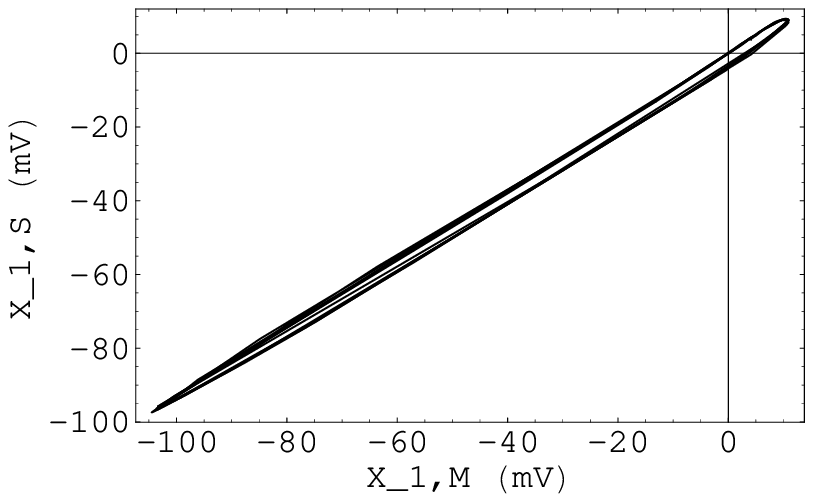}} \vskip 3ex
\begin{center} {\small{Fig.
8.6:}$\quad$ Phase locking of the action potentials in robust
synchronization state of Fig 8.4.}
\end{center}
\bigskip

\section{Conclusion of the chapter}

In this chapter, we have shown that there exists a nonlinear
unidirectional coupling such that two silent HH neurons attain
synchronized states in spite of parametric discrepancies. Results
show that synchronized spiking patterns of the action potentials
are displayed by the unidirectionally coupled system of HH
neurons. The synchronization coupling via the control action
(\ref{eq-u}) yields a synaptic-like control current term
containing as control parameter the convergence rate. Increases on
the control parameter have the effect to allow faster
synchronization convergence for the refractory dynamics. Regular
spiking patterns in synchronized state can be achieved for
regular-irregular desynchronized spiking sequences as shown in
Fig. 8.1. Because of measurements for the ionic channels
activation and inactivation are not physically realizable, a
robust adaptive scheme has been developed to yield the
synchronization dynamics of the HH neurons. A modified feedback
control law composed of a dynamic compensator and a linearizing
control law with uncertainties estimation has been implemented.
The adaptive scheme leads to robust synchronization dynamical
states of the action potentials as shown in Fig. 8.4. The ionic
channels activity is also synchronized. The dynamic compensator
allows to reconstruct the dynamics of the states $(z_{1},\eta)$
from measurements of the action potentials, and it requires only
one tuning parameter, $L_{0}$. Artificial devices experimentally
implemented in neuronal systems \cite{elson} are elucidating in
the unveiling of mechanisms underlying synchronization and control
parameters perform a very important role. The nonlinear control
theory could provide useful methods in studying synchronization
phenomena in single neurons and neural networks, even though
natural properties like intrinsic noise and synaptic conductances
must be regarded.
\bigskip

\newpage
{\bf Appendix A:} Internal dynamics of the synchronization error
system

\medskip

The synchronization error system (\ref{eqerr1})-(\ref{eqerr4}) is
already in canonical form. This means that the closed-loop system
has two subsystems: the first one, given by Eq. (\ref{eqerr1}), is
controllable while second one, given by Eqs.
(\ref{eqerr2})-(\ref{eqerr4}), is not affected by the
unidirectionally synchronization force $u$. Hence, to assure
asymptotic stability it is necessary to study the dynamics of
subsystem (\ref{eqerr2})-(\ref{eqerr4}). If subsystem
(\ref{eqerr2})-(\ref{eqerr4}) is (asymptotically) stable at
origin, then the closed-loop is said minimum-phase and, as a
consequence, the synchronization force $u$ leads the trajectories
of system (\ref{eq13}) to zero. Thus, we have that, for any time
$t>t_{0}\ge 0$, $x_{M}\equiv x_{S}$ and GS via Exact
Synchronization is achieved. The zero dynamics can be obtained by
setting $x_{1}=0$ \cite{isidori1} and considering the master
system dynamics,
\begin{eqnarray}
\dot x_{2} &=& -[\alpha_{n}(x_{1,M})+\beta_{n}(x_{1,M})]x_{2}, \nonumber\\
\dot x_{3} &=& -[\alpha_{m}(x_{1,M})+\beta_{m}(x_{1,M})]x_{3},\nonumber\\
\dot x_{4} &=&
-[\alpha_{h}(x_{1,M})+\beta_{h}(x_{1,M})]x_{4},\nonumber\\
\dot x_{1,M} &=& 1/C_{m_{M}} \left[ I_{ext_{M}}(t) \right.
 - g_{K_{M}}x^4_{2,M} \left( x_{1,M}-V_{K_{M}} \right)  \nonumber \\
&&- g_{Na_{M}} x_{3,M}^3 x_{4,M} \left( x_{1,M} -V_{Na_{M}}
\right) \left. - g_{l_{M}}\left( x_{1,M}-V_{l_{M}} \right)
\right],\quad\quad\quad\quad\quad\quad\;\;\;(\textrm{A}.1)\nonumber
\label{eq5a}\\
\dot x_{2,M} &=& \alpha_{n}(x_{1,M})\left( 1-x_{2,M} \right) -
 \beta_{n}(x_{1,M})x_{2,M}\, ,
 \label{eq6a}\nonumber\\
\dot x_{3,M} &=& \alpha_{m}(x_{1,M})\left( 1-x_{3,M} \right) -
 \beta_{m}(x_{1,M})x_{3,M}\, ,
 \label{eq7a}\nonumber\\
\dot x_{4,M} &=& \alpha_{h}(x_{1,M})\left( 1-x_{4,M} \right) -
 \beta_{h}(x_{1,M})x_{4,M} \, .
 \label{eq8a}\nonumber
\end{eqnarray}

Calculation of a linear approximation for the zero dynamics
(\textrm{A}.1) allows to obtain the corresponding eigenvalues in
order to determine the stability of the system. Let $\dot
x_{z}=f_{z}(x_{z},t)$ be the zero dynamics system where $x_{z}$ is
the state vector and $f_{z}(x_{z},t)$ is a smooth vector field.
The linear approximation $\dot x_{z}=Ax_{z}$ where $A$ is the
Jacobian matrix of the mapping $f_{z}$ evaluated at $x_{z}=0$,
yields the following result for the matrix $A$,
\begin{eqnarray}
A = \left( \begin{array}{ccccccc}
-0.183 & 0 & 0 & 0 & 0 & 0 & 0\\
0 & -4.224 & 0 & 0 & 0 & 0 & 0\\
0 & 0 & -0.117 & 0 & 0 & 0 & 0\\
0 & 0 & 0 & -0.3 & 0 & 0 & 0\\
0 & 0 & 0 & -0.0034 & -0.183 & 0 & 0\\
0 & 0 & 0 & -0.0154 & 0 & -4.224 & 0\\
0 & 0 & 0 & 0.0035 & 0 & 0 & -0.117 \end{array}
\right)\nonumber\\
(\textrm{A}.2)\nonumber
\end{eqnarray}
Due to the system (\textrm{A}.1) has all its eigenvalues in the
left-half complex plane the internal dynamics is locally
asymptotically stable and system (\ref{eqerr1})-(\ref{eqerr4}) is
minimum-phase.
\bigskip

\newpage
{\bf Appendix B:} Stability analysis for the synchronization error
system under the modified feedback control law

\medskip

Let $e \in R^2$ be the estimation error vector \cite{femat} with
states given by $e_{1}=z_{1}-\hat{z}_{1}$ and
$e_{2}=\eta-\hat{\eta}$. Then, the dynamics of the estimation
error is represented by the following system,
\begin{eqnarray}
\dot e=De + (0,\Gamma(z_{1},\eta,\nu,u))^T =\left(
\begin{array}{cc}
-L_{0}\kappa_{1}^{*} & 1 \\
-L_{0}^2\kappa_{2}^{*} & 0  \end{array} \right)e+\left(
\begin{array}{c}
0 \\
\Gamma(z_{1},\eta,\nu,u) \end{array}
\right)\quad\quad\;\;(\textrm{B}.1)\nonumber
\end{eqnarray}
Because of the trajectories of the synchronization error system
are contained in a chaotic attractor, the uncertain terms
$\eta(t)$ and $\Gamma(z_{1},\eta,\nu,u)$ are bounded functions.
Moreover, the matrix $D$ has all its eigenvalues in the left-half
complex plane, consequently, the dynamics of the estimation error
converges asymptotically to zero for any $L_{0}>L_{0}^{*}>0$, and
$(\hat{z}_{1},\hat{\eta})\rightarrow ({z}_{1},\eta)$. Therefore,
the closed loop system (\ref{eqex1}),
(\ref{eqest1})-(\ref{eqex-u2}) is asymptotically stable for
$L_{0}>L_{0}^{*}>0$.
\bigskip
\bigskip

\part{CONCLUSION}

\chapter{Final conclusion}
}
For the first part of this thesis the main original result is an
efficient factorization method for second order ordinary
differential equations (ODE) with polynomial nonlinearities. This
method allows us to find kink particular solutions for
reaction-diffusion (RD) equations and anharmonic oscillator
equations. In addition, application of SUSYQM-type factorization
techniques allows to find a pair of travelling wave solutions for
different RD equations with the same wave velocity. The method is
also applied to more complicated second order nonlinear equations
with interesting results. We believe that this factorization
scheme is easier and more efficient than other employed methods to
find exact particular solutions of second order ODE. Exact
solutions have been found for differential equations with
applications in nonlinear physics and biology, for instance, the
generalized and convective Fisher equations, the Duffing-van der
Pol oscillator equation and the generalized Burgers-Huxley
equation. An application to the biological dynamics of
microtubules (MTs) has been developed as a byproduct of
supersymmetric procedures. Possible interpretation of our results
may be related to the motion of impurities along the MTs or to the
structural discontinuities in the arrangement of tubulin
molecules. Another interesting result, in the context of
applications of supersymmetric factorization procedures in
physical systems, is a complex parametric extension of the
classical harmonic oscillator. This extension is based on a SUSYQM
procedure that has been previously used for Dirac equation in
relativistic particle physics. This result may have applications
in dissipative (absorptive) processes in physical optics as well
as in the physics of cavities. Also, an application to the
chemical physics of diatomic molecules using the same
supersymmetric factorization scheme is included. As a result an
exactly solvable nonhermitic quantum Morse problem is obtained.

\medskip
In the second part of the thesis it was shown that two noiseless
Hodgkin-Huxley (HH) neurons attain synchronized dynamical states
when a feedback action is implemented. Because there exist
uncertain states that cannot be accurately measured in practice
(for instance, the ionic channels activity), a robust approach
that guarantees the synchronization of the HH neurons is
implemented. Numerical results describing the synchronized
behavior of the membrane action potentials of the two neurons are
displayed.



\part{BIBLIOGRAPHY}

\end{document}